\documentclass[10pt,twocolumn,english,prb,aps,showpacs,amsmath,amssymb]{revtex4}
\usepackage{graphicx}
\usepackage[latin9]{inputenc}
\usepackage{url}
\usepackage{babel}
\usepackage{natbib}

\usepackage{dcolumn}
\usepackage{bm}

\begin{document}
\global\long\def\media#1{\left\langle #1\right\rangle }

\title{Exact solution of the 1D Hubbard model in the atomic limit\\with inter-site magnetic coupling}

\author{F. Mancini$^{1,2}$}
\author{E. Plekhanov$^{3}$}
\author{G. Sica$^{1,4,}$}
\email[Email: ]{gerardo.sica@physics.unisa.it}

\affiliation{$^{1}$ Dipartimento di Fisica ``E.R. Caianiello'', Universit\`{a} degli Studi di Salerno, I-84084 Fisciano (SA), Italy}
\affiliation{$^{2}$Unità CNISM di Salerno, Università degli Studi di Salerno, I-84084 Fisciano (Salerno), Italy}
\affiliation{$^{3}$ Consiglio Nazionale delle Ricerche (CNR-SPIN), 67100 L'Aquila, Italy}
\affiliation{$^{4}$ Department of Physics, Loughborough University, Loughborough LE11 3TU, United Kingdom}

\begin{abstract}
In this paper we present for the first time the exact solution in the narrow-band limit of the $1$D extended Hubbard model with nearest-neighbour spin-spin interactions described by an exchange constant $J$. An external magnetic field $h$ is also taken into account. This result has been obtained in the framework of the Green's functions formalism, using the Composite Operator Method. By means of this theoretical background, we have studied some relevant features such as double occupancy, magnetization, spin-spin and charge-charge correlation functions and derived a phase diagram for both ferro ($J>0$) and anti-ferro ($J<0$) coupling in the limit of zero temperature. We also report a study on density of states, specific heat, charge and spin susceptibilities. In the limit of zero temperature, we show that the model exhibits a very rich phase diagram characterized by different magnetic orders and by the coexistence of charge and spin orderings at commensurate filling. Moreover, our analysis at finite temperature of density of states and response functions shows the presence of low-temperature charge and spin excitations near the phase boundaries.
\end{abstract}



\maketitle

\section{Introduction}
For many years the Mott-Hubbard theory has been considered as the archetype for a class of Hamiltonian models aiming the description of correlations between electrons in the \emph{d}-bands of transition metal compounds. However, although the Hubbard model\cite{HubbardModel} and its reduction, the $t$-$J$ model\cite{tJModel}, have been successful in reproducing a plethora of different anomalous phenomena among which magnetic orders\cite{Ferromagnetism1,Ferromagnetism2,AntiFerromagnetism,Paramagnetism}, Metal Insulator Transition (MIT)\cite{MIT}, Spin Density Waves (SDW)\cite{SDW} and High-Temperature Superconductivity (HTSC)\cite{HighTcSC1,HighTcSC2,HighTcSC3}, their application to a large class of compounds is still controversial. The Mott-Hubbard theory, in its simplest formulation, leads to a band gap of the order $U$ ($\sim 7$-$10$eV in oxides) which is difficult to justify for charge-transfer insulators such as Co, Ni and Cu. Furthermore, it is also difficult to understand the metallicity of many sulfides among which NiS, CuS and CoS which would require a strong reduction of the on-site interaction $U$ to $1$-$2$eV \cite{Zaaen1985}.

These and many other issues led to the conviction that single-band Hubbard and $t$-$J$ models are not sufficient to catch all the relevant features due to electronic correlations. Therefore, with the aim of describing strong electron-electron correlations as well as interactions of electrons with other degrees of freedoms (such as lattice vibrations, light), several extensions of the bare Hubbard Hamiltonian have been proposed resulting in the introduction of the so-called Extended Hubbard Models (EHMs). Among these, in the recent years an increasing interest arose in the study of the $t$-$U$-$J$ model as the minimal model capable to reproduce the exchange correlations, widely believed to be the basis of the pairing mechanism in cuprates\cite{Dagotto1992,Rice1995,Arita1998,Japaridze1999,Daul2000}. Unlikely the Hubbard and $t$-$J$ models, in the $t$-$U$-$J$ Hamiltonian the exchange coupling $J$ is not related to the Hubbard $U$ as $J\approx4t^2/U$, allowing finite exchange correlations even in the presence of strong on-site couplings. Furthermore, contrarily to the $t$-$J$ model, an independent treatment of $J$ does not necessarily require the $U\rightarrow\infty$ limit in which charge fluctuations are heavily suppressed.

The $t$-$U$-$J$ model has been analyzed in 1D and 2D cases. In the 2D case the $t$-$U$-$J$ Hamiltonian has been used as the minimal model capable to describe the charge-transfer nature of cuprates. In particular, several studies have been done to understand the influence of the on-site Coulomb repulsion and spin-spin exchange on superconductivity \cite{Laughlin2002,Zhang2003}.

Motivated by the discovery of close proximity of magnetic and superconducting ordering in (TMTSF)$_{2}$X family of quasi one-dimensional Bechgaard salts\cite{BechgaardSalts}, analytical and numerical studies on the $t$-$U$-$J$ model have also been performed in the one-dimensional case. At half-filling, the ground-state phase diagram of the 1D $t$-$U$-$J$ model has been intensively studied for both ferro and anti-ferro magnetic couplings in the weak-coupling limit in which $U,J\ll t$. By means of the bosonization procedure, it has been pointed out that, in the presence of ferromagnetic interactions, the system is dominated by superconducting and spin-density-wave instabilities even in the presence of moderate values of the on-site Hubbard interaction\cite{Japaridze1999}.
On the contrary, in the presence of an anti-ferromagnetic exchange, bosonization and transfer-matrix renormalization group methods showed that at half-filling the ground state of the system is a Mott insulator characterized by spontaneous dimerization for $U\ll J$. A transition to a gapless spin liquid phase occurs at $U_c\approx J/2$ \cite{Dai2004}.
The 1D $t$-$U$-$J$ model at $n=1$ has also been studied with the inclusion of an easy-plane anisotropy in the exchange interaction in order to investigate the coexistence of triplet superconductivity and ferromagnetism in a class of quasi-one-dimensional materials \cite{Dziurzik2004,Ding2010}. It has been shown that, in the large bandwidth limit, magnetic correlations are enhanced by the presence of a repulsive Coulomb potential $U$ and a transverse spin-exchange interaction between electrons on nearest-neighbor (NN) sites. Therefore the coexistence of antiferromagnetism and triplet superconductivity is no longer observed except for small values of the Coulomb interaction \cite{Dziurzik2004}. Recently, it has been also shown that in the weak-coupling limit and for $n=1$, in the presence of an isotropic anti-ferromagnetic exchange, CDW and bond SDW phases are suppressed and the ground state exhibits an insulating behavior characterized by SDW and bond CDW phases\cite{Ding2010}.

Because of the complexity of the model, in spite of numerous attempts, there are no exact solutions for the $t$-$U$-$J$ Hamiltonian and only few results are known outside of the weak coupling regime or away from half-filling.
Within this context we present in this work the exact solution of the $1$D $t$-$U$-$J$ model in the atomic limit and in the presence of an external magnetic field $h$. It is important to stress that the $t$-$U$-$J$ Hamiltonian, even in the atomic limit, does not represent an abstract toy model since a number of quasi-one dimensional insulating compounds, like CsMnBr$_{3}$, Sr$_{3}$ZnIrO, CuGeO$_{3}$, Ca$_{3}$CoRhO$_{6}$ and Sr$_{3}$CuPt$_{x}$Ir$_{1-x}$O$_{6}$ show magnetic features that can be described by the introduction of an inter-site magnetic coupling\cite{CuGeO,CaCoRhO,SrCuPtIrO}. Moreover, because of charge-charge and spin-spin interaction terms, the $U$-$J$ Hamiltonian in the presence of an external magnetic field represents a key model for the study of phase transitions between different magnetic orders or a good starting point for perturbative approaches in terms of the hopping parameter $t$. Besides all the motivations given above, we would like also to stress that the model is interesting from the point of view of statistical mechanics, showing several phase transitions, two tricritical points, anomalous behaviors for all the system response functions near the phase boundaries. Furthermore, as we shall see in Sec.\ref{SubSec:Isomorphism} for the one-dimensional case, the $U$-$J$-$h$ model can also be mapped to a spin ladder Hamiltonian with effective inter-chain and intra-chain spin-spin interactions, allowing for the exact solution of a quasi two-dimensional spin system.

Hereafter we report a summary of the work. In the second Section we introduce the Green's functions formalism and present the exact solution of the model obtained using the Composite Operator Method \cite{Ave1,Ave2} which, in the last fifteen years, has been successfully applied to several models and materials among which: Hubbard \cite{Ave3,Ave4,Ave5,Ave6}, $p$-$d$ \cite{Ave7}, $t$-$J$ \cite{Ave8}, $t$-$t^\prime$-$U$ \cite{Ave9}, extended Hubbard ($t$-$U$-$V$) \cite{Ave10}, Kondo \cite{Ave11}, Anderson \cite{Ave12}, two-orbital Hubbard \cite{Ave13,Ave14}, exact solvable models \cite{Ave15,Ave16,Manc EuroPhysLett}, $J_1$-$J_2$ \cite{Ave17,Ave18}, Hubbard-Kondo \cite{Ave19}, and Cuprates \cite{Ave20}. We show that, by means of algebraic relations, the hierarchy of the equations of motion closes and the Green's functions can be expressed in terms of a finite number of parameters to be determined self-consistently. A collection of results, including single particle correlation functions, charge and spin susceptibilities, density of states and some thermodynamic quantities, obtained in the limit of zero temperature and for finite $T$ , is reported in the third and in the fourth Sections, respectively.

Our study in the limit of zero temperature shows that the model exhibits a very rich phase diagram in which different magnetic orders are generated by the competition between $J$, $h$ and $U$ in the whole range of the filling $n$. In particular, in the presence of an anti-ferromagnetic inter-site coupling, we observe two different types of ferromagnetic order for high values of the magnetic field. Coexistence of charge and spin ordering has also been observed for commensurate fillings. Finally, our study on density of states at $T\rightarrow0$ and thermodynamic quantities at finite temperature shows clear signatures of charge and spin excitations in all the studied thermodynamic functions: specific heat, spin- and charge-susceptibilities.

\section{The model - Composite fields and Green's function}\label{Sec:COM}
A simple generalization of the Hubbard model can be obtained by including
magnetic inter-site interactions. In this case, the Hubbard Hamiltonian
can be written as follows:
\begin{eqnarray}
H & = & \sum_{\boldsymbol{i,j}}\left(t_{\boldsymbol{ij}}-\mu\delta_{\boldsymbol{ij}}\right)c^{\dagger}(i)c(j)+U\sum_{\boldsymbol{i}}n_{\uparrow}(i)n_{\downarrow}(i)+\nonumber \\
 &  & +\frac{1}{2}\sum_{\boldsymbol{i}\neq\boldsymbol{j}}J_{\boldsymbol{i},\boldsymbol{j}}n_{3}(i)n_{3}(j)-h\sum_{\boldsymbol{i}}n_{3}(i)\;,\label{eq:Hubbard_MagnInExtField}
\end{eqnarray}
where $c(i)$ and $c^{\dagger}(i)$ are annihilation and creation
operators of electrons in the spinorial notation: $c^{\dagger}(i)=\left(\begin{array}{cc}
c_{\uparrow}^{\dagger}(i) & c_{\downarrow}^{\dagger}(i)\end{array}\right)$, satisfying canonical anti-commutation relations. The spinorial notation
will be used for all fermionic operators. The Heisenberg picture is used
$[i=(\boldsymbol{i},t)]$, $\boldsymbol{i}$ is a vector of the lattice;
$t_{\boldsymbol{i},\boldsymbol{j}}$ denotes the transfer integral
and describes hopping between different sites; $\mu$ is the chemical
potential. $n_{\sigma}(i)=c_{\sigma}^{\dagger}(i)c_{\sigma}(i)$ is
the number density operator of electrons at the site $\boldsymbol{i}$
with spin $\sigma$. The intensity of the local Coulomb interaction
is parameterized by $U$; $n_{3}(i)=n_{\uparrow}(i)-n_{\downarrow}(i)$
is the third component of the spin density operator; $J_{\boldsymbol{i},\boldsymbol{j}}$
is the exchange inter-site interaction; $h$ represents the strength
of the external magnetic field. In this work we restrict our analysis
to the narrow-band limit and consider only first neighbor interactions
by taking $J_{\boldsymbol{i},\boldsymbol{j}}=-2dJ\alpha_{\boldsymbol{i},\boldsymbol{j}}$,
where $d$ is the dimensionality of the system and $\alpha_{\boldsymbol{i},\boldsymbol{j}}$
is the projection operator on the NN sites. For a $d$-dimensional
cubic Bravais lattice of lattice constant $a$, the Fourier transform
of $\alpha_{\boldsymbol{i},\boldsymbol{j}}$ is $F.T.\left[\alpha_{\boldsymbol{i},\boldsymbol{j}}\right]=\frac{1}{d}\sum_{n=1}^{d}\cos(k_{n}a)$.
Then, the Hamiltonian (\ref{eq:Hubbard_MagnInExtField}) can be written
under the form:
\begin{equation}
H=\sum_{\boldsymbol{i}}\biggl[-\mu n(i)+UD(i)-hn_{3}(i)-dJn_{3}(i)n_{3}^{\alpha}(i)\biggr]\;,\label{eq:Hubbard_Intersite+StrCoupl+Nn}
\end{equation}
where $n(i)=c^{\dagger}(i)c(i)$ is the total density operator and
$D(i)=n_{\uparrow}(i)n_{\downarrow}(i)$$=$$\frac{1}{2}n(i)[n(i)-1]$ the double occupancy operator.
Hereafter, for a generic operator $\Phi(i)$ we use the following
notation: $\Phi^{\alpha}(i)\equiv\sum_{\boldsymbol{j}}\alpha_{\boldsymbol{ij}}\Phi(\boldsymbol{j},t)$.
We note that (\ref{eq:Hubbard_Intersite+StrCoupl+Nn}) is invariant
under the transformation: $\left(h\rightarrow-h\;,\; n_{\uparrow}\rightarrow n_{\downarrow}\right)$.
Also, under the particle-hole transformation, the chemical potential
scales as $\mu(2-n)=U-\mu(n)$.

To solve the Hamiltonian (\ref{eq:Hubbard_Intersite+StrCoupl+Nn}),
we shall use the formalism of Green's functions and equations of motion
\cite{Ave1}. As a first step, we show that there exists
a closed set of eigenoperators and eigenvalues of $H$. To this end,
we introduce the composite field operators:
\begin{equation}
\psi_{p}^{(\xi)}(i)=\xi(i)[n_{3}^{\alpha}(i)]^{p-1}\;,\;\psi_{p}^{(\eta)}(i)=\eta(i)[n_{3}^{\alpha}(i)]^{p-1}\;,
\end{equation}
where $\xi(i)=[1-n(i)]c(i)$ and $\eta(i)=n(i)c(i)$ are the Hubbard
operators responsible for the transitions $|0\rangle_{i}\leftrightarrow|\sigma\rangle_{i}$
at the site $\boldsymbol{i}$, and $|\sigma\rangle_{i}\leftrightarrow|\uparrow\downarrow\rangle_{i}$, at site $\boldsymbol{i}$, respectively.
These fields satisfy the equations of motion:
\begin{equation}
\begin{array}{cl}
i\frac{\partial}{\partial t}\psi_{p}^{(\xi)}(i) & =\left[\psi_{p}^{(\xi)}(i),H\right]=\\
 & =-(\mu+h\sigma_{3})\psi_{p}^{(\xi)}(i)-2J\sigma_{3}\psi_{p+1}^{(\xi)}(i)\;,\\
i\frac{\partial}{\partial t}\psi_{p}^{(\eta)}(i) & =\left[\psi_{p}^{(\eta)}(i),H\right]=\\
 & =-(\mu-U+h\sigma_{3})\psi_{p}^{(\eta)}(i)-2J\sigma_{3}\psi_{p+1}^{(\eta)}(i)\;,
\end{array}\label{eq:EqsOfMotion}
\end{equation}
where $\sigma_3$ is the third Pauli matrix. By taking higher-order time derivatives we generate a hierarchy of composite operators. However, on the basis of the algebraic relations:
\begin{equation}
\begin{array}{ll}
n^{(p)}(i)=n(i)+a_{p}D(i) & \quad[n_{3}(i)]^{2p}=n(i)-2D(i)\\
a_{p}=2^{p}-2 & \quad[n_{3}(i)]^{2p+1}=n_{3}(i)\\
D^{p}(i)=D(i) & \quad n(i)n_{3}(i)=n_{3}(i)\\
n^{p}(i)D(i)=2^{p}D(i) & \quad D(i)n_{3}(i)=0
\end{array}\;,\label{eq:AlgebraicRelations}
\end{equation}
for $p\geqslant1$, the following recursion formula can be established for the field $[n_{3}^{\alpha}(i)]^{p}$:
\begin{equation}
\begin{array}{l}
[n_{3}^{\alpha}(i)]^{2p-1}=\sum_{m=1}^{2d}A_{m}^{(p)}[n_{3}^{\alpha}(i)]^{2m-1}\;,\\
{}[n_{3}^{\alpha}(i)]^{2p}=\sum_{m=1}^{2d}A_{m}^{(p)}[n_{3}^{\alpha}(i)]^{2m}\;,
\end{array}\label{eq:RecursionRules}
\end{equation}
where $A_{m}^{(p)}$ are rational numbers, satisfying the sum rule
$\sum_{m=1}^{4d}A_{m}^{(p)}=1$, and, for $1\leqslant p\leqslant m$, $A_{m}^{(p)}=\delta_{pm}$.
For $p>m$ the expressions of the coefficients $A_{m}^{(p)}$ depend
on the coordination number $z=2d$. In the Appendix \ref{AppendixA} we report some
values.

We now define the composite operator:
\begin{equation}
\psi(i)=\left(\begin{array}{c}
\psi_{\uparrow}^{(\xi)}(i)\\
\psi_{\uparrow}^{(\eta)}(i)\\
\psi_{\downarrow}^{(\xi)}(i)\\
\psi_{\downarrow}^{(\eta)}(i)
\end{array}\right)\;,
\end{equation}
where $\psi_{\sigma}^{(\xi)}(i)$ and $\psi_{\sigma}^{(\eta)}(i)$
are multiplet operators of rank $2z+1$:
\begin{equation}
\begin{array}{l}
\psi_{\sigma}^{(\xi)}(i)=\left(\begin{array}{c}
\xi_{\sigma}(i)\\
\xi_{\sigma}(i)[n_{3}^{\alpha}(i)]\\
\xi_{\sigma}(i)[n_{3}^{\alpha}(i)]^{2}\\
\vdots\\
\xi_{\sigma}(i)[n_{3}^{\alpha}(i)]^{2z+1}
\end{array}\right)\;,\\
\psi_{\sigma}^{(\eta)}(i)=\left(\begin{array}{c}
\eta_{\sigma}(i)\\
\eta_{\sigma}(i)[n_{3}^{\alpha}(i)]\\
\eta_{\sigma}(i)[n_{3}^{\alpha}(i)]^{2}\\
\vdots\\
\eta_{\sigma}(i)[n_{3}^{\alpha}(i)]^{2z+1}
\end{array}\right)\;.
\end{array}
\end{equation}
On the basis of the equations of motion (\ref{eq:EqsOfMotion}) and
by means of the recursion rules (\ref{eq:RecursionRules}), it is
easy to see that these fields are eigenoperators of the Hamiltonian
(\ref{eq:Hubbard_Intersite+StrCoupl+Nn}):
\begin{equation}
\begin{array}{l}
i\frac{\partial}{\partial t}\psi_{\sigma}^{(\xi)}(i)=\left[\psi_{\sigma}^{(\xi)}(i),H\right]=\varepsilon_{\sigma}^{(\xi)}\psi_{\sigma}^{(\xi)}(i)\\
\\
i\frac{\partial}{\partial t}\psi_{\sigma}^{(\eta)}(i)=\left[\psi_{\sigma}^{(\eta)}(i),H\right]=\varepsilon_{\sigma}^{(\eta)}\psi_{\sigma}^{(\eta)}(i)
\end{array}\;,\label{eq:FieldEqs}
\end{equation}
where $\varepsilon_{\sigma}^{(\xi)}$ and $\varepsilon_{\sigma}^{(\eta)}$
are the energy matrices which can be calculated by means of the equations
of motion (\ref{eq:EqsOfMotion}) and the recursion rule (\ref{eq:RecursionRules}), whose eigenvalues, $E_{\sigma}^{(\xi,n)}$ and $E_{\sigma}^{(\eta,n)}$, with $n\in\{1,4d+1\}$, determine the quasiparticle excitation spectrum. Explicit expressions of $\varepsilon_{\sigma}^{(s)}$
and $E_{\sigma}^{(s)}$ ($s=\xi,\eta$) for $z=2$ are given in Appendix \ref{AppendixB}.

The knowledge of a complete set of eigenoperators and eigenvalues
of the Hamiltonian allows for an exact expression of the retarded
Green's function (GF):
\begin{equation}
G_{\sigma}^{(s)}(t-t^\prime)=\media{R\left[\psi_\sigma^{(s)}(\textbf{i},t)\psi_\sigma^{(s)\dagger}(\textbf{i},t^\prime)\right]}
\end{equation}
and consequently, by using the spectral theorem, of the correlation function (CF):
\begin{equation}
C_{\sigma}^{(s)}(t-t^{\,\prime})=\media{\psi_{\sigma}^{(s)}(\boldsymbol{i},t)\psi_{\sigma}^{(s)\dagger}(\boldsymbol{i},t^{\,\prime})}\;.
\end{equation}
In the above equations $s=(\xi,\eta)$, $\media{\dots}$ denotes
the quantum-statistical average over the grand canonical ensemble while $R$ represents the retarded operator with $\langle R[\psi(x)\psi^\dagger(y)]\rangle=\theta(t_x-t_y)\langle\{\psi(x),\psi^\dagger(y)\}\rangle$.
By means of the field equations (\ref{eq:FieldEqs}), this Green's
function satisfies the equation:
\begin{equation}
\left[\omega-\varepsilon_{\sigma}^{(s)}\right]G_{\sigma}^{(s)}(\omega)=I_{\sigma}^{(s)}\;,\label{eq:EqForG}
\end{equation}
where $I_{\sigma}^{(s)}$ is the normalization matrix:
\begin{equation}
I_{\sigma}^{(s)}=\media{\left\{ \psi_{\sigma}^{(s)}(\boldsymbol{i},t),\psi_{\sigma}^{(s)\dagger}(\boldsymbol{i},t)\right\} }\;.
\end{equation}
The solution of Eq.(\ref{eq:EqForG}) gives:
\begin{eqnarray}
 &  & G_{\sigma}^{(s)}(\omega)=\sum_{n=1}^{2z+1}\frac{\rho_{\sigma}^{(s,n)}}{\omega-E_{\sigma}^{(s,n)}+i\delta}\;,\label{eq:GF}\\
 &  & C_{\sigma}^{(s)}(\omega)=\pi\sum_{n=1}^{2z+1}\rho_{\sigma}^{(s,n)}T_{\sigma}^{(s,n)}\delta\left(\omega-E_{\sigma}^{(s,n)}\right)\;,\label{eq:CF}
\end{eqnarray}
where $T_{\sigma}^{(s,n)}=1+\tanh\left(\beta E_{\sigma}^{(s,n)}/2\right)$,
$\beta=1/k_{B}T$ and the spectral density matrices $\rho_{\sigma}^{(s,n)}$
are given by:
\begin{equation}
\rho_{\sigma,ab}^{(s,n)}=\Omega_{\sigma,an}^{(s)}\sum_{c=1}^{2z+1}\left[\Omega_{\sigma,nc}^{(s)}\right]^{-1}I_{\sigma,cb}^{(s)}\;.\label{eq:SpectralDensityMatrix}
\end{equation}
$\Omega_{\sigma}^{(s)}$ is the matrix whose columns are the eigenvectors
of the matrix $\varepsilon_{\sigma}^{(s)}$. Calculations of the matrices
$\Omega_{\sigma}^{(s)}$ are reported in Appendix \ref{AppendixB}. It is worth
noting that $\Omega_{\sigma}^{(\xi)}=\Omega_{\sigma}^{(\eta)}$.
It can be shown that the $\rho_{\sigma}^{(s,n)}$ satisfy the sum
rule $\sum_{n=1}^{2z+1}\rho_{\sigma,ab}^{(s,n)}=I_{\sigma,ab}^{(s)}$.
The relevant elements of $\rho_{\sigma}^{(s,n)}$ are given in Appendix
\ref{AppendixB} for $z=2$. It is important to stress that the expressions (\ref{eq:GF})
and (\ref{eq:CF}) are exact and, in principle, give an
exact solution of the Hamiltonian (\ref{eq:Hubbard_Intersite+StrCoupl+Nn})
\emph{for any dimension of the system}. However, in order to obtain quantitative
information of the physical properties of the system we need to have
a complete knowledge of the Green's and correlation functions. As
(\ref{eq:GF}) and (\ref{eq:CF}) show, both of them depend on
two quantities: the energy matrix and the normalization matrix. The energy matrices are exactly known (see Appendix \ref{AppendixB}). Let us then concentrate
the attention on the normalization matrix. At first, we note that
by using the recursion rules (\ref{eq:RecursionRules}), the matrix
elements $I_{\sigma;ab}^{(s)}$ can be expressed in terms of the elements
of the first raw $I_{\sigma;1,p}^{(s)}$ ($p=1,\dots,2z+1$):
\begin{equation}
\begin{array}{l}
I_{\sigma;1,p}^{(\xi)}=\kappa^{(p)}-\lambda_{-\sigma}^{(p)}\\
I_{\sigma;1,p}^{(\eta)}=\lambda_{-\sigma}^{(p)}
\end{array}\;(p=1,\dots,2z+1)\;,
\end{equation}
where $\kappa^{(p)}=\media{\left[n_{3}^{\alpha}(i)\right]^{p-1}}$
and $\lambda_{\sigma}^{(p)}=\media{n_{\sigma}(i)\left[n_{3}^{\alpha}(i)\right]^{p-1}}$.
Thus, the correlation functions depend on the external parameters
$n=\media{n(i)}$, $T$, $U$, $J$, $h$ and on the internal parameters:
$\mu$, $\kappa^{(p)}$, $\lambda_{\sigma}^{(p)}$. It is easy to
show that the CFs obey the following self-consistent equations:
\begin{equation}
C_{\sigma;1,p}^{(\xi)}+C_{\sigma;1,p}^{(\eta)}=\kappa^{(p)}-\lambda_{\sigma}^{(p)}\;(p=1,\dots,2z+1)\;,
\end{equation}
where $C_{\sigma}^{(s)}=\media{\psi_{\sigma}^{(s)}(i)\psi_{\sigma}^{(s)\dagger}(i)}$.
The number of these equations is not sufficient to determine all the
internal parameters, and one needs other equations. This problem will be considered in the next Section, where a self-consistent scheme, capable of computing the internal parameters, will be formulated for the one-dimensional case.


\subsection{Self-consistent equations}
The previous analysis shows that the complete solution of the model
requires the knowledge of the parameters $\mu$, $\kappa^{(p)}$,
and $\lambda_{\sigma}^{(p)}$. These quantities may be computed by
using algebra constraints and symmetry requirements.

Let us consider a one-dimensional system and fix one site, say $\boldsymbol{i}$,
at some arbitrary point of the chain. We split the Hamiltonian (\ref{eq:Hubbard_Intersite+StrCoupl+Nn}) as the sum of two terms\cite{Manc EuroPhysLett}:
\begin{equation}
\begin{array}{l}
H\equiv H_{0}^{(i)}+H_{I}^{(i)}\\
H_{I}^{(i)}=-2Jn_{3}(i)n_{3}^{\alpha}(i)
\end{array}\;,
\end{equation}
and introduce the $H_{0}^{(i)}$-representation: the statistical average
of any operator can be expressed as:
\begin{equation}
\media O=\frac{\media{Oe^{-\beta H_{I}(i)}}_{0,i}}{\media{e^{-\beta H_{I}(i)}}_{0,i}}\;,\label{eq:StatisticalAverage}
\end{equation}
where $\media{\dots}_{0,i}$ stands for the thermal average with respect
to the reduced Hamiltonian $H_{0}^{(i)}$: $\media{\dots}_{0,i}=Tr\left\{ \dots e^{-\beta H_{0}(i)}\right\} /Tr\left\{ e^{-\beta H_{0}(i)}\right\} $. Because of translational invariance, hereafter the dependence on the index $\textbf{i}$ will be omitted.
As it is shown in Appendix \ref{AppendixB}, the parameters $\kappa^{(p)}$ and $\lambda_{\sigma}^{(p)}$
can be written as functions of three parameters $X_{1}=\media{n^{\alpha}(i)}_{0}$,
$X_{2}=\media{n_{3}^{\alpha}(i)}_{0}$ and $X_{3}=\media{D^{\alpha}(i)}_{0}$,
in terms of which one finds a solution of the model. In the $H_0$-representation, by exploiting
the translational invariance along the chain:
\begin{equation}
\begin{array}{l}
\media{n(i)}=\media{n^{\alpha}(i)}\\
\media{n_{3}(i)}=\media{n_{3}^{\alpha}(i)}\\
\media{D(i)}=\media{D^{\alpha}(i)}
\end{array}\;,\label{eq:TranslationalInvariance}
\end{equation}
one obtains three equations, reported in Appendix \ref{AppendixB}: \eqref{eq:SCeq1}-\eqref{eq:SCeq3}, allowing to determine $X_{1}$, $X_{2}$ and $X_{3}$ in terms of $\mu$. The chemical potential is fixed by means of the equation $\langle n(i)\rangle=n$ that gives:
\begin{eqnarray}
F_{4} & \equiv & (1-n)\langle e^{-\beta H_{I}^{(i)}}\rangle_{0}+G_{1}-1=0\;,\label{eq:SCequationForN}
\end{eqnarray}
where $n$ is the particle number per site, and
\begin{eqnarray}
\langle e^{-\beta H_{I}^{(i)}}\rangle_0 & = & 1+2aG_{2}X_{2}\left[1+b(X_{1}-2X_{3})\right]+\nonumber \\
 &  & +(G_{1}-2G_{3})\bigr[a^{2}X_{2}^{2}+2b\left(X_{1}-2X_{3}\right)+\nonumber \\
 &  & +b^{2}(X_{1}-2X_{3})^{2}\bigr]\;.
\end{eqnarray}
The final expressions for $G_1$, $G_2$ and $G_3$ are reported in Appendix \ref{AppendixB}: \eqref{eq:GiCoefficients}. Thus, (\ref{eq:SCeq1})-(\ref{eq:SCeq3}) and (\ref{eq:SCequationForN})
constitute a system of coupled equations ascertaining the four parameters
$\mu$, $X_{1}$, $X_{2}$ and $X_{3}$ in terms of the external parameters
of the model $n$, $U$, $J$, $h$, $T$. Once these quantities are
known, all local properties of the model can be computed. In particular,
the magnetization $m=\frac{1}{2}\media{n_{3}(i)}$ and the double occupancy $D$ are given by:
\begin{eqnarray}
m & = & \frac{1}{2\langle e^{-\beta H_{I}^{(i)}}\rangle_{0}}\{X_{2}+(G_{1}-2G_{3})X_{2}[b+\nonumber\\
 &  & +(a^{2}+b+b^{2})(X_{1}-2X_{3})]+aG_{2}[X_{2}^{2}(1+b)+\nonumber \\
 &  & +(X_{1}-2X_{3})+b(X_{1}-2X_{3})^{2}]\}\;,\\
D & = & \media{D(i)}=\frac{G_{3}}{\langle e^{-\beta H_{I}^{(i)}}\rangle_{0}}\;.
\end{eqnarray}
It is worth noting that from Eqs.\eqref{eq:TranslationalInvariance}
and \eqref{eq:SCeq1} we can derive the following exact relation between
the particle number and the double occupancy
\begin{equation}
1=gD+n\;,
\end{equation}
where $g=(1-G_{1})/G_{3}$. The local charge, spin and double occupancy
correlation functions are given by:
\begin{eqnarray}
\media{n(i)n^{\alpha}(i)} & = & \frac{1}{\langle e^{-\beta H_{I}^{(i)}}\rangle_{0}}\{G_{1}X_{1}+aG_{2}X_{2}[1+X_{1}+\nonumber \\
 &  & +2b(X_{1}-2X_{3})]+(G_{1}-2G_{3})\nonumber \\
 &  & [a^{2}X_{2}^{2}+b(1+X_{1})(X_{1}-2X_{3})+\nonumber \\
 &  & +b^{2}(X_{1}-2X_{3})^{2}]\}\;,\\
\media{n_{3}(i)n_{3}^{\alpha}(i)} & = & \frac{1}{\langle e^{-\beta H_{I}^{(i)}}\rangle_{0}}\{G_{2}X_{2}+G_{2}X_{2}[b+(a^{2}+\nonumber \\
 &  & +b+b^{2})(X_{1}-2X_{3})]+a(G_{1}-2G_{3})\nonumber \\
 &  & [(1+b)X_{2}^{2}++(X_{1}-2X_{3})\nonumber \\
 &  & [1+b(X_{1}-2X_{3})]]\}\;,\\
\media{D(i)D^{\alpha}(i)} & = & \frac{G_{3}X_{3}}{\langle e^{-\beta H_{I}^{(i)}}\rangle_{0}}\;.
\end{eqnarray}
The internal energy per site is given by:
\begin{equation}
E=UD-J\media{n_{3}(i)n_{3}^{\alpha}(i)}-2hm\;.\label{eq:InternalEnergyPerSite}
\end{equation}
Specific heat $C$, charge $\chi_{c}$ and spin $\chi_{s}$ susceptibilities are given
by:
\begin{equation}
C=\frac{dE}{dT}\;,\;\chi_{c}=T\frac{\partial n}{\partial\mu}\quad,\quad\chi_{s}=\frac{\partial m}{\partial h}\;.\label{eq:Susceptibilities_Definition}
\end{equation}

\subsection{Half-filling case}
When $n=1$, equation (\ref{eq:SCequationForN}) gives $G_{1}=1$.
Recalling (\ref{eq:GiCoefficients}):
\begin{equation}
e^{\beta(2\mu+h-U)}=e^{\beta h}\;\Rightarrow\;\mu=U/2\;,
\end{equation}
in agreement with the particle-hole requirement $\mu(2-n)=U-\mu(n)$.

Then, the self-consistent equations (\ref{eq:SCeq1})-(\ref{eq:SCeq3})
take the form:
\begin{eqnarray}
F_{1} & \equiv & (1-X_{1})[1+aG_{2}X_{2}+\nonumber \\
 &  & +b(1-2G_{3})(X_{1}-2X_{3})=0\;,\label{eq:SCeq1(n=00003D1)}\\
F_{2} & \equiv & G_{2}-X_{2}+(1-2G_{3})X_{2}\{2a-b+\nonumber \\
 &  & -\left[b+(a-b)^{2}\right](X_{1}-2X_{3})+\nonumber \\
 &  & +G_{2}\bigl\{ aX_{2}^{2}(a-b-1)+(X_{1}-2X_{3})\cdot\nonumber \\
 &  & \cdot\left[2b-a+b(b-a)(X_{1}-2X_{3})\right]\bigr\}=0\;,\label{eq:SCeq2(n=00003D1)}\\
F_{3} & \equiv & G_{3}-X_{3}\bigl[1+aG_{2}X_{2}+b(1-2G_{3})\cdot\nonumber \\
 &  & \cdot(X_{1}-2X_{3})\bigr]=0\;.\label{eq:SCeq3(n=00003D1)}
\end{eqnarray}
The first equation has the solutions:
\begin{equation}
\begin{array}{l}
X_{1}=1\\
X_{1}=-\frac{(1+aG_{2}X_{2})}{b(1-2G_{3})}+2X_{3}
\end{array}\;.
\end{equation}
For the second solution, substituting in (\ref{eq:SCeq3(n=00003D1)})
we obtain $G_{3}=0$ which is clearly not satisfied. Therefore, for
$n=1$ we have $X_{1}=1$, and we remain with two parameters $X_{2}$,
$X_{3}$ to be determined by means of the equations:
\begin{eqnarray*}
F_{2} & \equiv & G_{2}-X_{2}+(1-2G_{3})X_{2}\{2a-b+\\
 &  & -\left[b+(a-b)^{2}\right](1-2X_{3})+\\
 &  & +G_{2}\left\{ aX_{2}^{2}(a-b-1)+(1-2X_{3})\cdot\right.\\
 &  & \left.\cdot\left[2b-a+b(b-a)(1-2X_{3})\right]\right\} =0\;,\\
F_{3} & = & G_{3}-X_{3}\bigl[1+aG_{2}X_{2}+b(1-2G_{3})\cdot\\
 &  & \cdot(1-2X_{3})\bigr]=0\;.
\end{eqnarray*}

\subsection{Density of states}
By noting that the cross GFs $\media{R\left[\psi^{(\xi)}(\textbf{i},t)\psi^{(\eta)\dagger}(\textbf{i},t')\right]}$
vanish, the electronic density of states (DOS) for spin $\sigma$
is given by
\begin{eqnarray}
N_{\sigma}(\omega) & = & \left(-\frac{1}{\pi}\right)Im\left[G_{\sigma,11}^{(\xi)}(\omega)+G_{\sigma,11}^{(\eta)}(\omega)\right]=\nonumber\\
& = & \sum_{n=1}^{2z+1}\biggl[\rho_{\sigma,11}^{(\xi,n)}\delta\left(\omega-E_{n}^{(\xi)}\right)+\nonumber\\
&  & \qquad+\rho_{\sigma,11}^{(\eta,n)}\delta\left(\omega-E_{n}^{(\eta)}\right)\biggr]\;.\label{eq:DOS}
\end{eqnarray}
Therefore we have:
\begin{equation}
\int_{-\infty}^{+\infty}d\omega N_{\sigma}(\omega)=\sum_{n=1}^{2z+1}\biggl[\rho_{\sigma,11}^{(\xi,n)}+\rho_{\sigma,11}^{(\eta,n)}\biggr]=1\;.
\end{equation}

\subsection{Isomorphism with a two-level Ising model}\label{SubSec:Isomorphism}
Let us consider the following transformation:
\begin{equation}
  n_\uparrow(i)=\frac{1}{2}\left[1+S_1(i)\right]\;,\;n_\downarrow(i)=\frac{1}{2}\left[1+S_2(i)\right]\;.
\end{equation}
From the properties of $n_\uparrow(i)$ and $n_\downarrow(i)$, it is clear that $S_1(i)$ and $S_2(i)$ are two spin variables with $S_{1,2}(i)=\pm1$. According to this transformation the Hamiltonian \eqref{eq:Hubbard_Intersite+StrCoupl+Nn} reads as:
\begin{equation}\small
  \begin{split}
   H&=J_{//}\sum_{\boldsymbol{i}}\Bigl(S_1(i)S^\alpha_1(i)+S_2(i)S^\alpha_2(i)\Bigr)+\\
    &\;+J_{\perp}\sum_{\boldsymbol{i}}S_1(i)S_2(i)+J_{\perp}^\alpha\sum_{\boldsymbol{i}}S_1(i)S^\alpha_2(i)+\\
    &\;-\sum_{\boldsymbol{i}}\left(h_1S_1(i)+h_2S_2(i)\right)\;,
  \end{split}
\end{equation}
and describes two interacting spin chains in the presence of an external magnetic field. Here $J_{//}=-J/4$ is the NN intra-chain interaction, $J_{\perp}=U/4$ and $J_{\perp}^\alpha=J/2$ are NN and next-NN inter-chain interactions, respectively. Importantly, in both chains the spins are coupled with an effective chain-dependent magnetic field $h_n=U/4-\mu/2+(-1)^nh/2$ ($n=1,2$). In particular, at half-filling we have $\mu=U/2$ and an alternating magnetic field $h_n=(-1)^nh/2$.

\subsection{Summary}
Summarizing, in this Section we have shown that the atomic Hubbard
model with magnetic inter-site interactions is in principle exactly
solvable for any dimension of the system by using the Green's function
formalism and the equation of motion method. The central point is the fact that there exists a closed set of eigenoperators of the Hamiltonian (\ref{eq:Hubbard_Intersite+StrCoupl+Nn}) {[}see
Eqs. (\ref{eq:FieldEqs}){]} allowing to determine exact expressions
of the Green's and correlation functions. These expressions are determined
in terms of few local correlators $\kappa^{(p)}$ and $\lambda_{\sigma}^{(p)}$.
For the one-dimensional case, by using algebraic relations of the
relevant operators and symmetry properties, these correlators can
be expressed in terms of the chemical potential $\mu$ and of the
basic parameters $X_{1}$, $X_{2}$, and $X_{3}$. These four quantities
are determined as functions of the external parameters $n$, $U$,
$J$, $h$, $T$ by solving the system of equations (\ref{eq:SCeq1})-(\ref{eq:SCeq3(n=00003D1)})
and (\ref{eq:SCequationForN}). Once these quantities are known, all properties of the model can be computed. In the next Section we shall present a comprehensive study of the zero-temperature phase diagram and report the distribution of the state density by identifying the different phases in the plane ($U$,$h$). We shall also study the behavior of several quantities such as specific heat, charge and spin susceptibilities, and density of states, in the limit of vanishing $T$. A detailed analysis on the temperature dependence of all the aforementioned quantities is reported in Sec.\ref{Sec:TfiniteResults}.

\section{Zero temperature limit}\label{Sec:TzeroResults}
In this Section we present several results obtained in the limit of
zero temperature. Hereafter we will put $|J|=1$ in order to fix the
energy scale and analyze both ferro ($J=1$) and antiferro ($J=-1$)
inter-site magnetic couplings.

\subsection{Phase diagram}
In the limit of zero temperature we expect to find charge and spin long-range orders. To characterize all the possible configurations, we have solved numerically the set of equations \eqref{eq:SCeq1}-\eqref{eq:SCeq3}, \eqref{eq:SCequationForN} and studied the trends of each relevant correlation function depending on the values of $n$, $J$, $h$ and $U$. For the sake of simplicity, due to the invariance under $h\rightarrow-h$, $n_{\uparrow}\rightarrow n_{\downarrow}$ transformation, we restrict our analysis to $h\geqslant0$ . One can distinguish two different cases: $J=1$, $J=-1$.

\subsubsection{Ferromagnetic inter-site coupling}
\begin{figure}[htp]
\begin{centering}
\includegraphics[width=0.5\columnwidth]{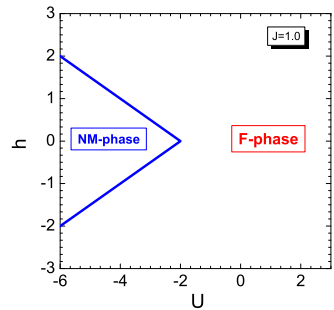}
\par\end{centering}
\caption{(Color online) $2$D phase diagrams for $J=1$ at $T=0$.\label{fig:PhDiagJp}}
\end{figure}

The phase diagram at zero temperature for positive values of $J$ is shown in Fig. \ref{fig:PhDiagJp}, where we consider the ($U$,$h$) plane. Two different phases can be identified.

\begin{figure}[htp]
\begin{centering}
\includegraphics[width=1\columnwidth]{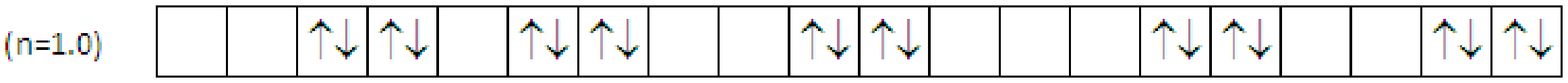}
\par\end{centering}
\caption{One of the possible spin and charge configurations for the NM-phase
at $T=0$, $J=1$. $\uparrow$and $\downarrow$ represent the two
possible spin states.\label{fig:NM_configuration}}
\end{figure}

\textit{NM-phase}. This phase, which we call \textquotedblleft{}non-magnetic\textquotedblright{}
(NM) phase, is observed for $U<U_{c}$ where $U_{c}=-2(J+h)$ and
for $0\leqslant n\leqslant2$. It is characterized by the absence of magnetization
and spin-spin correlations; there is no dependence on the external
magnetic field $h$. The attractive local potential $U$ prevails
on both the magnetic field and the ferromagnetic coupling; there are
no sites singly occupied and the double occupancy $D$ is an increasing
function of the filling. The chemical potential depends only on $U$;
the electrons are not correlated and the two-site charge and double
occupancy correlation functions are the squares of the corresponding
one-site ones:

\begin{equation}
\begin{array}{lcl}
\mu=U/2 &  & \langle n(i)n^{\alpha}(i)\rangle=n^{2}\\
\langle D(i)\rangle=n/2 &  & \langle n_{3}(i)n_{3}^{\alpha}(i)\rangle=0\\
\langle m(i)\rangle=0 &  & \langle D(i)D^{\alpha}(i)\rangle=(n/2)^{2}
\end{array}\;,\;\forall\;0\leqslant n\leqslant2\;.\label{eq:Features_NM(Jpositive)}
\end{equation}
Recalling \eqref{eq:InternalEnergyPerSite}, it is immediate to see
that the internal energy per site has the value $E_{NM}=nU/2$. A
typical configuration occurring in this phase is a mixture of doubly
occupied and empty sites and is shown in Fig. \ref{fig:NM_configuration}.

\begin{figure}[htp]
\begin{centering}
\includegraphics[width=1\columnwidth]{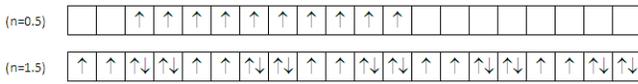}
\par\end{centering}
\caption{Some of the possible spin and charge configurations for the F-phase
at $T=0$, $J=1$.\label{fig:F_configuration}}
\end{figure}

\textit{F-phase.} This phase, called \textquotedblleft{}ferromagnetic\textquotedblright{}
(F) phase, is observed for $U>U_{c}$ and for $0\leqslant n\leqslant2$. It
is characterized by a dominant ferromagnetic order and finite values
of magnetization and spin-spin correlation functions. In the region
$0\leqslant n\leqslant1$ the double occupancy is zero; all the spins are aligned
and the magnetization reaches its saturation value $\langle m(i)\rangle=n/2$;
the spin-spin correlation function is positive and increases with
$n$. The chemical potential does not depend on the filling and is
a function of $J$ and $h$:
\begin{equation}
\begin{array}{lll}
\mu=-J-h & \langle n(i)n^{\alpha}(i)\rangle=n\\
\langle D(i)\rangle=0 & \langle n_{3}(i)n_{3}^{\alpha}(i)\rangle=n & \;,0\leqslant n\leqslant1\;.\\
\langle m(i)\rangle=n/2 & \langle D(i)D^{\alpha}(i)\rangle=0
\end{array}\label{eq:FeaturesFphase1}
\end{equation}
At $n=1$ the chemical potential exhibits a discontinuity and jumps
at $\mu=U/2$ . In the region of
filling $n>1$ the chemical potential takes the value $\mu=J+h+U$,
in agreement with the scaling law $\mu(2-n)=U-\mu(n)$. The double
occupancy increases linearly with $n$; correspondingly, the magnetization
and the spin-spin correlation function $\langle n_{3}(i)n_{3}^{\alpha}(i)\rangle$
decrease:

\begin{equation}
\begin{array}{lll}
\mu=J+h+U & \langle n(i)n^{\alpha}(i)\rangle=3n-2\\
\langle D(i)\rangle=n-1 & \langle n_{3}(i)n_{3}^{\alpha}(i)\rangle=2-n & \;,1\leqslant n\leqslant2\;.\\
\langle m(i)\rangle=1-n/2 & \langle D(i)D^{\alpha}(i)\rangle=n-1
\end{array}\label{eq:FeaturesFphase2}
\end{equation}
The internal energy has the value:
\begin{equation}\label{eq:Energy_Fphase}
E_{F}=\begin{cases}
-n(J+h) & ,\;0\leqslant n\leqslant1\\
U(n-1)-(2-n)(J+h) & ,\;1\leqslant n\leqslant2
\end{cases}\;.
\end{equation}
A typical configuration occurring in this phase is shown in Fig. \ref{fig:F_configuration}
for the cases $n<1$ and $n>1$. We see that:
\begin{equation}\label{eq:E_NMF}
E_{F}-E_{NM}=\begin{cases}
-n\left(J+h+\frac{U}{2}\right) & ,\;0\leqslant n\leqslant1\\
-(2-n)\left(J+h+\frac{U}{2}\right) & ,\;1\leqslant n\leqslant2
\end{cases}\;,
\end{equation}
therefore, regardless of the specific value of filling $n$, there
is a critical value of the local potential $U_{c}=-2(J+h)$ which
separates the two phases.

\begin{figure}[htp]
\begin{centering}
\includegraphics[width=1\columnwidth]{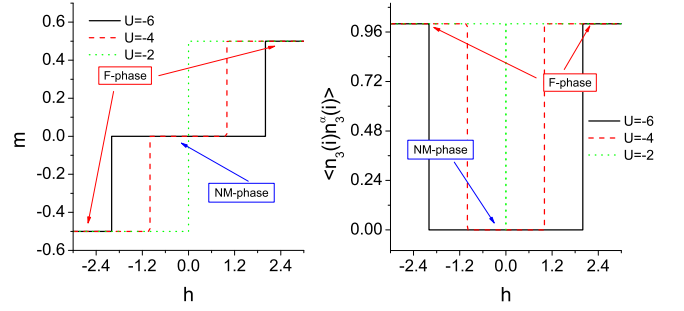}
\par\end{centering}
\caption{(Color online) Signature of NM-F phase transition in magnetization
(left) and spin-spin correlation function (right) plotted as functions
of the magnetic field $h$, for different values of $U$ at $n=1$
and $J=1$.\label{fig:Signature_NM-F(Jp)}}
\end{figure}

Signatures of the NM-F phase transition occurring in magnetization and spin-spin
correlation function are reported in Fig. \ref{fig:Signature_NM-F(Jp)}.

It is immediate to see that for $U<-2$, there is a phase transition
from the NM phase to the F phase at $h=h_{c}=-J-U/2$. According to
\eqref{eq:Features_NM(Jpositive)}, both magnetization and spin-spin
correlation functions are zero for $h<h_{c}$ . Vice-versa, when $h>h_{c}$,
$\langle m(i)\rangle$ and $\langle n_{3}(i)n_{3}^{\alpha}(i)\rangle$
assume finite values according to \eqref{eq:FeaturesFphase1}-\eqref{eq:FeaturesFphase2}.
For $U\geqslant-2$ instead, there is no phase transition: the system is
always in the F phase. In particular, for zero field there is a magnetic
order: the spin-spin correlation function is equal to $n$ while the
magnetization jumps from $-n/2$ to $n/2$ crossing $h=0$.

\subsubsection{Antiferromagnetic inter-site coupling}
\begin{figure}[htp]
\centering{}\includegraphics[width=1\columnwidth]{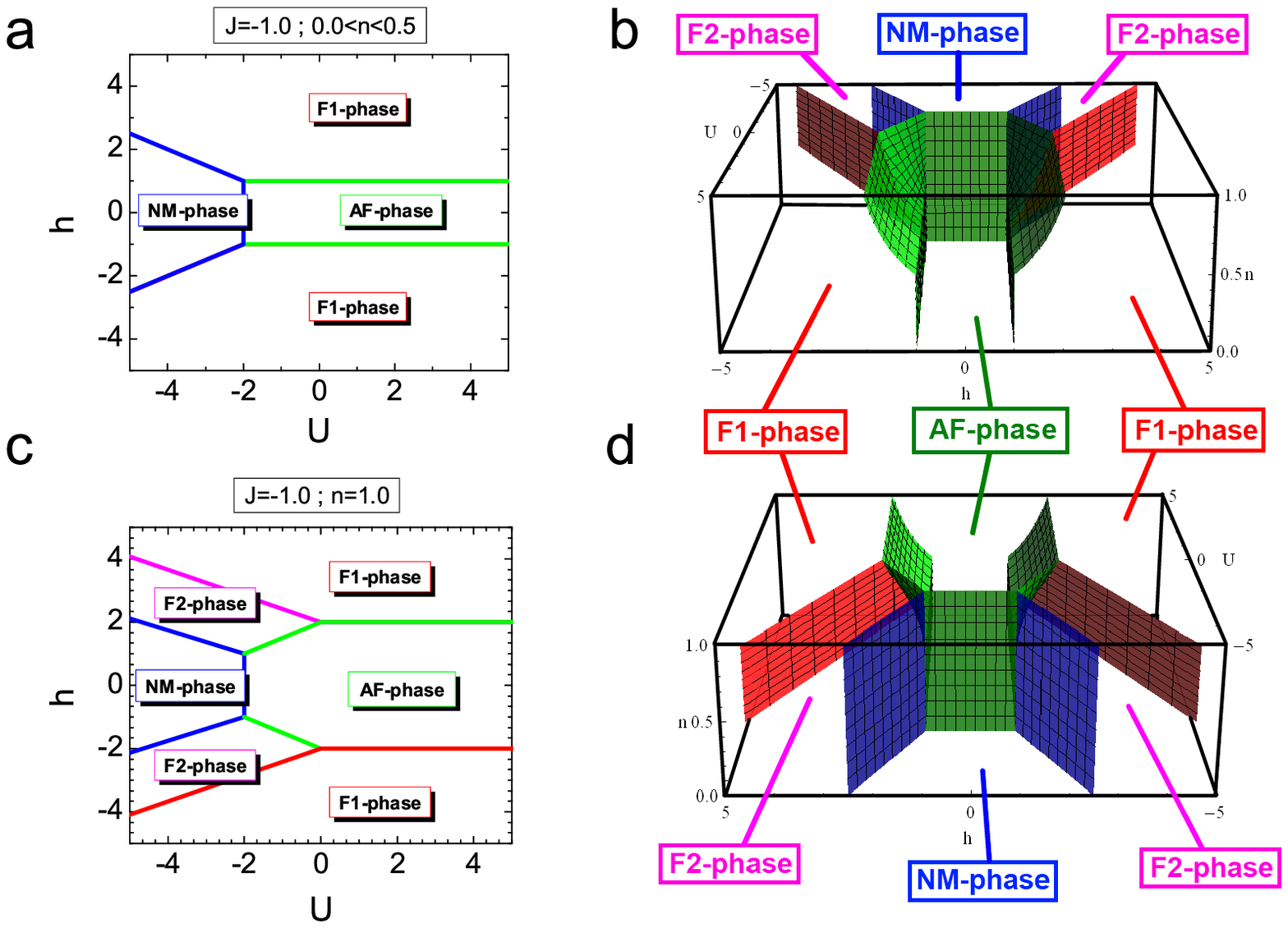}
\caption{(Color online) $2$D and $3$D phase diagrams for $J=-1.0$ at $T=0.$\label{fig:PhDiag_hU(Jn)}}
\end{figure}

The phase diagram at zero temperature for negative values of $J$ is shown in Fig. \ref{fig:PhDiag_hU(Jn)}, where we consider the $2$D ($U$,$h$) and the $3$D ($U$,$h$,$n$)
cases.

\begin{figure}[htp]
\centering{}\includegraphics[width=1\columnwidth]{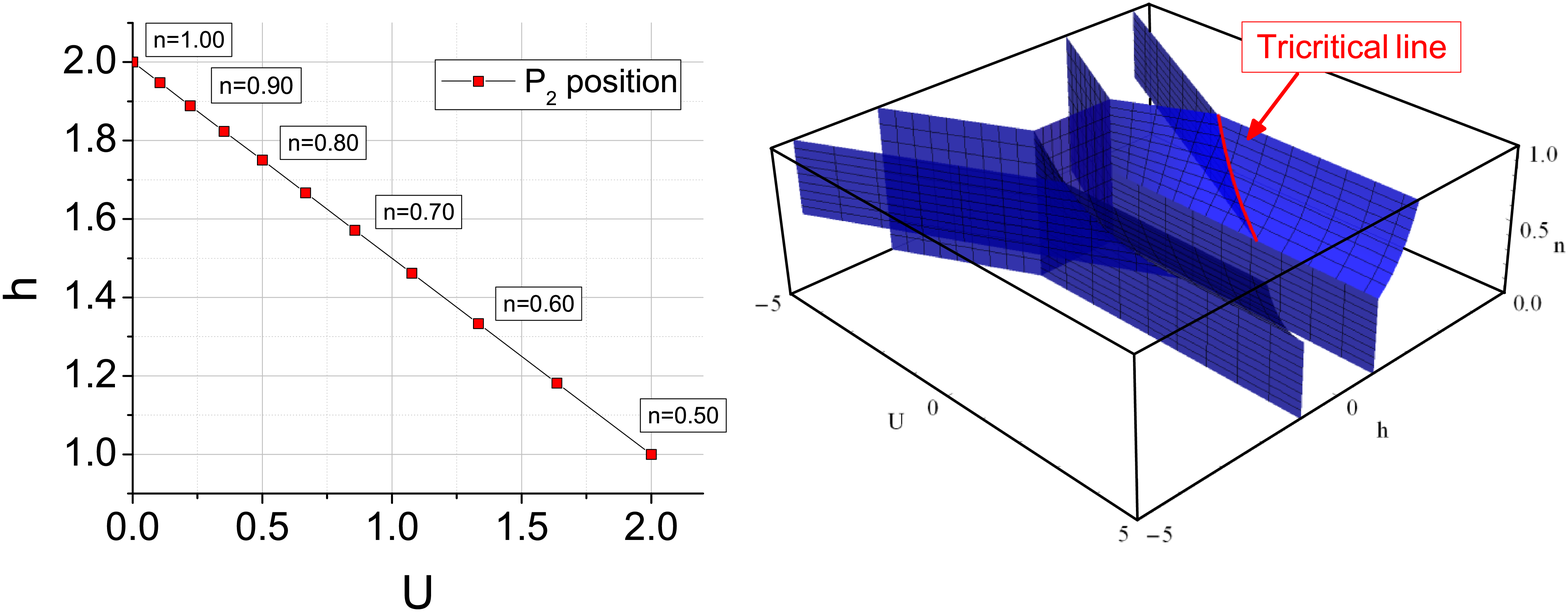}
\caption{(Color online) Position of $P_{2}$ 3-critical point in the h-U plane
(left) and in the 3D phase diagram (right).\label{fig:3criticalPoint}}
\end{figure}

In the regions $0\leqslant n\leqslant0.5$ and $1.5\leqslant n\leqslant2$ the phase diagram does not depend on $n$ (as shown in Figs. \ref{fig:PhDiag_hU(Jn)}$a$
and \ref{fig:PhDiag_hU(Jn)}$b$) and is characterized by three different
phases which join at the tri-critical point $P_{1}$=$\left\{ U=-2,h=1\right\} $.
In the region $0.5\leqslant n\leqslant 1.5$ the phase diagram depends on n and is characterized by four different phases and two tricritical points: $P_1$ and $P_2$, as shown in Figs. \ref{fig:PhDiag_hU(Jn)}$c$ and \ref{fig:PhDiag_hU(Jn)}$d$. It is worth noting that while $P_1$ is fixed, the position of $P_2$ in the $U$-$h$ plane changes as a function of $n$, as reported in Fig. \ref{fig:3criticalPoint}. We can also note that, changing the filling, $P_1$ and $P_2$ never coincide and remain always well separated from each other.

\textit{NM-phase.} Also for negative values of $J$ a \textquotedblleft{}non-magnetic\textquotedblright{} (NM) phase is observed, characterized by the absence of magnetization and spin-spin correlations. All the features of this
phase have already been reported in Eqs.\eqref{eq:Features_NM(Jpositive)}.

\textit{AF-phase.} For $U>-2J$ and for low values of the magnetic
field $h$ an antiferromagnetic (AF) order is observed in the entire
region of filling. This phase is characterized by the absence of magnetization and negative values of spin-spin correlation function $\langle n_{3}(i)n_{3}^{\alpha}(i)\rangle$:



\begin{equation}
  \begin{tabular}{c|c|c|c}
    & $0\leqslant n<1$ & $n=1$ & $1<n\leqslant2$ \\
    \hline
    $\mu$                                & $J$ & $U/2$ & $U-J$ \\
    $\langle m(i)\rangle$                & $0$ & $0$   & $0$   \\
    $\langle D(i)\rangle$                & $0$ & $0$   & $n-1$ \\
    $\langle n(i)n^\alpha(i)\rangle$     & $n$ & $1$   & $3n-2$\\
    $\langle n_3(i)n_3^\alpha(i)\rangle$ & $-n$& $-1$  & $n-2$ \\
    $\langle D(i)D^\alpha(i)\rangle$     & $0$ & $0$   & $n-1$ \\
    $E_{AF}$                             & $nJ$& $J$   & $U(n-1)-J(n-2)$\\
  \end{tabular}
\end{equation}

\begin{figure}[htp]
\begin{centering}
\includegraphics[width=1\columnwidth]{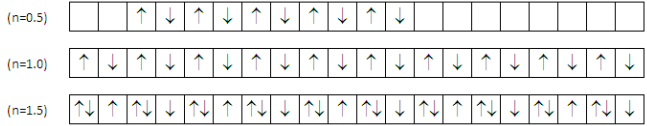}
\par\end{centering}
\caption{Some of the possible spin and charge configurations for the AF-phase
at $T=0$, $J=-1$.\label{fig:AF_configuration}}
\end{figure}

For $0\leqslant n\leqslant1$, as the filling increases, the electrons start
to singly occupy the available empty sites aligning their spins opposite
with respect to their nearest neighbors; correspondingly, the magnetization
and the double occupancy are zero. The half-filling configuration corresponds to a perfect
Neel state. For $n>1$, by increasing the filling, some of the sites
become doubly occupied, while the magnetization remains zero. A typical configuration which takes place
in this phase is shown in Fig. \ref{fig:AF_configuration} for the
cases $n=0.5$, $n=1$ and $n=1.5$.

\textit{F1-phase.} When the external magnetic field is strong enough
to dominate with respect to $J$, a ferromagnetic behavior (F1) is
observed despite of the presence of an anti-ferromagnetic coupling.
In this phase, the effect of the magnetic field induces the spins
to be all aligned and the magnetization reaches its saturation value.
Due to the presence of an anti-ferromagnetic coupling, contrarily to what has been said for the F-phase, NN spin-spin
and charge-charge correlation functions remain zero until $n=0.5$. At $n=0.5$ a charge order state is observed with
a checkerboard structure.

\begin{equation}\small
  \begin{tabular}{c|c|c|c}
    & $0\leqslant n\leqslant0.5$                   & $0.5\leqslant n\leqslant1.0$ & $n=1$ \\
    \hline
    $\mu$                                & $h$ & $h-2J$  & $U/2$ \\
    $\langle m(i)\rangle$                & $n/2$ & $n/2$ & $1/2$ \\
    $\langle D(i)\rangle$                & $0$ & $0$    & $0$ \\
    $\langle n(i)n^\alpha(i)\rangle$     & $0$ & $2n-1$    & $1$\\
    $\langle n_3(i)n_3^\alpha(i)\rangle$ & $0$& $2n-1$   & $1$ \\
    $\langle D(i)D^\alpha(i)\rangle$     & $0$ & $0$    & $0$ \\
    $E_{F1}$                             & $-nh$& $-J(2n-1)-nh$ & $-J-h$\\
  \end{tabular}
\end{equation}
\begin{equation}\small
  \begin{tabular}{c|c|c}
    & $1<n\leqslant1.5$ & $1.5<n\leqslant2$ \\
    \hline
    $\mu$                                & $-h+2J+U$ & $-h+U$ \\
    $\langle m(i)\rangle$                & $1-n/2$   & $1-n/2$\\
    $\langle D(i)\rangle$                & $n-1$    & $n-1$  \\
    $\langle n(i)n^\alpha(i)\rangle$     & $2n-1$    & $4(n-1)$\\
    $\langle n_3(i)n_3^\alpha(i)\rangle$ & $3-2n$    & $0$ \\
    $\langle D(i)D^\alpha(i)\rangle$     & $0$       & $2n-3$ \\
    $E_{F1}$                             & $U(n-1)-J(3-2n)+$ & $U(n-1)+$\\
                                         & $-h(2-n)$         & $-h(2-n)$\\
  \end{tabular}
\end{equation}



\begin{figure}[htp]
\centering{}\includegraphics[width=1\columnwidth]{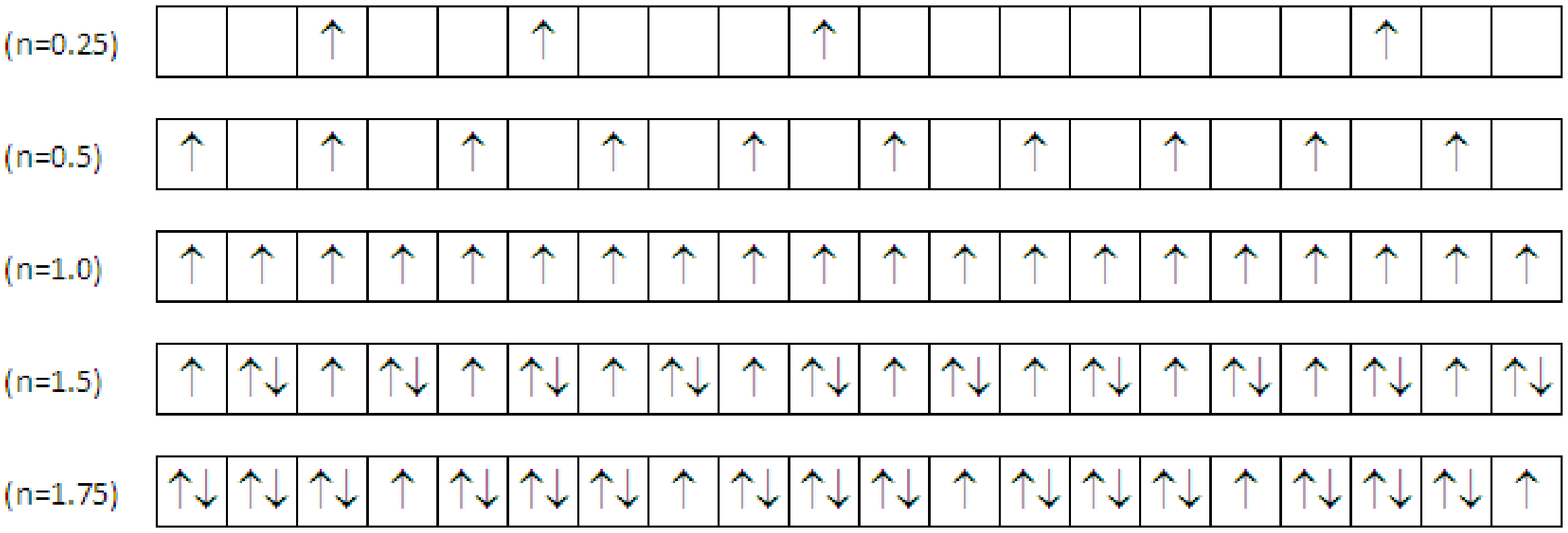}
\caption{Some of the possible spin and charge configurations for the F1-phase
at $T=0$, $J=-1$.\label{fig:F1_configuration}}
\end{figure}

On the contrary, as reported in Fig. \ref{fig:F1_configuration}, for $0.5<n<1.5$
the spin-spin correlation function becomes finite; $\langle n_{3}(i)n_{3}^{\alpha}(i)\rangle$
increases with n up to half-filling, then decreases and vanishes for
$n\geqslant1.5$. For $0.5<n\leqslant1.0$, $D=0$ and
the sites are singly occupied by electrons with parallel spins. Exactly
at half-filling, all the sites are singly occupied with polarized spins, therefore the magnetization assumes its maximum value. For $1.0\leqslant n<1.5$,
the double occupancy becomes finite but the NN correlation
function $\langle D(i)D^{\alpha}(i)\rangle$ is always zero; exactly
at $n=1.5$ another checkerboard structure with a pattern
composed of alternating singly and doubly occupied sites is observed. For $n\geqslant1.5$,
$\langle D(i)\rangle$ and $\langle D(i)D^{\alpha}(i)\rangle$ increase
up to the maximum value, while the magnetization goes to zero as $n\to 2$.

\begin{figure}[htp]
\centering{}\includegraphics[width=1\columnwidth]{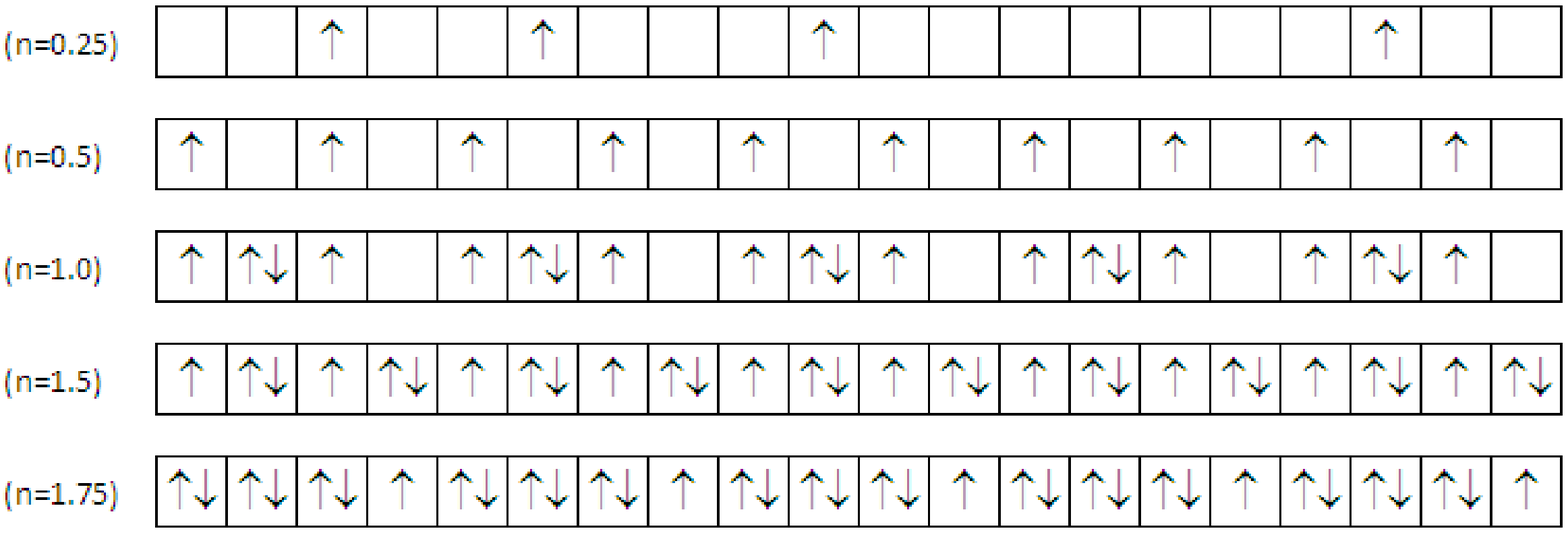}
\caption{Some of the possible spin and charge configurations for the F2-phase
at $T=0$, $J=-1$.\label{fig:F2_configuration}}
\end{figure}

\begin{figure}[htp]
\begin{centering}
\includegraphics[width=1\columnwidth]{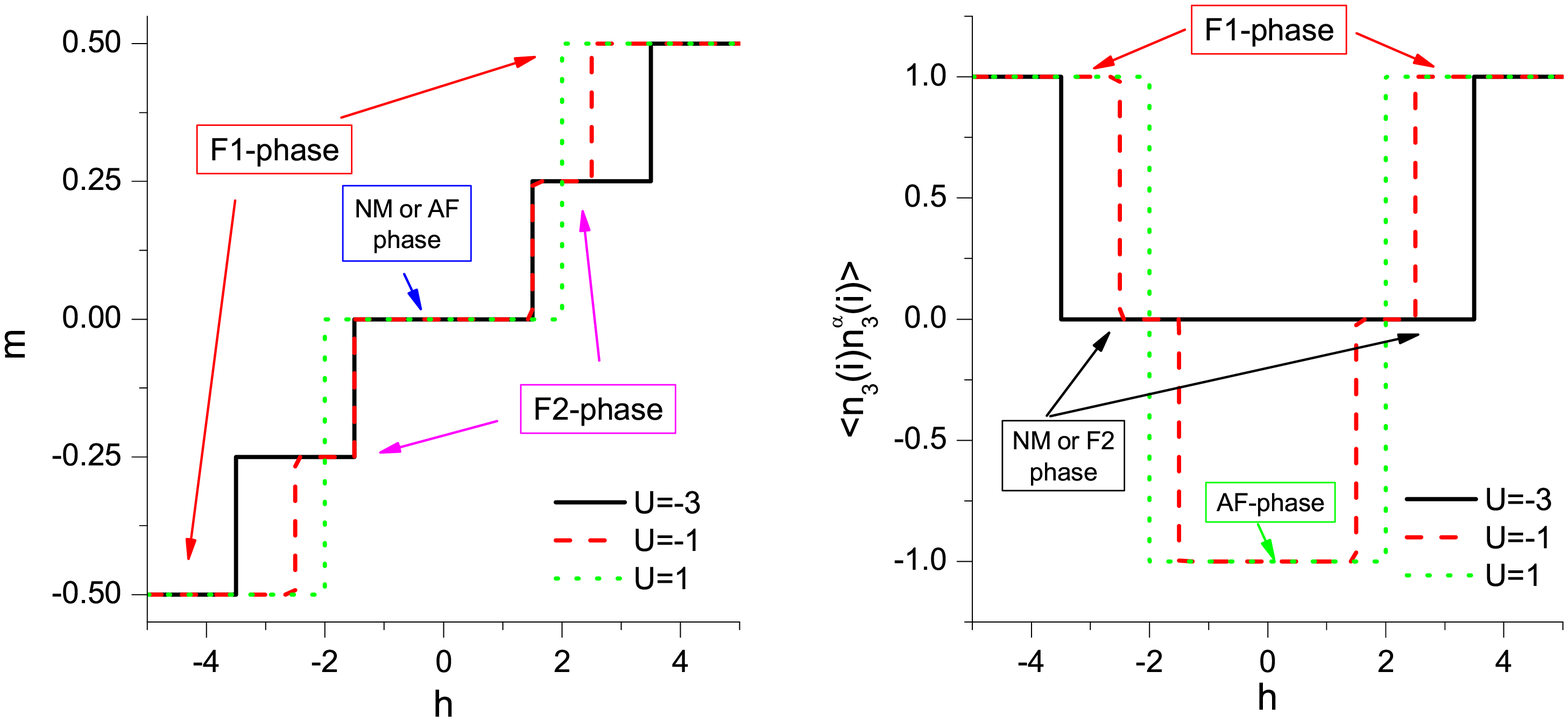}
\par\end{centering}
\caption{(Color online) Signatures of phase transition in magnetization (left)
and spin-spin correlation function (right) plotted as functions of
the magnetic field $h$, for different values of $U$ at $n=1$ and
$J=-1$.\label{fig:Signature_Jn}}
\end{figure}

\textit{F2-phase.} As shown in Fig. \ref{fig:PhDiag_hU(Jn)},
in the range in which $h$ and $J$ are comparable, for $U<0$ and for $0.5<n<1.5$
the model exhibits an anomalous ferromagnetic phase (F2) induced by the competition
between the magnetic field and the antiferromagnetic inter-site coupling.
This phase is characterized by the absence of NN spin-spin correlations
and the presence of a constant magnetization.

\begin{equation}
\small\begin{array}{lllc}
\mu=U/2 &  & \langle n(i)n^{\alpha}(i)\rangle=2n-1\\
\langle D(i)\rangle=n/2-1/4 &  & \langle n_{3}(i)n_{3}^{\alpha}(i)\rangle=0 & \;.\\
\langle m(i)\rangle=1/4 &  & \langle D(i)D^{\alpha}(i)\rangle=0\\
E_{F2}=U\left(\frac{n}{2}-\frac{1}{4}\right)-\frac{h}{2}
\end{array}\normalsize\label{eq:Features_F2}
\end{equation}
As shown in Fig. \ref{fig:F2_configuration}, in this phase both $\langle D(i)\rangle$
and $\langle n(i)n^{\alpha}(i)\rangle$ are finite while $\langle n_{3}(i)n_{3}^{\alpha}(i)\rangle$
remains zero. At $n=0.5$, doubly occupied sites appear in between
two singly occupied ones; by increasing $n$, the number of doubly
occupied sites increases, while the number of singly occupied sites
remains constant. This explains why the magnetization does not change.
The $n=1$ configuration is characterized by a particular pattern in which one or
more clusters, composed of the following pattern: ($\sigma$,$\uparrow\downarrow$,$\sigma$,$0$), are repeated periodically along the chain. Some signatures of NM-F1-F2 and AF-F1-F2 phase transitions occurring in magnetization and spin-spin correlation function are reported in Fig. \ref{fig:Signature_Jn}.

As shown in Fig. \ref{fig:Signature_Jn}, for $U\leqslant-2$,
changing $h$ one crosses three different phases: NM, F1 and F2. In this
range (solid curves) the magnetization assumes three different values:
$0$ in the NM-phase, $1/4$ in the F2-phase and $n/2$ in the F1-phase.
The spin-spin correlation function jumps from zero (in NM and F2
phases) to one (in the F1-phase). Instead, at $-2\leqslant U\leqslant0$ (dashed curves), NM-phase is replaced by AF-phase and the spin-spin correlation
function becomes negative when $|h|\leqslant1.5$. Finally, for positive
values of $U$ (solid curves) the F2-phase is not observed: the magnetization
jumps from zero (AF-phase) to $n/2$ (F1-phase) while the spin-spin
correlation function goes from negative (AF-phase) to positive (F1-phase)
values.

\subsection{Charge and spin excitations}

The presence of different phases and long-range orders at $T\to0$ can also be predicted by looking at charge and spin excitations in system's response functions such as charge and spin susceptibilities. As shown in the previous Subsection, in each phase at $T=0$ all single particle properties and correlation functions depend only on $n$ but not on the model parameters ($U$, $J$, $h$). Therefore charge and spin susceptibilities are expected to have a constant value in all the phases with divergencies or discontinuities localized at the phase boundaries where small variations of the model parameters can imply the transition from a charge/spin ordering to another. In particular, while $\chi_s(T=0)=0$ in each phase, for the charge susceptibility we have instead:

\begin{equation}
  \begin{tabular}{l|ll}
    \textrm{Phase}&$\lim_{T\rightarrow0}\chi_c$&\\
    \hline
    \textrm{NM}&$n(2-n)$&$0\leqslant n\leqslant1$\\
    \textrm{F,AF}&$\infty$&$0<n<1$\\
                 &$0$&$n=0\textrm{, or, }n=1$\\
    \textrm{F1}&$n(n-1)(2n-1)$&$0\leqslant n\leqslant0.5$\\
    \textrm{F2}&$2n(2-n)-3/2$&$0.5\leqslant n\leqslant1$\\
  \end{tabular}
\end{equation}

\begin{figure}[htp]
\begin{tabular}{cc}
\includegraphics[width=0.4\columnwidth]{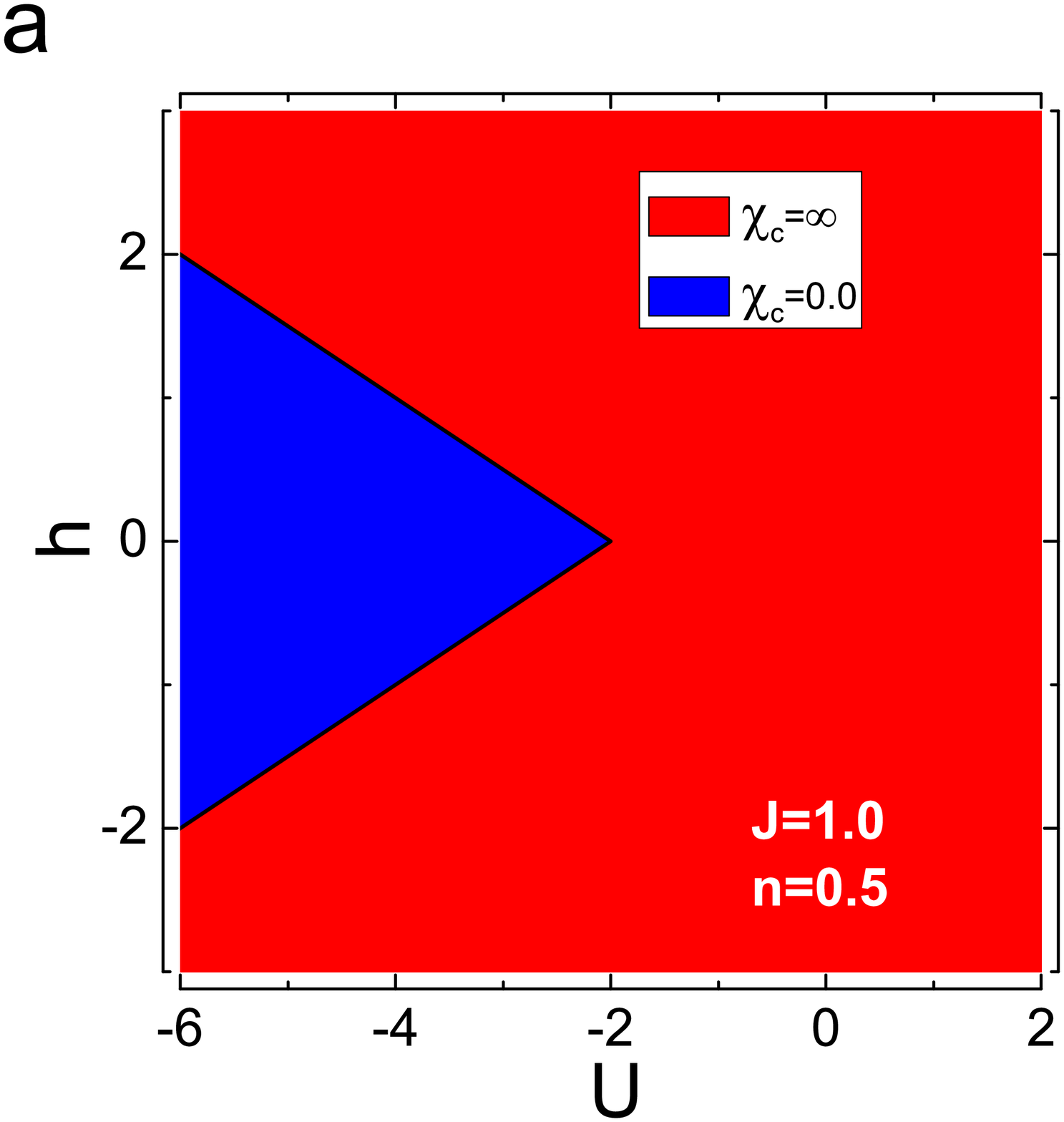}
\includegraphics[width=0.4\columnwidth]{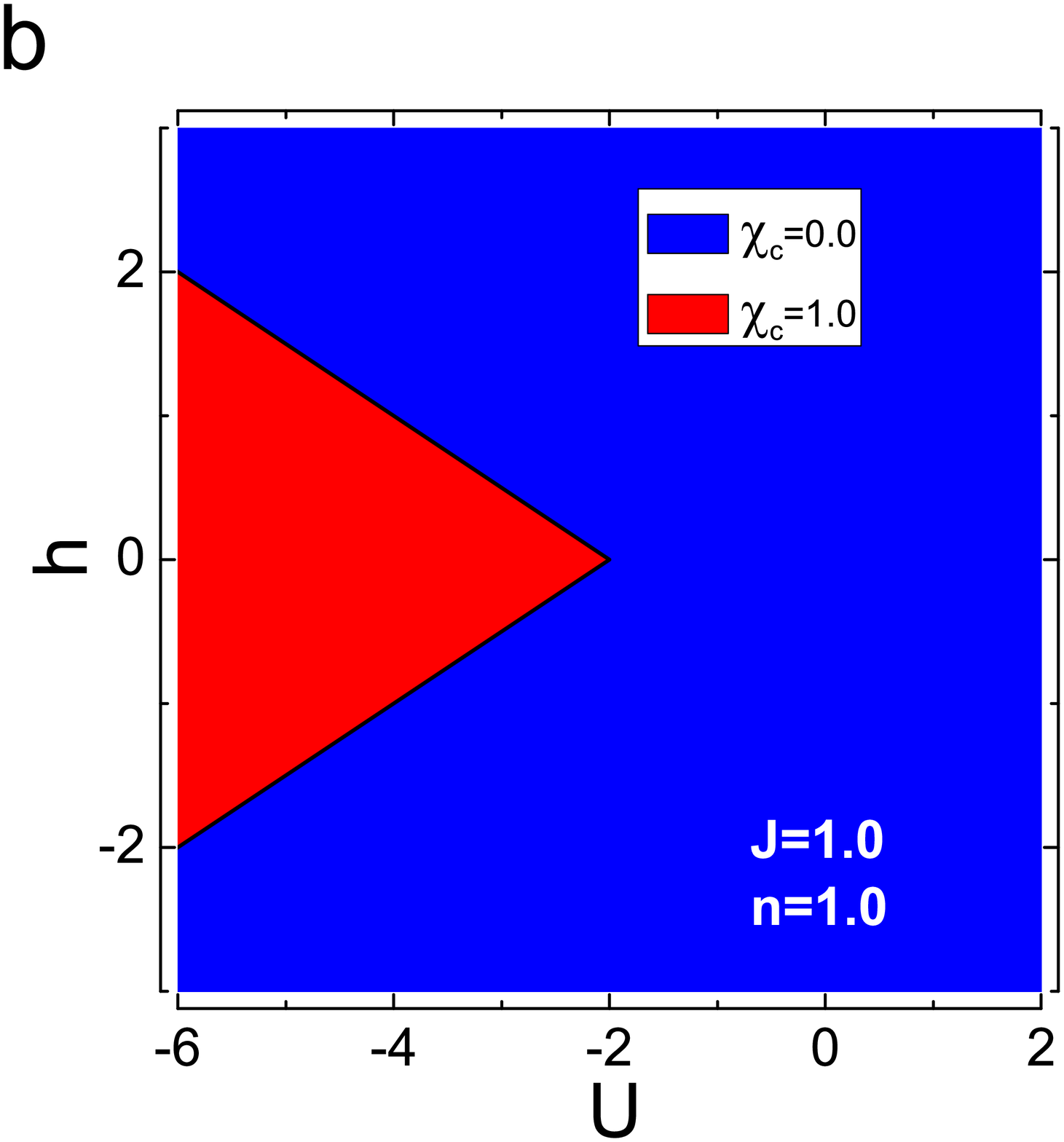}\\
\includegraphics[width=0.4\columnwidth]{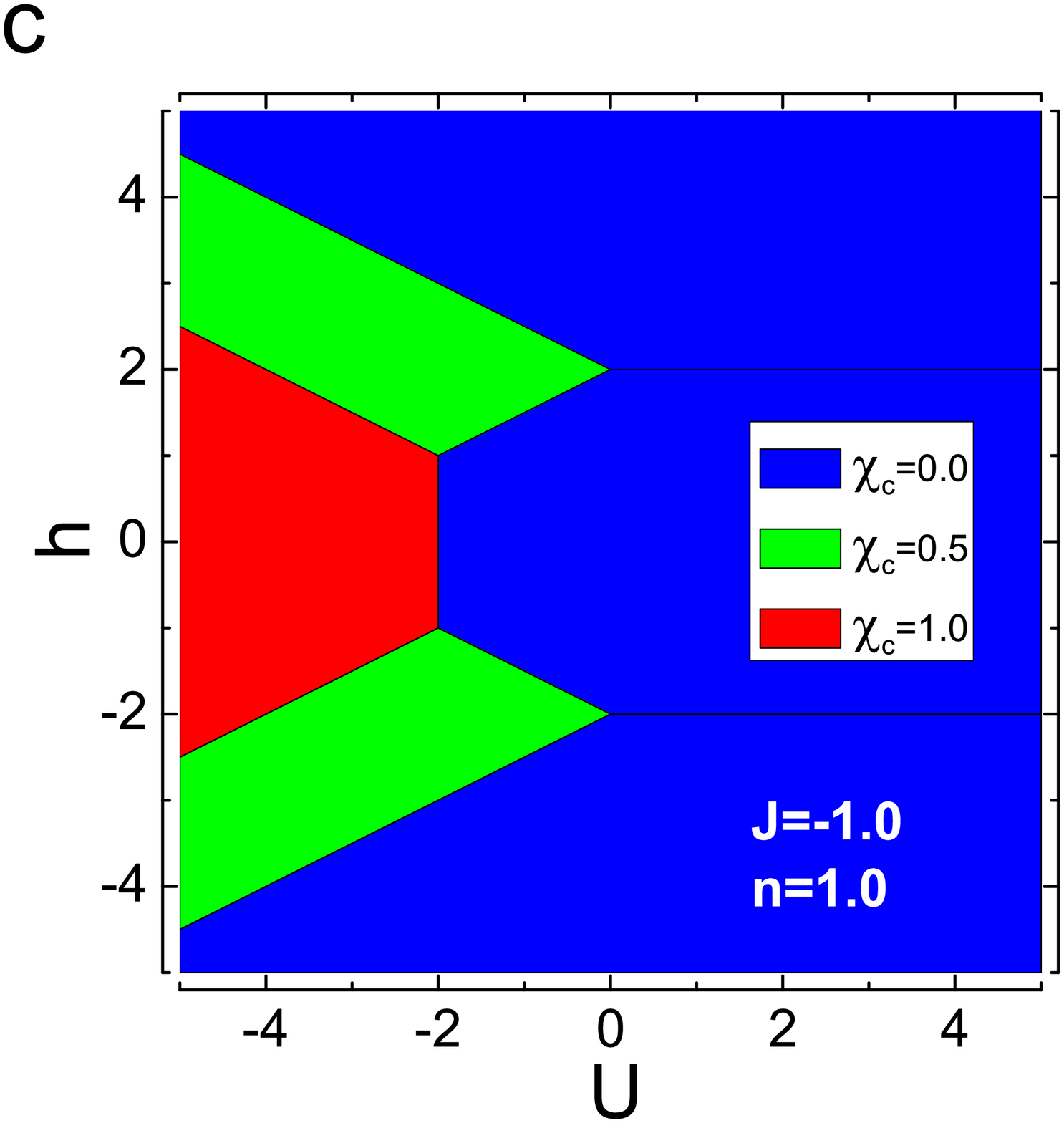}
\includegraphics[width=0.4\columnwidth]{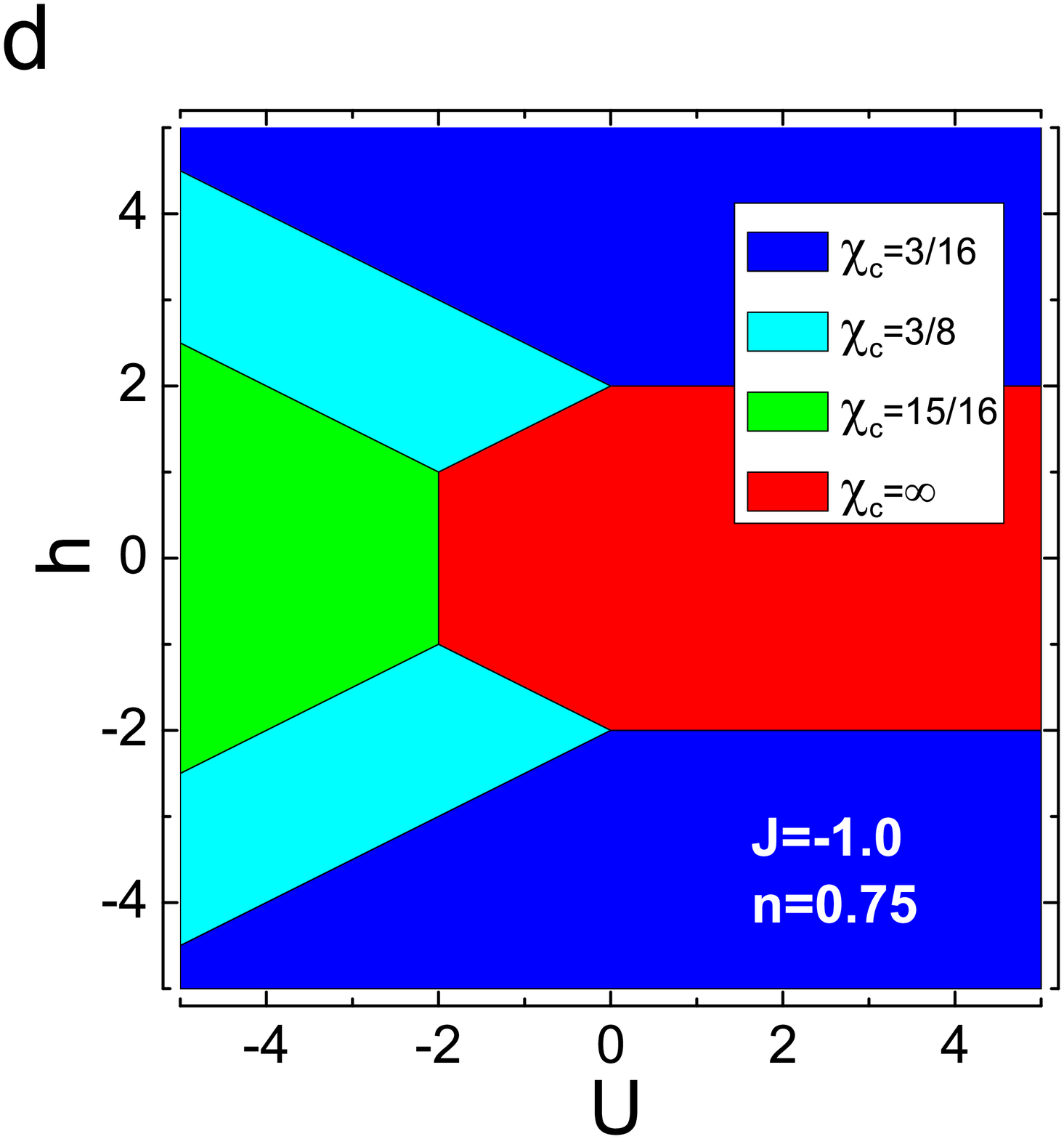}
\end{tabular}
\caption{(Color online) Contour-plot of the charge susceptibility at $T=0.001$
as a function of the external magnetic field $h$ and the local Coulomb
potential $U$ for: $(a)$ $J=1$, $n=0.5$; $(b)$ $J=1$, $n=1$;
$(c)$ $J=-1$, $n=0.75$; $(d)$ $J=-1$, $n=1$.\label{fig:ChiC_ZeroT}}
\end{figure}

\begin{figure}[htp]
\includegraphics[width=1.0\columnwidth]{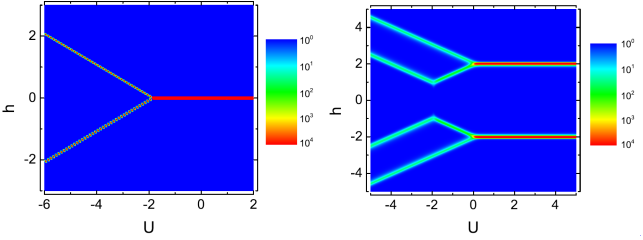}
\caption{(Color online) 3D plot of spin susceptibility in the limit of zero
temperature ($T=0.001$) at $n=1$ as a function of the external magnetic
field $h$ and local Coulomb potential $U$ for $J=1$ (left) and
$J=-1$ (right).\label{fig:ChiS_T0JpJn}}
\end{figure}

As reported in Fig. \ref{fig:ChiC_ZeroT} and Fig. \ref{fig:ChiS_T0JpJn},
the zero temperature behavior of charge/spin susceptibilities shows a discontinuity/divergence crossing the phases characterized by different charge/spin orderings, reproducing the boundaries of the phase diagram of the model for both $J=1$ and $J=-1$.

Importantly, it is worth noting that the phase diagram of the model can be reconstructed by joining together all the information coming from both charge and spin susceptibilities since different phases can exhibit the same charge ordering but different spin orderings. In this regard, we note for example that at $n=1$ the charge susceptibility is not able to distinguish between AF and F2 phases since both of them at $n=1$ are characterized by the same charge ordering with all sites singly occupied. Similarly, spin susceptibility is not able to distinguish between NM and AF phases since both of them are characterized by the absence of magnetization.

\begin{figure}[htp]
\includegraphics[width=1.0\columnwidth]{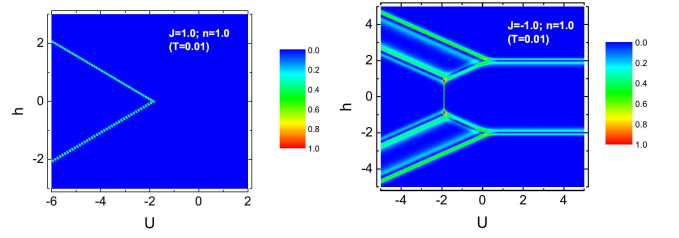}
\caption{(Color online) Contour-plot of the specific heat at low temperature
at half filling for $J=1$ (left) and $J=-1$ (right) as a function
of the external magnetic field $h$ and local Coulomb potential $U$.\label{fig:Cv_ZeroT}}
\end{figure}

The same analysis can also be done in terms of the specific heat $C(T)$ whose behavior in the $T\rightarrow0$ limit, as we shall see in detail in the next Section, is characterized by low-temperature peaks in the proximity of the phase boundaries. This analysis does not allow to distinguish between charge and spin excitations; however, as shown in Fig.\ref{fig:Cv_ZeroT}, it provides a good estimate of all the phase boundaries of the model.

\subsection{Density of states}
\begin{figure}[htp]
\begin{centering}
\includegraphics[width=1\columnwidth]{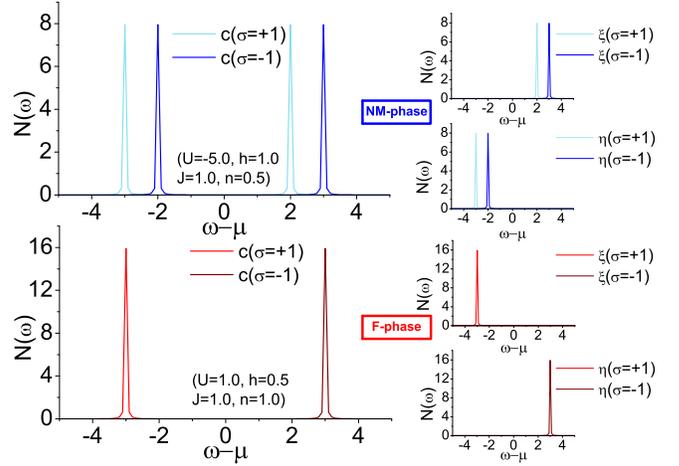}
\par\end{centering}
\caption{(Color online) Density of states in the limit of zero temperature
for the NM and F phases ($J=1$) . We report the total density of states ($c(i)=\xi(i)+\eta(i)$) contributions for both spin up ($\sigma=+1$) and spin down ($\sigma=-1$). The contributions due only to $\xi(i)$ and $\eta(i)$ fields are also reported in the insets.\label{fig:DOS_Jpositive}}
\end{figure}

\begin{figure}[htp]
\begin{centering}
\includegraphics[width=1\columnwidth]{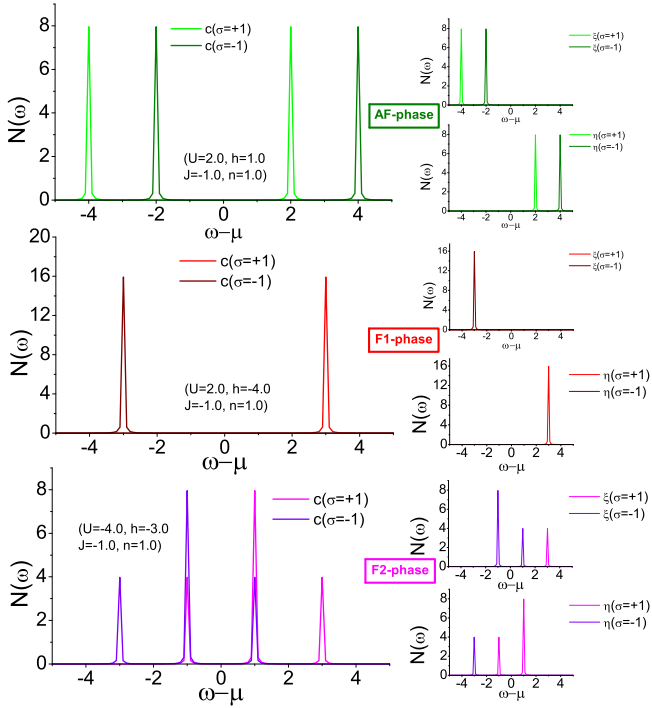}
\par\end{centering}
\caption{(Color online) Density of states in the limit of zero temperature
for the AF, F1 and F2 phases ($J=-1$). We report the contributions
for both spin up ($\sigma=+1$) and spin down ($\sigma=-1$). In the
insets the contributions due to $\xi(i)$ and $\eta(i)$ fields are
reported.\label{fig:DOS_Jnegative}}
\end{figure}

To complete the zero temperature analysis, we report in this Subsection
the results obtained for the density of states in the limit of zero
temperature. As shown in Eq.\eqref{eq:DOS}, the density of states is
expressed as a superposition of several delta functions, each of them
centered at the energy levels $E_{n}^{(a)}$ and weighted by the spectral
functions $\rho^{(a,n)}$ . In the limit of zero temperature, most
of the weights vanish and only few energies, corresponding to the
ground state and first excited states, give a contribution. We report
in Fig. \ref{fig:DOS_Jpositive} and Fig. \ref{fig:DOS_Jnegative}
the density of states calculated for each phase that appears in the
$J=1$ and $J=-1$ phase diagrams at half-filling and $T=0.001$.

For ferromagnetic inter-site coupling, as shown in Fig. \ref{fig:NM_configuration},
the NM-phase at $n=1$ is characterized by the presence of either doubly occupied or empty sites. Accordingly, the low-laying excited states contain singly occupied sites with spins aligned along $h$. Therefore, as shown in Fig. \ref{fig:DOS_Jpositive}, the density of states has a peak above the Fermi level induced by $\xi_{\uparrow}(i)$. The
F-phase at $n=1$ is composed of singly occupied sites with spins pointing towards $h$. On the contrary, the low-laying excited states admit a number of doubly occupied sites, resulting in a peak above the chemical potential induced by $\eta_\downarrow$, as shown in Fig. \ref{fig:DOS_Jpositive}.

The same considerations can also be done in the case of negative $J$
(see Fig. \ref{fig:DOS_Jnegative}) so that the nature of the peaks around
the Fermi level can be easily predicted by observing the zero-temperature
configuration of each phase. It is worth noting that, only in the F2-phase, both the contributions of $\xi$ and $\eta$ are allowed thanks to the fact that any F2-configuration comprises either polarized spins or double occupancies (Fig. \ref{fig:F2_configuration}).


Our analysis of the density of states also allows to get some insight about the gap
$\Delta$ that separates the ground state from the first excited one.
It can be seen that, as long as the system remains in the same phase, a finite gap exists according to the results obtained for the specific heat in the limit of zero temperature (Fig.\ref{fig:Cv_ZeroT}).
On the contrary, moving from a phase to another, $\Delta$ closes as one approaches the phase boundaries and low-temperature excitations appear in the specific heat signalling a level-crossing between the ground and the first excited state.


\subsection{Comparison with variational approach}\label{SubSec:Comparison}
In closing this Section, to further emphasize the crucial role played by our exact solution as a guideline for testing analytical and numerical approximations, we compare our results with those obtained in Ref.\onlinecite{Kapcia2012}, where the same model, but in absence of magnetic field, has been studied within a variational and a mean-field approach. First of all, the absence of magnetic field introduces a symmetry between the ferro ($J>0$) and the antiferro ($J<0$) case. Owing to this reason, only the case $J>0$ has been studied in Ref.\onlinecite{Kapcia2012}, where the ground state phase diagram exhibits only the two phases NM and F (or AF for $J<0$). We note that, as a consequence of the mean-field approach, the ground state energy in the F phase (called EF Ref.\onlinecite{Kapcia2012}) reported in Ref.\onlinecite{Kapcia2012} does not agree with ours, given in Eq.\eqref{eq:Energy_Fphase}. In particular EF of Ref.\onlinecite{Kapcia2012} depends on the square of the filling and this implies that the transition line between the NM and F phases depend on $n$ in strong contrast with our exact calculations reported in Eq.\ref{eq:E_NMF}. Another important difference is that the thermodynamic instability of the NM phase, predicted in Ref.\onlinecite{Kapcia2012}, is not observed in our exact solution.

\section{Finite temperature results}\label{Sec:TfiniteResults}
In this Section we present the study of the model \eqref{eq:Hubbard_Intersite+StrCoupl+Nn} at finite temperature. As already done in the previous Section, we put $|J|=1$ and analyze both $J=1$ and $J=-1$ cases. We report the temperature dependence of single-particle (chemical potential, double occupancy, magnetization) and thermodynamic (entropy, specific heat, charge and spin susceptibility) quantities for different values of $n$, $U$, $h$ and $J$. Because of the one-dimensionality of the model, there is no long-range order at finite temperature. However, in this Section we maintain the reference to the $T=0$ phases, described in Section III, in the sense that we label each set of parameter ($n$, $U$, $h$, $J$) according to the corresponding phase observed at zero temperature.

\begin{figure}[htp]
\begin{centering}
\includegraphics[width=1\columnwidth]{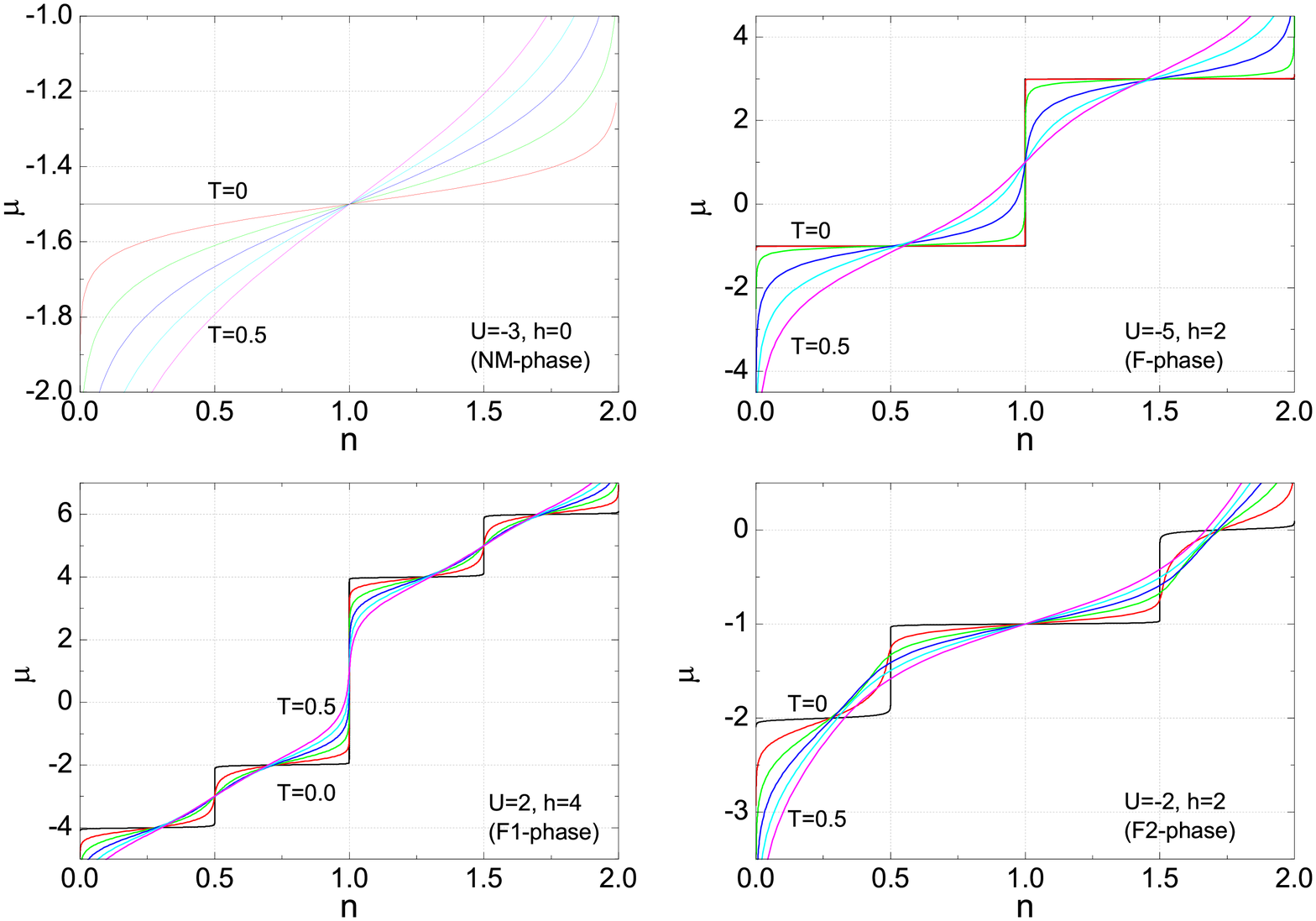}
\par\end{centering}
\caption{(Color online) Chemical potential plotted as a function of the filling for different temperatures. Different values of $h$ and $U$ are related to NM (top, left), F (top, right), F1 (bottom, right) and F2 (bottom, left) phases.\label{fig:ChemPotT_JpJn}}
\end{figure}

\emph{Chemical potential}. The behavior of the chemical potential $\mu$ as a function of the particle density $n$ is reported in Fig. \ref{fig:ChemPotT_JpJn} for different values of $T$, $J$, $U$, and $h$. In the magnetic phases, an interesting feature is the presence of crossing points in the chemical potential curves, when plotted versus the filling for different temperatures. More precisely, one observes crossing points at the commensurate fillings which also are turning points where the derivative $d\mu/dT$ changes sign.

\begin{figure}[htp]
\begin{centering}
\includegraphics[width=1\columnwidth]{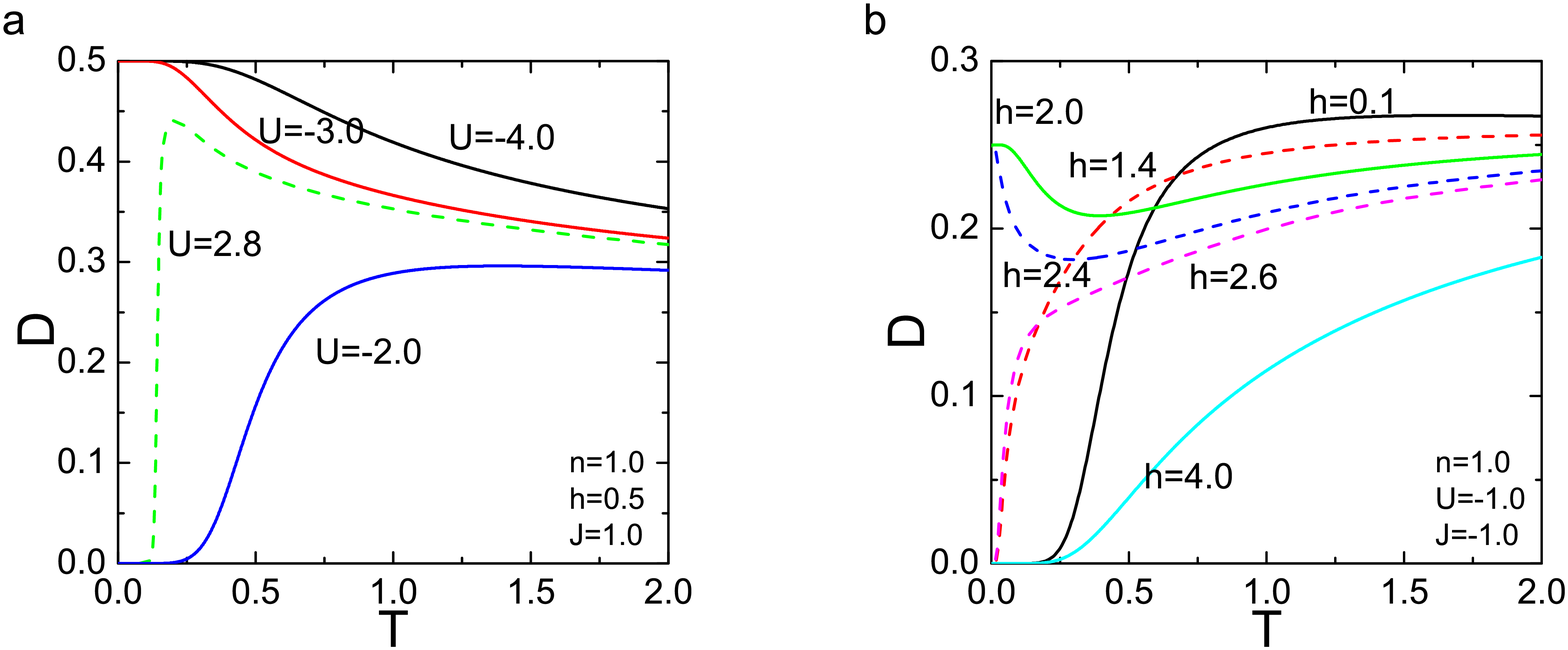}
\par\end{centering}
\caption{(Color online) The double occupancy as a function of temperature for $J=1$, $h=0.5$, $n=1$, $-4\leqslant U\leqslant -2$ (left panel) and $J=-1$, $U=-1.0$, $n=1$, $0.0\leqslant h\leqslant4.0$ (right panel).\label{fig:DT_JpJn}}
\end{figure}

\emph{Double occupancy}. We report in Fig. \ref{fig:DT_JpJn} the temperature dependence of the double occupancy at $n=1$ and different values of $U$ and $h$, for both $J=1$ and $J=-1$ cases.

In the NM phase (Fig. \ref{fig:DT_JpJn}a, $U=-4$ and $U=-3$), the attractive local potential $U$ prevails and at zero temperature $D$ reaches its maximum value $D=n/2$; as the temperature increases, the thermal excitations tend to break the doublons (the pairs of electrons with anti-parallel spins residing on the same site); several configurations with singly occupied sites become available resulting in a depression of the double occupancy. $D$ appears always a decreasing function of the temperature, going to $n^2/4$ in the limit $T\to\infty$.

In the F phase (Fig. \ref{fig:DT_JpJn}a, $U=-2.8$ and $U=-2$), the magnetic interactions prevail on the attractive local potential $U$ and $D$ vanishes at $T=0$. For $U>U_c$ two behaviors can be distinguished. When the temperature is raised above zero, $D$ increases: because of the presence of an attractive on-site $U$ potential, the thermal fluctuations favor the formation of doublons, and the relative contribution of the excited states with large average $D$ increases. By further increasing $T$, the double occupancy $D$ reaches a maximum, then decreases. The temperature where $D$ exhibits a maximum decreases with $U$ and goes to zero for $U\rightarrow U_c$. When $U$ is very close to $U_c$ (e.g. $U=-2.9$) at the temperatures of the order of the gap between the first excited and the ground state, the two become quasi-degenerate, and $D$ acquires a contribution from the latter. In the limit of $U\rightarrow U_c$, close to $T=0$ the double occupancy exhibits a jump from zero to $n/2$.
	
Similar considerations can be done for the $J=-1$ case. As reported in Fig. \ref{fig:DT_JpJn}b, in AF ($h=0.1$ and $h=1.4$) and F1 ($h=2.6$ and $h=4.0$) phases double occupancy is suppressed by the presence of magnetic orders at $T=0$. By increasing the temperature, $D$ monotonically increases. On the contrary, in the F2-phase ($h=2.0$ and $2.4$), despite the presence of a ferromagnetic order, one has a finite double occupancy at $T=0$ $[D(0)=(2n-1)/4]$, resulting in a non-monotonic $T$-dependence. In all phases $D$ tends to $n^2/4$ in the limit of $T\rightarrow\infty$ .

\begin{figure}[htp]
\begin{centering}
\includegraphics[width=1\columnwidth]{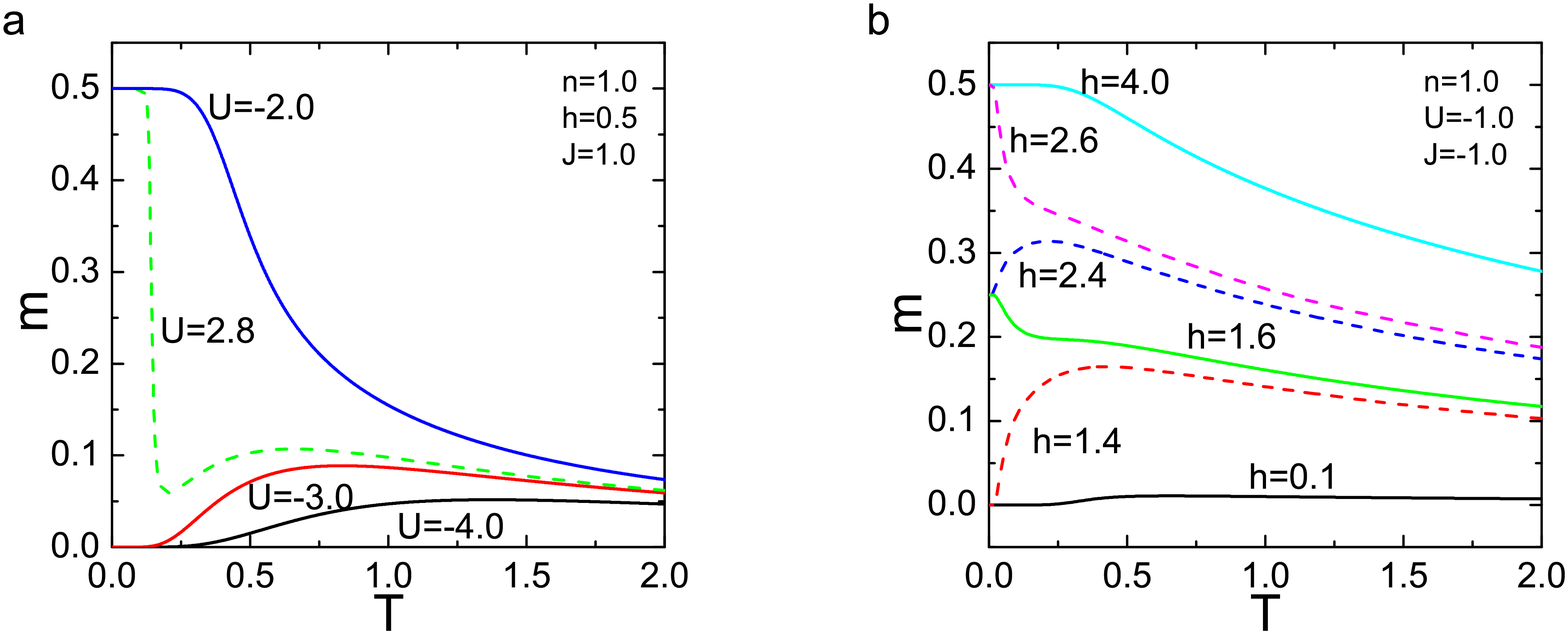}
\par\end{centering}
\caption{(Color online) The magnetization as a function of temperature for $J=1$, $h=0.5$, $n=1$, $-4\leqslant U\leqslant-2$ (left panel) and $J=-1$, $U=-1.0$, $n=1$, $0.0\leqslant h\leqslant 4.0$ (right panel).\label{fig:mT_JpJn}}
\end{figure}

\emph{Magnetization}. We report in Fig. \ref{fig:mT_JpJn} the temperature dependence of the magnetization at different values of $U$ and $h$, for both $J=1$ and $J=-1$ cases at $n=1$. In the NM phase (Fig. \ref{fig:mT_JpJn}a, $U=-4$ and $U=-3$) the ground state is composed of double occupied sites only (see Fig. \ref{fig:NM_configuration}).
By increasing $T$, excitations destroy doublons and a finite magnetization appears.
The increase of the magnetization with temperature is contrasted by thermal fluctuations which eventually destroy $m$ at $T\rightarrow\infty$.
In the F-phase (Fig. \ref{fig:mT_JpJn}a, $U=-2$ and $U=-2.8$) the spins in the ground state are fully polarized. Therefore $m(T=0)=n/2$. Increasing the temperature, doubly occupied sites progressively emerge, lowering the magnetization that also exhibits a broad maximum at intermediate temperatures associated with higher energy excited states.

Apart from the NM-phase, in the case of antiferromagnetic coupling, three different behaviors of magnetization can be distinguished which correspond to three different phases.
As reported in Fig. \ref{fig:F1_configuration}, in the F1-phase at $T=0$ all the spins are aligned in the field direction and $m(T=0)$ assumes its saturation value. Thermal fluctuations destroy the magnetization which decreases monotonically upon the increase of temperature.
In the AF phase the average magnetization per site is zero at $T=0$ because of the antiferromagnetic ordering. At finite $T$, the magnetization acquires contributions from the excited states which are magnetic only at finite $h$. Therefore, for $h\neq0$ the magnetization exhibits a maximum at an intermediate temperature $T^\ast$ and goes gradually to zero as $T\rightarrow\infty$. As shown in Fig. \ref{fig:F2_configuration}, in the F2-phase ($h=1.6$ and $h=2.4$) the ground state is characterized by the coexistence of singly and doubly occupied sites under the constraint that $m=1/4$ independently on the value of $n\in[0.5,1.5]$.
Upon the increase of $T$, thermal excitations break some of the doublons and misalign the spins. Hence there is a high competition among multiple energy scales: $T$, $h$, $J$, $U$ and the magnetization may exhibit one or several peaks, each of them corresponding to a distinct energy scale.

\begin{figure}[htp]
\begin{centering}
\includegraphics[width=1\columnwidth]{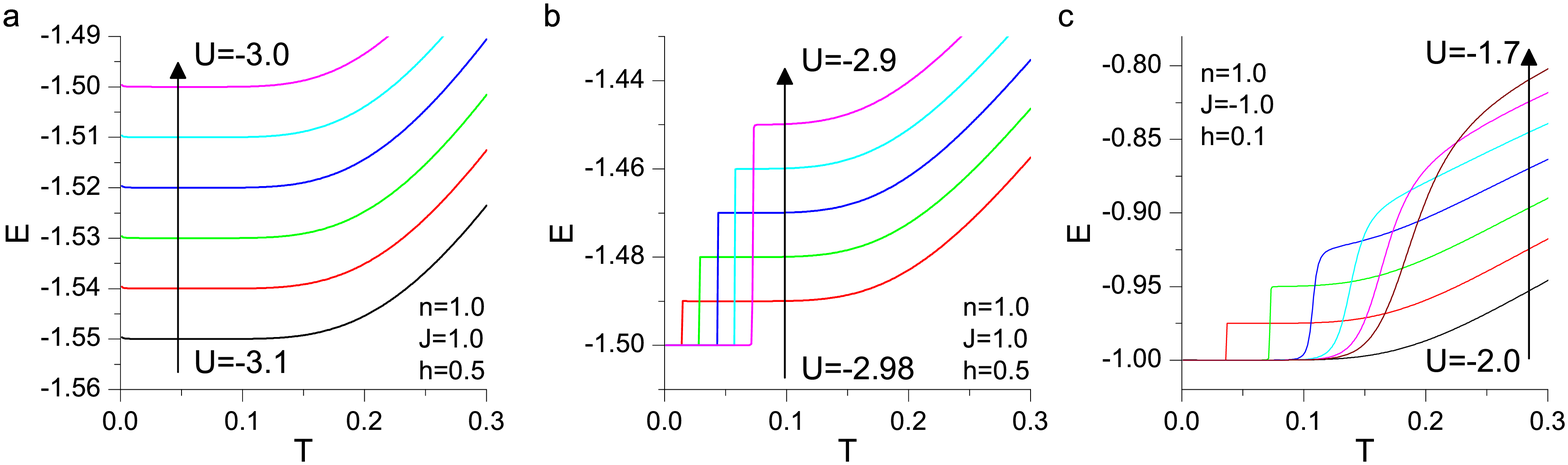}
\par\end{centering}
\caption{(Color online) Internal energy as a function of temperature for NM-phase (panel a) and F-phase (panel b) at $J=1$ and AF-phase (panel c) at $J=-1$.\label{fig:ET_JpJn}}
\end{figure}

\emph{Internal energy}. In the NM phase, as shown in Fig. \ref{fig:ET_JpJn}a, the internal energy monotonically decreases by lowering $T$ and tends to the value $E_{NM}=nU/2$ as $T\rightarrow0$. In Fig. \ref{fig:ET_JpJn}b, the temperature dependence of the internal energy $E$ is shown for the F-phase, by considering various values of $U$ and $n=1$. E decreases with $T$ and tends to the value $E_F=-n(J+h)$ as $T\rightarrow0$. When $U$ is close to $U_c$, $E$ exhibits a discontinuity at a certain temperature, which goes to zero as $U$ tends to $U_c$.

For the case $J=-1$, interesting features for the internal energy are observed (see Fig. \ref{fig:ET_JpJn}c) only for the AF-phase in proximity of the transition to the NM-phase. For all the other phases, the energy is found to be a continuous and increasing function of the temperature. The significant variations observed in the $E(T)$ dependence lead to the presence of very low temperature features in the specific heat, as we shall discuss below.

\begin{figure}[htp]
\begin{centering}
\includegraphics[width=1\columnwidth]{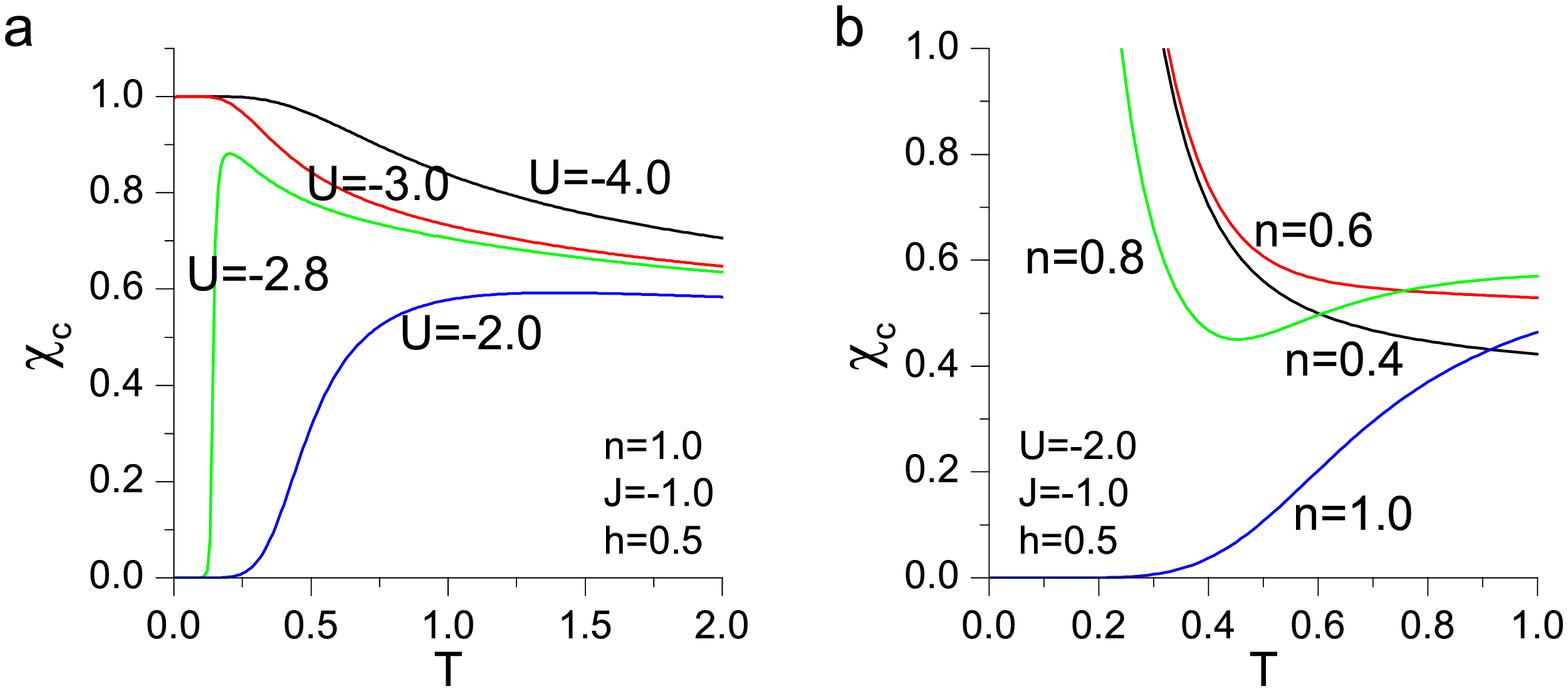}
\par\end{centering}
\caption{(Color online) The charge susceptibility as a function of temperature at $J=1$, $h=0.5$ for: $n=1$, $-4\leqslant U\leqslant -2$ (left panel) and $U=-2$ and $0.4\leqslant n\leqslant1$ (right panel).\label{fig:ChiCT_Jp}}
\end{figure}

\emph{Charge susceptibility}. We report in Fig. \ref{fig:ChiCT_Jp} the temperature dependence of the charge susceptibility $\chi_c$ at different values of $n$, $U$ and $h$ for the $J=1$ case. In the NM phase (Fig. \ref{fig:ChiCT_Jp}a, $U=-4$ and $U=-3$), $\chi_c$ takes the value $\chi_c=n(2-n)$ at $T=0$, independently on the value of $U$, and decreases by increasing $T$. For a given temperature, $\chi_c$ decreases by increasing $U$ and by decreasing the filling. In the F phase at $n=1$ (Fig. \ref{fig:ChiCT_Jp}a, $U=-2.8$ and $U=-2$) $\chi_c$ vanishes at $T=0$, increases with $T$ up to a maximum and tends to a finite value in the limit $T\rightarrow\infty$. When $U$ approaches $U_c$, the maximum becomes more pronounced and tends to one as $T\rightarrow0$. As shown in Fig. \ref{fig:ChiCT_Jp}b, in the F phase $\chi_c$ has a different behavior away from half filling. The charge susceptibility increases by lowering $T$ and tends to diverge as $T\rightarrow0$; when $U$ approaches $U_c$, the tendency to diverge becomes much stronger. In the limit of high temperatures, for both the NM and F phases, the charge susceptibility tends to a constant value which does not depend on $U$ but only on $n$ according to the law: $\lim_{T\rightarrow\infty}\chi_c=n(2-n)/2$.

\begin{figure}[htp]
\begin{centering}
\includegraphics[width=1\columnwidth]{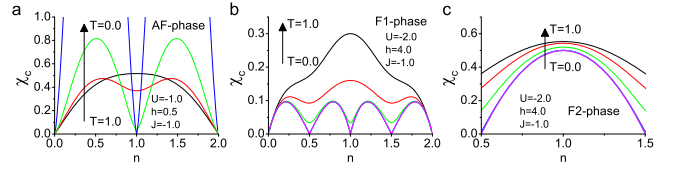}
\par\end{centering}
\caption{(Color online) The charge susceptibility as a function of the filling and different temperatures for AF-phase (left panel) and F1 (central panel) and F2 phases (right panel).\label{fig:ChiCT_Jn}}
\end{figure}

In the case of anti-ferromagnetic coupling ($J=-1$), the AF phase exhibits the same temperature and filling dependence already described for the F phase. As reported in Fig. \ref{fig:ChiCT_Jn}a, for any $n\neq1$ the charge susceptibility increases as the temperature decreases and diverges for $T\rightarrow0$. At half-filling $\chi_c$ has a peak which gradually moves towards $T=0$ as the system approaches the phase boundary.

On the contrary, as reported in Fig.\ref{fig:ChiCT_Jn}b and Fig.\ref{fig:ChiCT_Jn}c, in F1 and F2 phases there are no divergencies in the charge susceptibility at any value of filling and temperature. In the F1-phase (Fig.\ref{fig:ChiCT_Jn}b) $\chi_c(T=0)$ is zero at any value of the filling which corresponds to the presence of ordered phases at $T=0$ ($n=0$,$1/2$,$1$,$3/2$,$2$). In the F2-phase (Fig.\ref{fig:ChiCT_Jn}c) $\chi_c$ has a maximum at $n=1$, then decreases as one approaches to $n=1/2$ and $n=3/2$. These values of filling are situated at the phase boundary between the F1 and F2 phases, therefore $\chi_c(T=0)=0$ in agreement with the above discussion. It is worth noting that, similarly to the $J=1$ case, $\lim_{T\rightarrow\infty}\chi_c=n(2-n)/2$  in all the phases.

\begin{figure}[htp]
\begin{tabular}{cc}
\includegraphics[width=0.5\columnwidth]{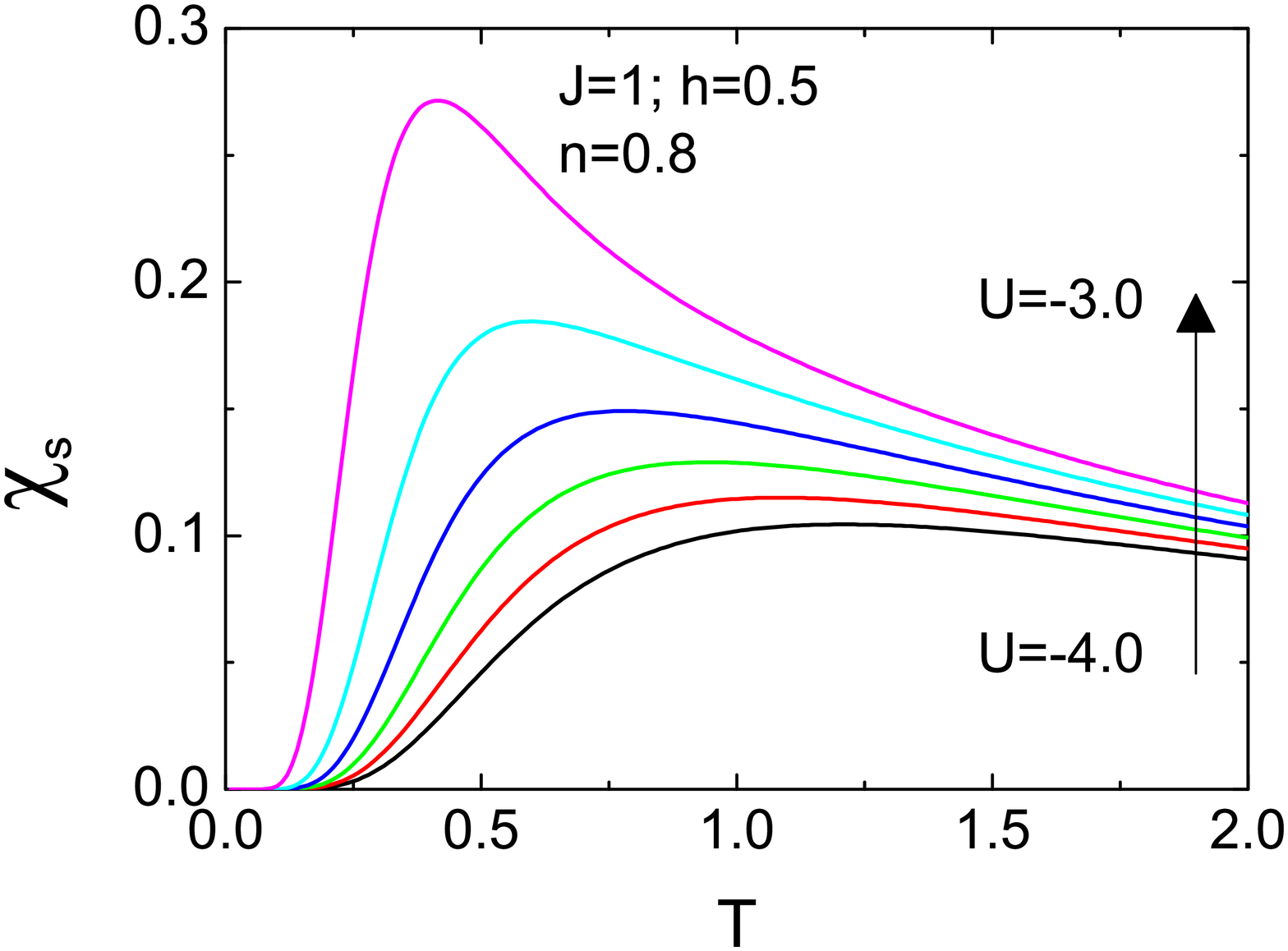}
\includegraphics[width=0.5\columnwidth]{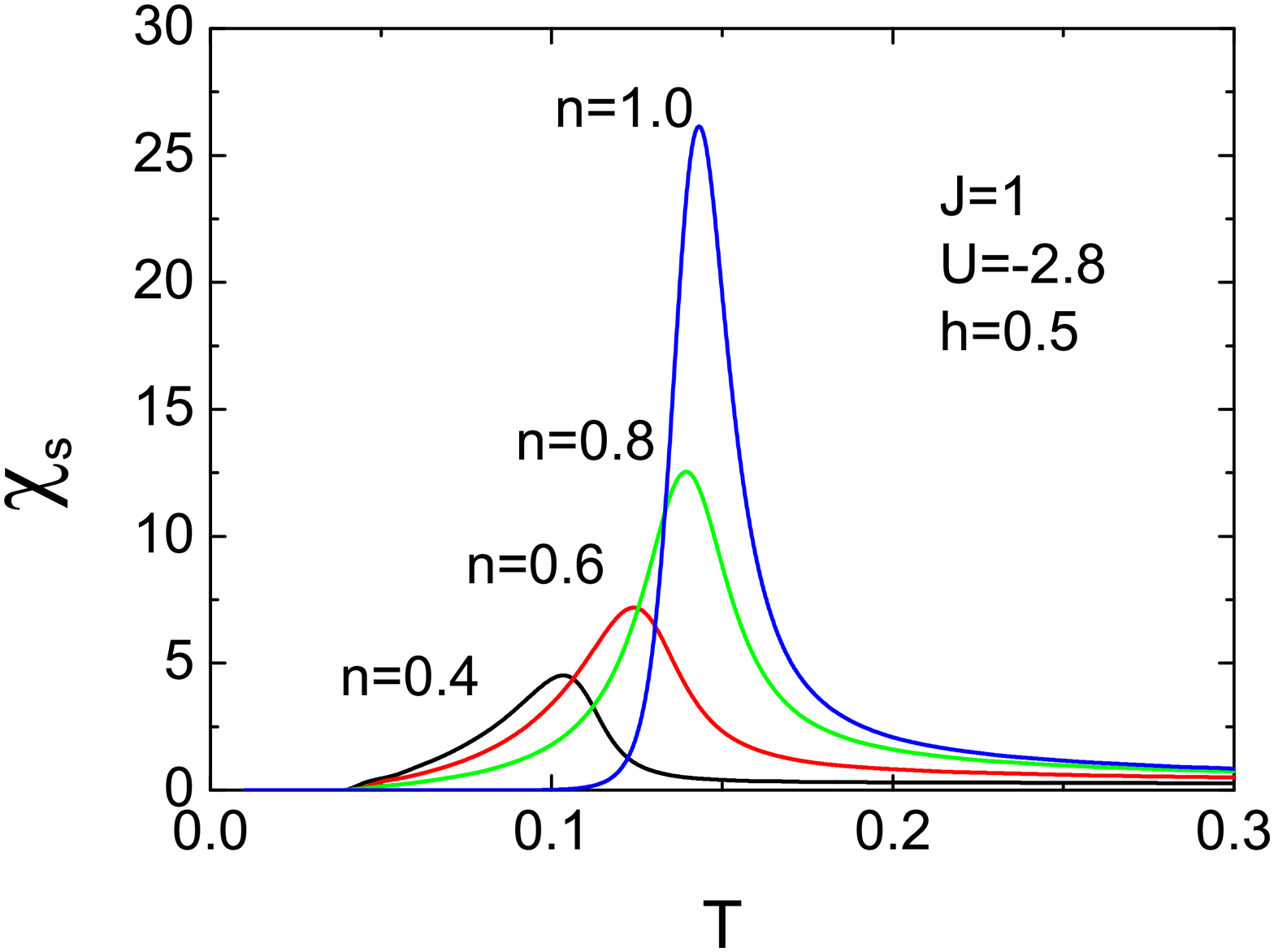}
\end{tabular}
\caption{(Color online) The spin susceptibility $\chi_s$ as a function of $T$ for $J=1$, $h=0.5$; left panel: $n=0.8$ and various values of $U$; right panel $U=-2.8$ and various values of $n$.\label{fig:ChiST_Jp}}
\end{figure}

\emph{Spin susceptibility}. In Fig. \ref{fig:ChiST_Jp}a,  the temperature dependence of the spin susceptibility $\chi_s$ is shown for the NM phase, by considering $J=1$, $h=0.5$, $n=0.8$ and various values of $U$. $\chi_s$ vanishes at $T=0$, increases by increasing $T$ up to a maximum temperature $T_{\max}$, then decreases. This behavior is observed for all values of the filling. For a given temperature, $\chi_s$ decreases by decreasing $U$ and by decreasing the filling. The temperature $T_{max}$ decreases by increasing $U$. In Fig. \ref{fig:ChiST_Jp}b, the temperature dependence of the spin susceptibility is shown for the F-phase, by considering various values of $n$ and $U=-2.8$. $\chi_s$ vanishes at $T=0$, increases by increasing $T$ up to a maximum temperature $T_s$, then decreases. When $U$ approaches $U_c$, the temperature $T_s$ moves toward zero and $\chi_s$ tends to diverge.

\begin{figure}[htp]
\begin{centering}
\includegraphics[width=1\columnwidth]{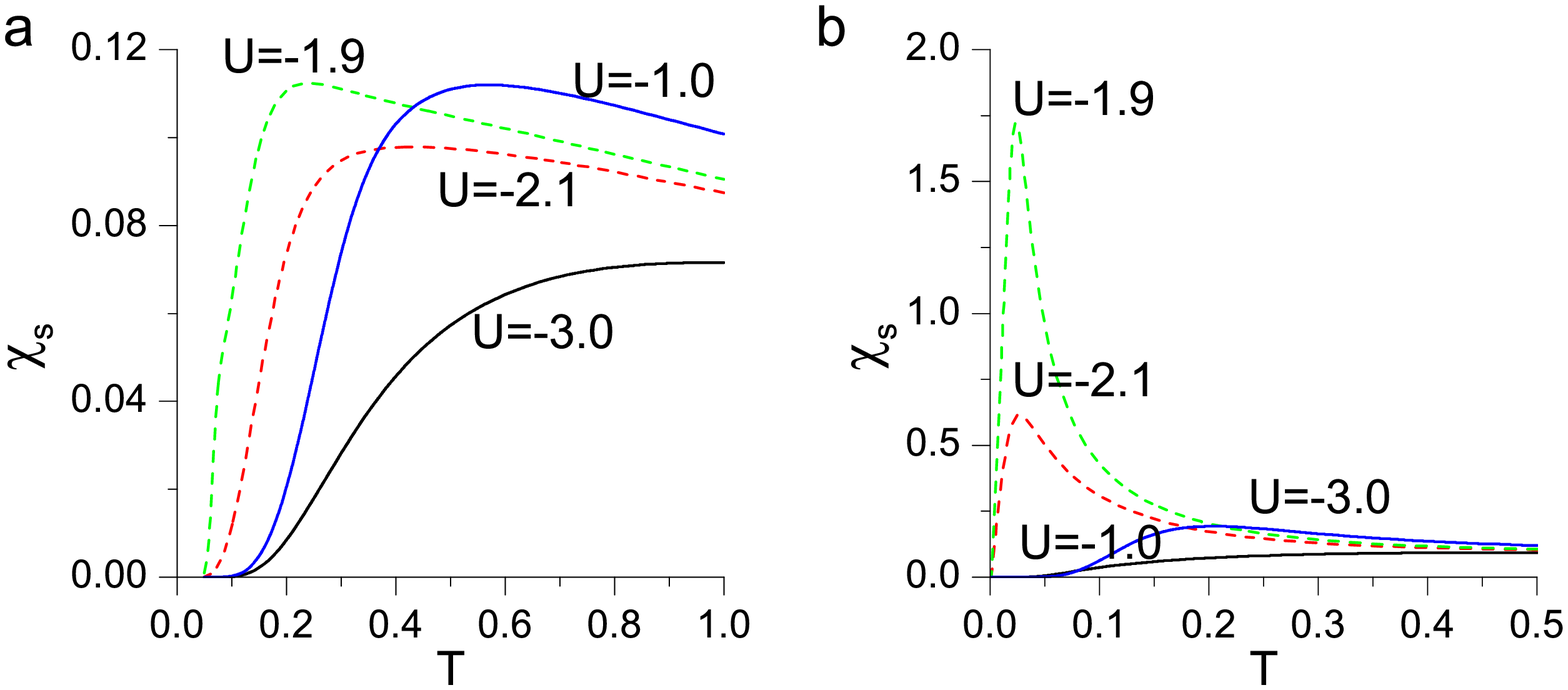}
\par\end{centering}
\caption{(Color online) The spin susceptibility $\chi_s$ as a function of $T$ for $J=-1$, $n=1.0$ and various values of $U$ for $h=3.0$ (left panel) and $h=0.5$ (right panel).\label{fig:ChiST_Jn}}
\end{figure}

\begin{figure}[htp]
\begin{centering}
\includegraphics[width=1\columnwidth]{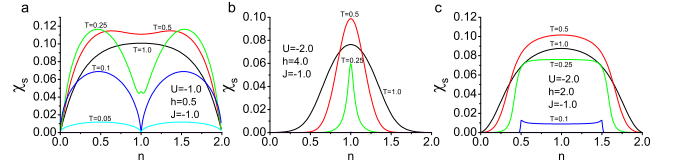}
\par\end{centering}
\caption{(Color online) The spin susceptibility as a function of the filling and different temperatures for AF-phase (panel $a$), F1-phase (panel $b$) and F1 ($0\leqslant n\leqslant 1/2$, $3/2\leqslant n\leqslant2$) - F2 ($1/2\leqslant n\leqslant3/2$) phases (panel $c$).\label{fig:ChiSnT_Jn}}
\end{figure}

In the case of anti-ferromagnetic coupling ($J=-1$), $\chi_s(T)$ in the AF-phase (see Fig. \ref{fig:ChiST_Jn}a, $U=-1.9$ and $U=-1.0$) exhibits the same features already analyzed for NM and F phases with a maximum that moves to $T=0$ as the phase boundary is approached. The behavior of $\chi_s(T)$ in the F1 and F2 phases is shown in Fig. \ref{fig:ChiST_Jn}b. Low-temperature peaks, whose heights increase by approaching the transition line, are observed in the F1 phase and are due to the contribution from the low-laying magnetic excited states.

Concerning the filling dependence, we can note that in all the phases $\chi_s(T=0)=0$ in the whole range of filling. By increasing the temperature the spin susceptibility also increases with different temperature dependencies according to the particular choice of the external parameters. In the AF phase, as shown in Fig.\ref{fig:ChiSnT_Jn}a, $\chi_s\neq0$ at any $n\neq 1$ at low-temperatures, with a maximum at $n=1/2$, $3/2$ moving to $n=1$ with increasing $T$. In the F1-phase instead, as shown in Fig.\ref{fig:ChiSnT_Jn}b, $\chi_s(T\neq 0)$ exhibits a very narrow low-temperature peak, centered at $n=1$, whose dispersion increases with increasing temperature. This feature also clearly appears in the crossover from F1 to F2 phase. In fact, as reported in Fig.\ref{fig:ChiSnT_Jn}c, $\chi_s(T)$ remains finite in a wide range of temperatures for $1/2\leqslant n\leqslant3/2$ (F2-phase) while rapidly decreases with decreasing temperature in the F1-phase for $0\leqslant n\leqslant1/2$ and $3/2\leqslant n\leqslant2$.

It is worth noting that, independently on the sign of the spin exchange $J$, in the limit of high temperatures the spin susceptibility follows the Curie law with a coefficient which does not depend on $U$ but only on $n$: $\lim_{T\rightarrow\infty}\chi_s=n(2-n)/4T$. Therefore we have $\lim_{T\rightarrow\infty}\chi_c/\chi_s=2T$, in which the factor $2$ is due to the spin multiplicity.

\begin{figure}[htp]
\begin{tabular}{cc}
\includegraphics[width=0.5\columnwidth]{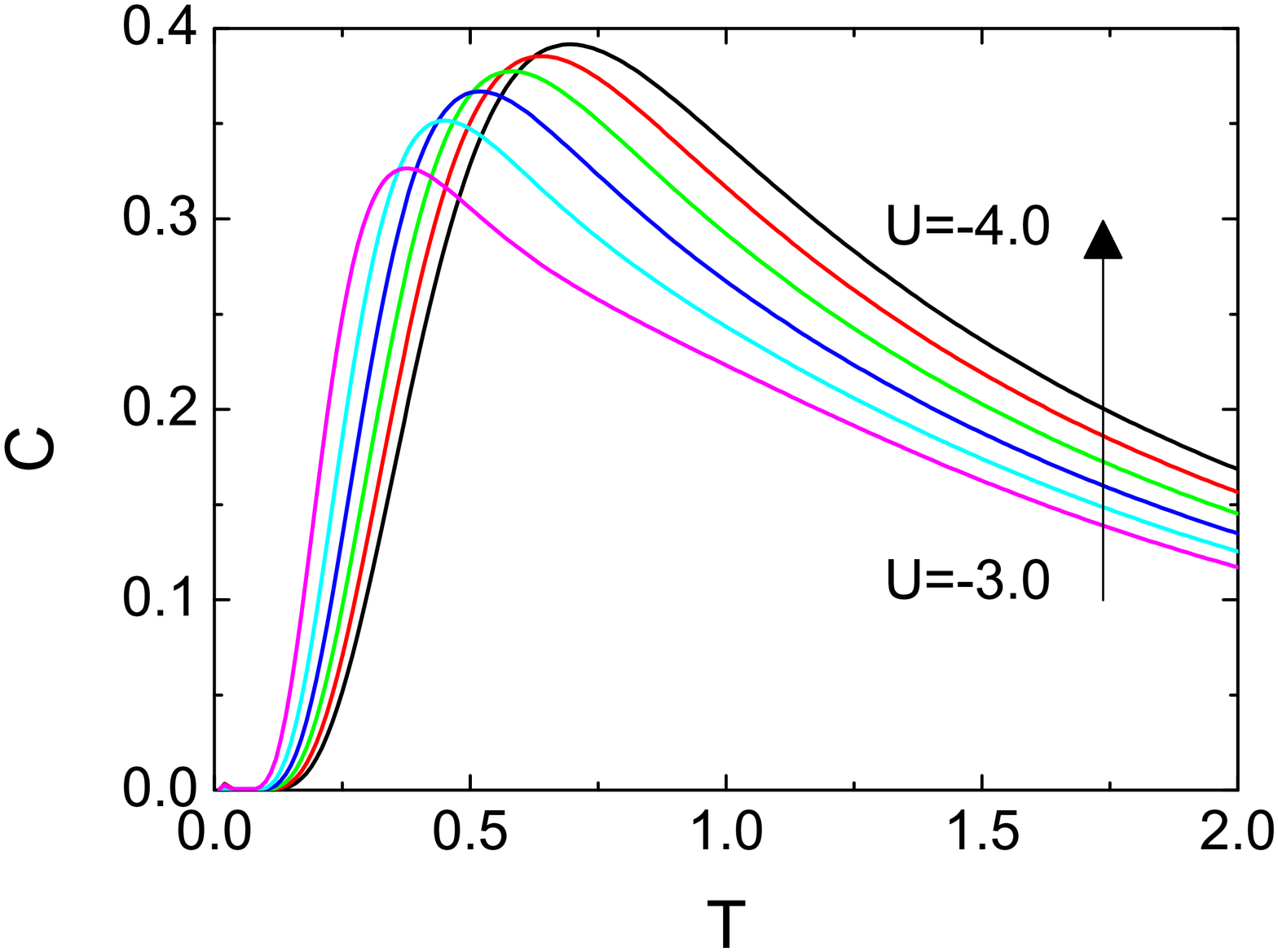}
\includegraphics[width=0.5\columnwidth]{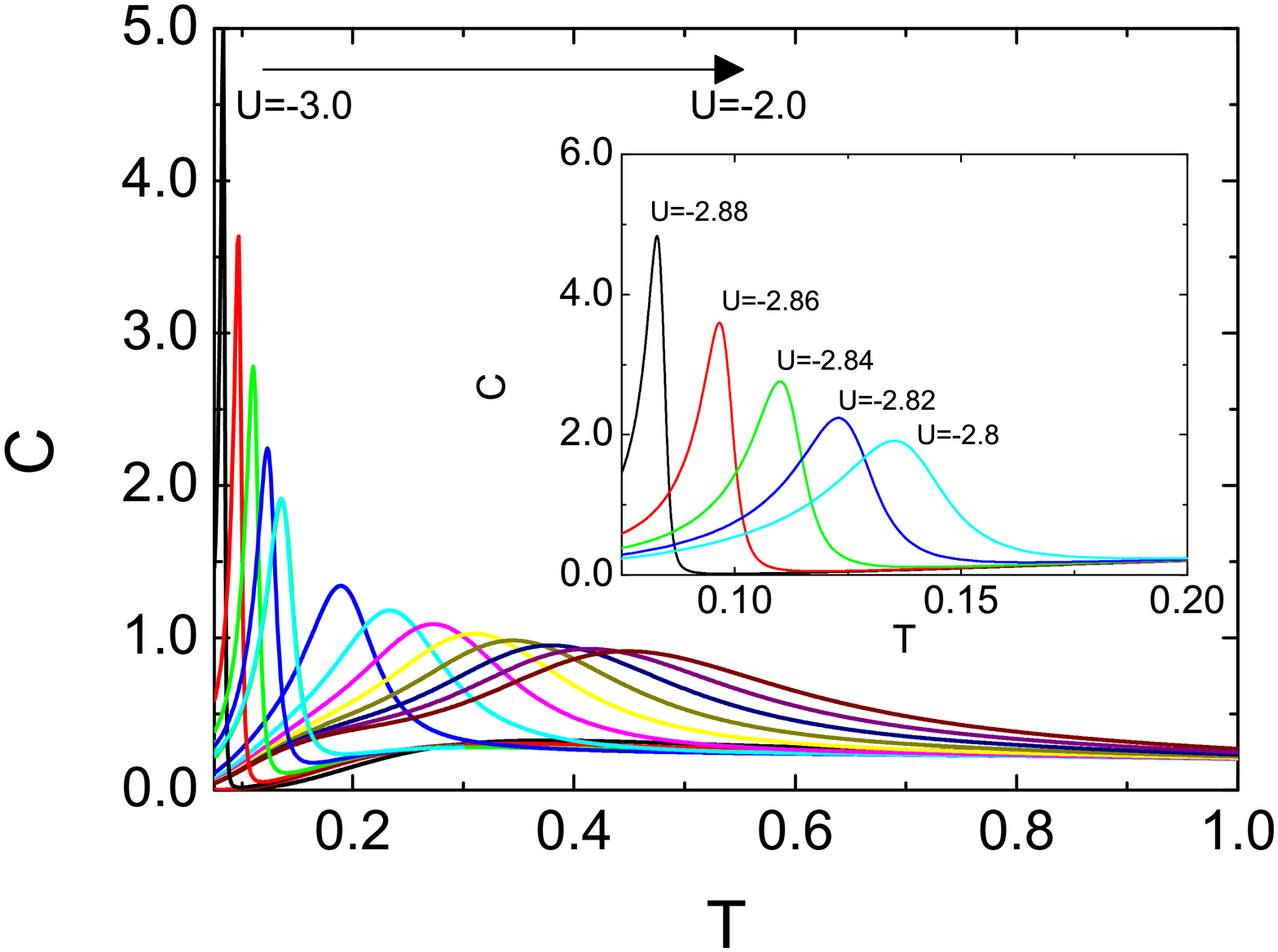}
\end{tabular}
\caption{(Color online) Specific heat as a function of temperature for $J=1$, $h=0.5$, $n=0.8$, $-4\leqslant U\leqslant-3$ (left panel) and $-3\leqslant U\leqslant-2$ (right panel).\label{fig:CvT_Jp}}
\end{figure}

\emph{Specific heat}. In the NM phase the specific heat exhibits a peak at high temperature (see Fig. \ref{fig:CvT_Jp}a). The position of this peak, $T_2$, and its intensity, $h_2$, decrease by increasing $U$. $T_2$ decreases with almost a linear law. At finite temperature, the possible excitations are transitions to configurations where some sites are singly occupied, with spins aligned in the direction of the magnetic field. The field $\psi_\sigma^{(\eta)}$ is responsible for these transitions and this explains why $T_2$ is linear in $U$. It is worth noting that the temperature $T_2$, at which the specific heat has a maximum, is close $T_m/2$, where $T_m$ is the position of the peak in the magnetization.

\begin{figure}[htp]
\begin{centering}
\includegraphics[width=1\columnwidth]{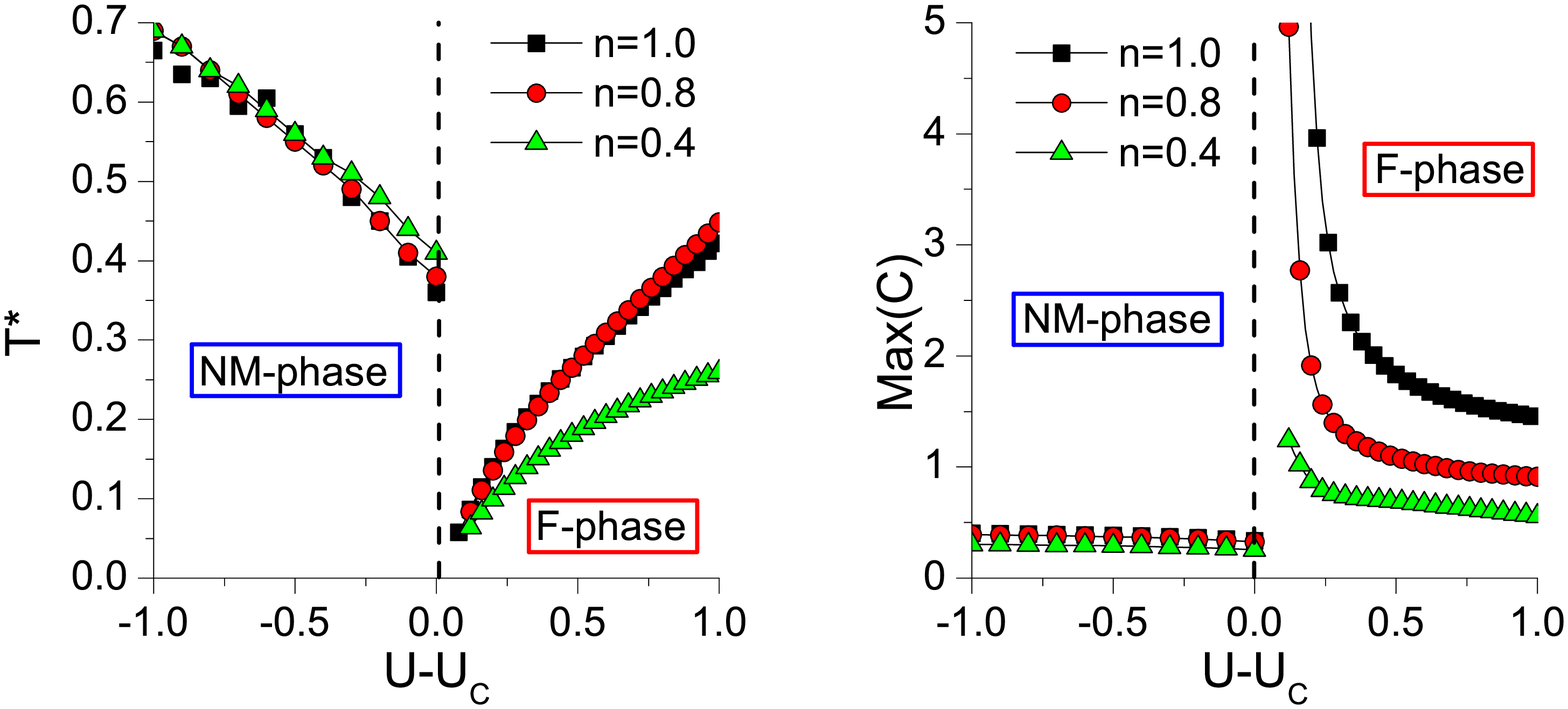}
\par\end{centering}
\caption{(Color online) Positions ($T_1$ in F-phase and $T_2$ in NM-phase) and intensities ($h_1$ in F-phase and $h_2$ in NM-phase) where the specific heat has maximas as functions of $U$ for $J=1$, $h=0.5$, and different values of the filling.\label{fig:PeaksCv_Jp}}
\end{figure}

The specific heat in the F-phase is shown in Fig. \ref{fig:CvT_Jp}b. In general, in the F-phase $C$ exhibits lower ($T_1$) and higher ($T_2$) temperature peaks, associated to transitions involving the ground state and the low-laying excited states. Approaching the boundaries, the ground and the first excited states become quasi degenerate. Hence a temperature of the order of the gap between these two states will be sufficient to induce thermal excitations, resulting in the presence of a low-temperature peak that moves towards $T=0$. On the contrary, position and intensities of high-temperature excitations, responsible for a second lower and broadened peak, remain almost unchanged.
In fact, as shown in Fig. \ref{fig:PeaksCv_Jp}b, the intensity $h_2$ of the higher temperature peak remains almost constant when $U$ varies. Contrarily, $h_1$ rapidly increases as $U$ approaches $U_c$ and diverges in the limit $U\rightarrow U_c$. This occurs since the degeneracy of the first excited state (belonging to the NM-phase) is infinite in the thermodynamic limit with respect to the degeneracy of the ground state (F-phase).

\begin{figure}[htp]
\begin{centering}
\includegraphics[width=1\columnwidth]{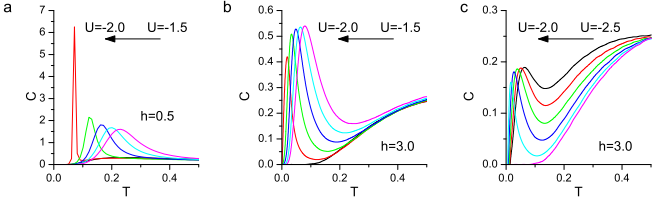}
\par\end{centering}
\caption{(Color online) Specific heat as a function of temperature at $J=-1$ for AF (left panel), F1 (central panel) and F2 (right panel) phases at $n=1.0$.\label{fig:CvT_Jn}}
\end{figure}

Most of the features observed in the F-phase persist for $J=-1$ in the AF-phase. As shown in Fig.~\ref{fig:CvT_Jn}a, the specific heat exhibits a low-temperature peak whose height (dispersion) increases (decreases) as one approaches the phase boundary. As reported in Fig. \ref{fig:AF_configuration}, the AF-configuration at $n=1$ is composed exclusively of singly-occupied sites. Therefore, the low-laying features correspond to transitions from the ground states to some excited states with finite $D$ or $m\neq0$ as confirmed by the analysis of the temperature dependence of double occupancy (Fig.\ref{fig:DT_JpJn}) and magnetization (Fig.\ref{fig:DT_JpJn}). On the contrary, F1 and F2 phases are characterized by both low and high temperatures peaks whose heights remain quite constant when the phase boundaries are approached. While the position of the former remains roughly constant, the latter moves towards $T=0$.

\begin{figure}[htp]
\begin{centering}
\includegraphics[width=1\columnwidth]{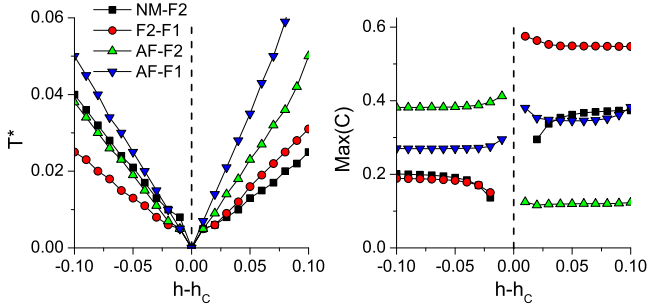}
\par\end{centering}
\caption{(Color online) Low-temperature features of the specific heat at different transition lines for $J=-1$. The maximum of the specific heat (right panel) and its position $T^{\ast}$ (left panel) are reported as a function of $h-h_{c}$ where $h_{c}$ is the critical value of the external magnetic field $h$ at the phase transition point.\label{fig:PeaksCv_Jn}}
\end{figure}

\begin{figure}[htp]
\begin{centering}
\includegraphics[width=1\columnwidth]{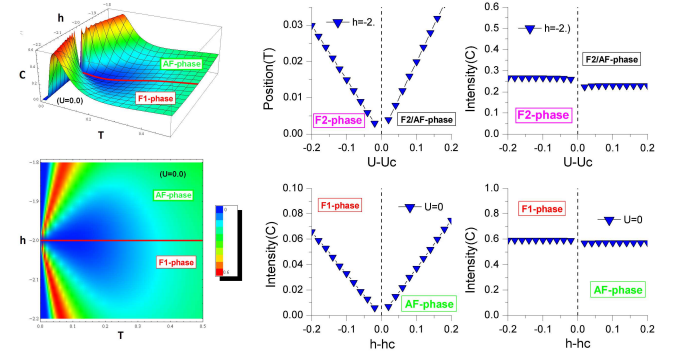}
\par\end{centering}
\caption{(Color online) Low temperature specific heat features at $P_{2}$ tricritical point for $n=1$ and $J=-1$. Contour-plot and 3Dplot on the left report specific heat at as a function of $h$ and $T$ for $U=0$. The four plots on the right report position and intensity of low temperature peaks moving to the phase transition point fixing $h$ and changing $U$ (top) and vice versa (bottom).\label{fig:CvJn_3criticalPointLowT}}
\end{figure}

We report in Fig.\ref{fig:PeaksCv_Jn} a detailed analysis of position and height of the low temperature peaks in the specific heat for all possible phase transition occurring at $J=-1$. As already pointed out, the position $T^\ast$ of low temperature peaks is a linear function of the temperature while, contrarily to what has been observed  for the $J=1$ case, the height remains constant. Furthermore, it is worth noting that, at the transition point, there is a jump in the height of the low temperature peak which can be traced back to the degeneracy ratio between the ground and the first excited states. As reported in Fig. \ref{fig:CvJn_3criticalPointLowT}, this last finding, that represents a common fingerprint for all the phase transitions analyzed so far, is not observed at the tricritical points because the first and the second excited states become degenerate with respect to the ground one in this very special case.

\begin{figure}[htp]
\begin{tabular}{cc}
\includegraphics[width=0.5\columnwidth]{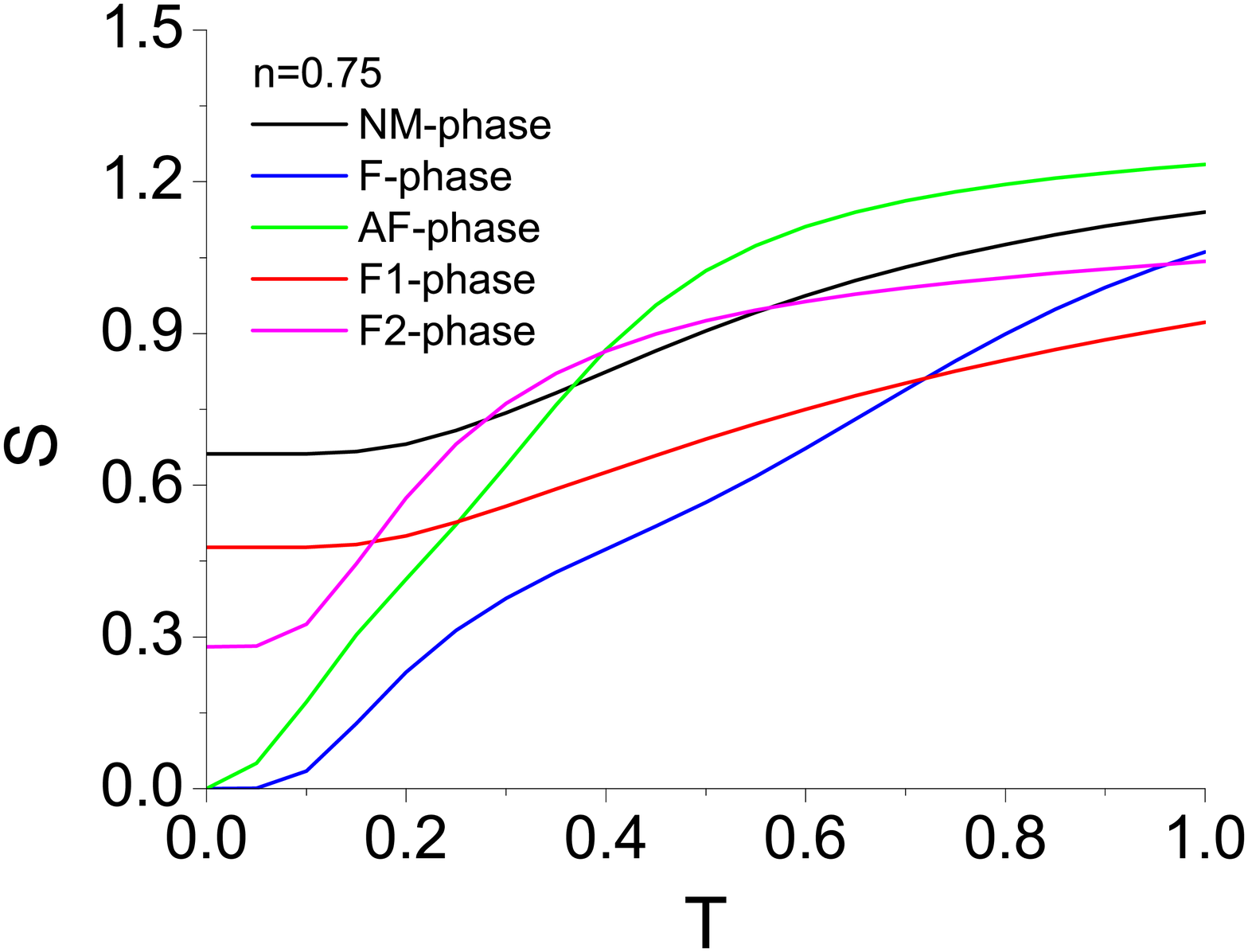}
\includegraphics[width=0.5\columnwidth]{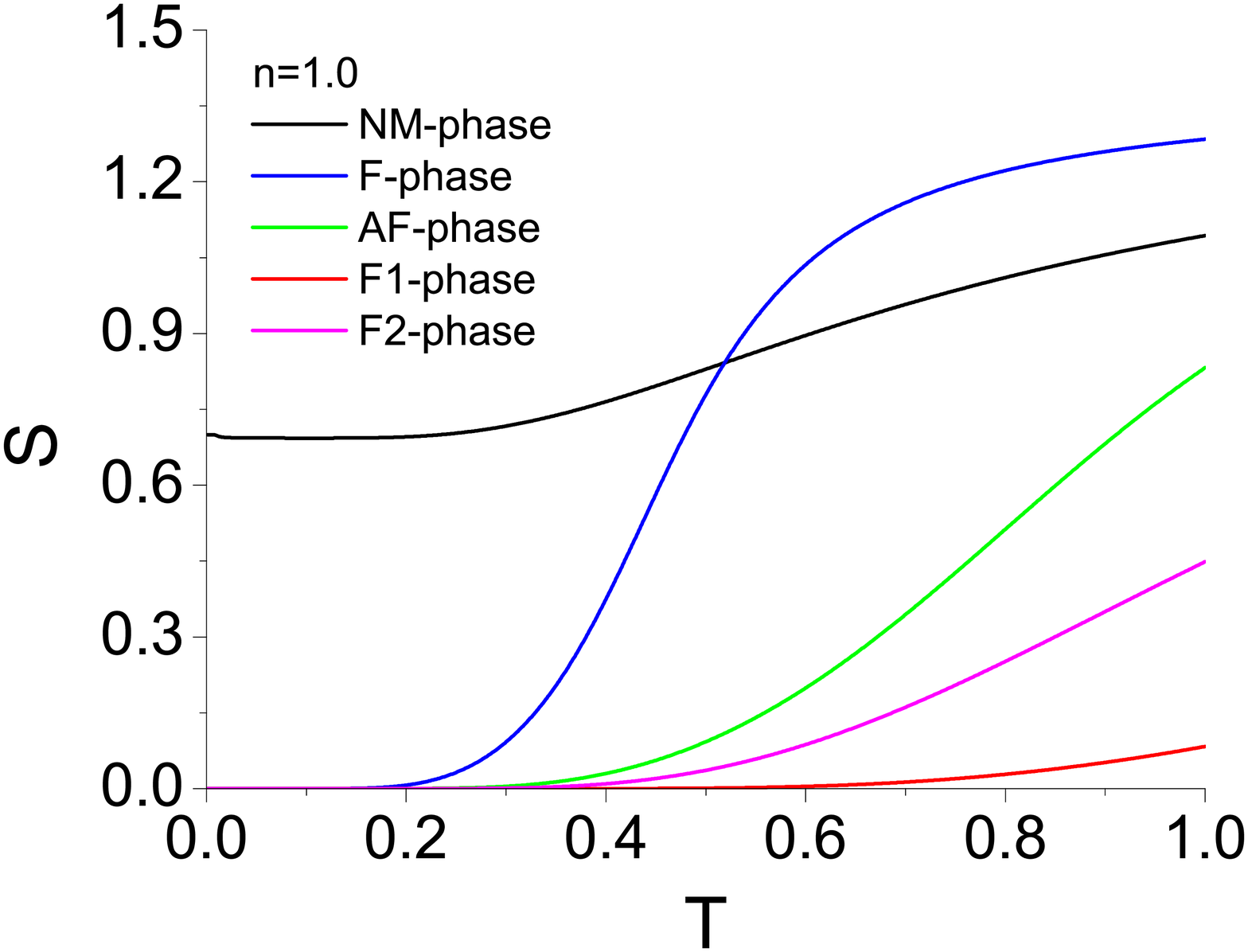}
\end{tabular}
\caption{(Color online) Temperature dependence of the entropy plotted for different values of the external parameters corresponding to different phases at $T=0$.\label{fig:Entropy3}}
\end{figure}

\begin{figure}[htp]
\begin{centering}
\includegraphics[width=1\columnwidth]{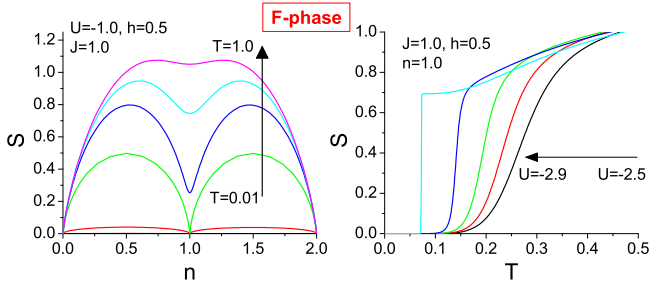}
\par\end{centering}
\caption{(Color online) The entropy $S$ at $J=1$, $h=0.5$ as a function of the filling and for different values of $T$ (left panel) and at $n=1$ as a function of $T$ for different values of $U$ (right panel).\label{fig:Entropy1}}
\end{figure}

\begin{figure}[htp]
\begin{centering}
\includegraphics[width=1\columnwidth]{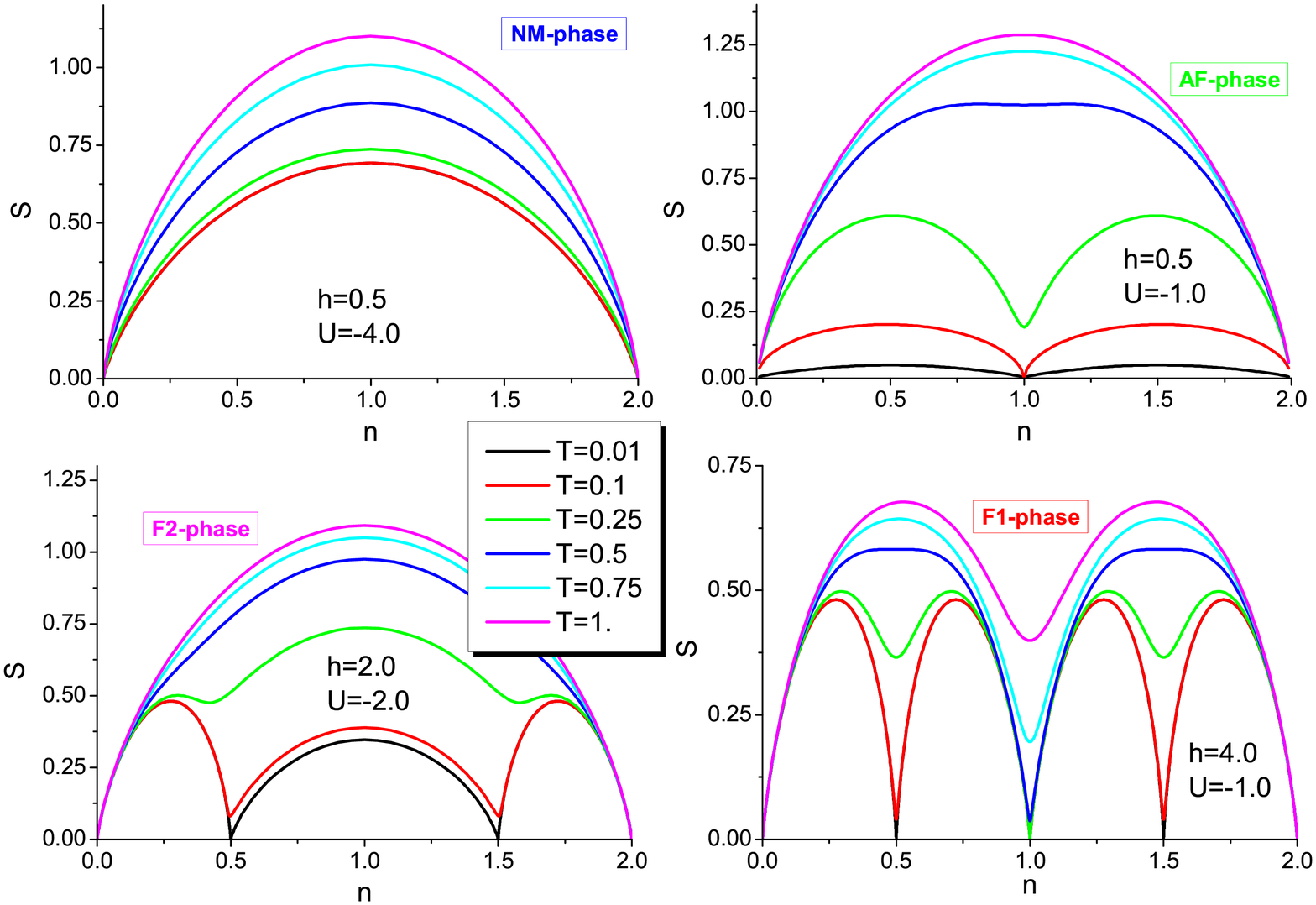}
\par\end{centering}
\caption{(Color online) Entropy as a function of the filling for different temperatures and four sets of the external parameters $U$ and $h$ corresponding to NM (top-left), AF (top-right), F1 (bottom-right) and F2 (bottom-left) phases observed at $J=-1$ and $T=0$.\label{fig:Entropy2}}
\end{figure}

\emph{Entropy}. The entropy can be calculated by means of two equivalent formulas:
\begin{equation}\label{eq:EntropyDef}
\small\begin{array}{ll}
  &S(T,n)=S(0,n)+\int_0^T\frac{C(T^\prime,n)}{T^\prime}dT^\prime\;,\\
  &S(T,n)=-\int_0^n\frac{\partial\mu(T,n^\prime)}{\partial T}dn^\prime\;.
\end{array}
\end{equation}
The first expression requires the knowledge of the entropy at zero temperature which can be computed by means of the formula $S(0,n)=k_B\ln(\Omega)$, where $k_B$ is the Boltzmann's constant and $\Omega$ is the degeneracy of the ground state.
Discarding few simple cases, it is generally not easy to compute $\Omega$ and, in the case when the degeneracy of the ground state is not known a priori, the only way to calculate the entropy is the use of the second equation in \eqref{eq:EntropyDef}. The study of the entropy plays a crucial role in the characterization and the identification of possible ordered phases in which we expect to have $S\rightarrow0$ in the thermodynamic limit. Filling and temperature dependence of the entropy for both $J=1$ and $J=-1$ cases are reported in Figs. \ref{fig:Entropy3}-\ref{fig:Entropy2}.

For $J=1$, as shown in Fig. \ref{fig:Entropy3} for different values of the filling, a finite ground state degeneracy is observed only in the NM-phase, as signalled by the value of the entropy in the $T\rightarrow0$ limit. In the F-phase, because of the presence of a spin ordering at $T=0$, the entropy goes to zero at any filling as one would expect. Importantly, as shown in Fig.\ref{fig:Entropy1}, when $U$ is close to $U_c$, $S$ rapidly decreases showing a zero-temperature jump in the limit of $U\rightarrow U_c$.

For the case $J=-1$, the filling dependence of the entropy is shown in Fig. \ref{fig:Entropy2} for different values of the external parameters $U$, $h$ and $T$, corresponding to the phases observed at $T=0$. As the temperature increases, the properties of the system at finite $T$ can be described as a superposition of an increasing number of different configurations. In contrast, at low temperatures the entropy as a function of filling decreases rapidly near those values of $n$ associated with charge and/or spin orderings. At these points, in fact, the state of the system can be uniquely described by a finite number of configurations (see Figs.\ref{fig:F_configuration}, \ref{fig:AF_configuration}, \ref{fig:F1_configuration}, and \ref{fig:F2_configuration} for F, AF, F1 and F2-phase respectively) whose contribution to the entropy is expected to vanish in the thermodynamic limit.

\section{Concluding Remarks}
In this paper, we have evidenced how the use of the Composite Operator Method, based on Green's function and equations of motion formalism, leads to the exact solution of the $t$-$U$-$J$ model in the narrow-band limit. In this limit the system can be described by a closed set of composite eigenoperators allowing for an exact solution that holds in principle for one, two and three dimensions. We have considered the cases of both ferro ($J>0$) and antiferro ($J<0$) couplings with an external longitudinal magnetic field. We have shown that a complete rigorous solution of the model does exist in the one-dimensional case in which Green's and correlation functions can be expressed in terms of four parameters, determined self-consistently. Within this framework, we have reported a detailed analysis of zero and finite temperature properties for both ferro and antiferro exchange coupling $J$. We have found that in the $T\rightarrow0$ limit the system
exhibits a very rich phase diagram characterized by long-range charge and spin orderings. In addition to the standard ferromagnetic and anti-ferromagnetic phases, that occur for $J=1$ and $J=-1$, respectively, in the case of antiferromagnetic coupling the competition between $J$, $U$ and $h$ results in the presence of two ferromagnetic phases. We have reported a detailed analysis of each phase in terms of all the relevant single-particle correlators (magnetization, double occupancy, density-density and spin-spin CFs), thermodynamic quantities (free energy, entropy) and density of states. Furthermore, we have also reported that the presence of zero-temperature long-range orderings clearly characterizes the temperature dependence of response functions such as specific heat, charge and spin susceptibilities, leading to the presence of low-temperature charge and/or spin excitations.

Summarizing, we have presented an exact and comprehensive study of the one-dimensional $t$-$U$-$J$-$h$ model, in the narrow-band limit. Our results show that this study is of substantial interest from many points of view. The model itself can be considered as a key-model for the study of phase transitions between different charge and spin orderings, generated by the competition among the different energy scales ($U$, $J$ and $h$). The present exact solution is an important test for various approximate approaches. Our exact solution of the model can be used as a starting point for a perturbative expansion in powers of the hopping term in which the exact analytical knowledge of the Green's function will be of crucial importance. This work is currently in progress. Furthermore, we also remark the importance of the model from the point of view of statistical mechanics, motivated by the presence of several phase transitions, bicritical and tricritical points and anomalous behaviors in all the system response functions near the phase boundaries. Moreover, because of the isomorphism discussed in Sec.\ref{SubSec:Isomorphism} for the one-dimensional case, the analytical framework developed in Sec.\ref{Sec:COM} also provides the exact solution for a two-leg ladder  spin system with effective inter-chain and intra-chain spin-spin interactions in the presence of an external magnetic field.

\appendix
\section{Algebraic relations, spectral densities, energy and normalization matrices}\label{AppendixA}

\subsection{$A_{1}^{(p)}$ coefficients}\label{AppendixA1}
By means of the algebraic relations (\ref{eq:AlgebraicRelations})
it is straightforward to show that the filed $n_{3}^{\alpha}(i)=\frac{1}{z}\sum_{z=1}^{z}n_{3}(i_{k})$,
where $i_{k}$ ($k=1,\dots,z$) are the first neighbor site of $\boldsymbol{i}$,
satisfies the following recursion formula:
\begin{eqnarray}
\left[n_{3}^{\alpha}(i)\right]^{2p-1} & = & \sum_{m=1}^{2d}A_{m}^{(p)}\left[n_{3}^{\alpha}(i)\right]^{2m-1}\;,\\
\left[n_{3}^{\alpha}(i)\right]^{2p} & = & \sum_{m=1}^{2d}A_{m}^{(p)}\left[n_{3}^{\alpha}(i)\right]^{2m}\;,
\end{eqnarray}
where $A_{m}^{(p)}$ are rational numbers, satisfying the sum rule:
\begin{equation}
\sum_{m=1}^{4d}A_{m}^{(p)}=1\;.
\end{equation}
For $1\leqslant p\leqslant m$, $A_{m}^{(p)}=\delta_{pm}$. For $p>m$ the expressions
of the coefficients $A_{m}^{(p)}$ depend on the coordination number
$z=2d$. For $z=2$ we have:
\begin{eqnarray}
A_{1}^{(p)} & = & \frac{1}{3}\left(2^{4-2p}-1\right)\;,\\
A_{2}^{(p)} & = & \frac{4}{3}\left(1-2^{2-2p}\right)\;.
\end{eqnarray}
In Tab.\eqref{tab:Am(p)_coefficients} we report some values for $z=2$
and $z=4$.

\begin{table}
\begin{centering}
\begin{tabular}{c|cc||cccc}
 & \multicolumn{2}{c||}{$(z=2)$} & \multicolumn{4}{c}{$(z=4)$}\tabularnewline
$p$ & $A_{1}^{(p)}$ & $A_{2}^{(p)}$ & $A_{1}^{(p)}$ & $A_{2}^{(p)}$ & $A_{3}^{(p)}$ & $A_{4}^{(p)}$\tabularnewline
\hline
$1$ & $1$ & $0$ & $1$ & $0$ & $0$ & $0$\tabularnewline
$2$ & $0$ & $1$ & $0$ & $1$ & $0$ & $0$\tabularnewline
$3$ & $-\frac{1}{4}$ & $\frac{5}{4}$ & $0$ & $0$ & $1$ & $0$\tabularnewline
$4$ & $-\frac{5}{16}$ & $\frac{21}{16}$ & $0$ & $0$ & $0$ & $1$\tabularnewline
$5$ & $-\frac{21}{64}$ & $\frac{85}{64}$ & $-\frac{9}{1024}$ & $\frac{205}{1024}$ & $-\frac{273}{256}$ & $\frac{15}{8}$\tabularnewline
$6$ & $-\frac{85}{256}$ & $\frac{341}{256}$ & $-\frac{135}{8192}$ & $\frac{3003}{8192}$ & $-\frac{3685}{2048}$ & $\frac{627}{256}$\tabularnewline
\end{tabular}\caption{Values of the coefficients $A_{m}^{(p)}$ for $p=1,\dots,6$ and $z=2,4$.
\label{tab:Am(p)_coefficients}}
\par\end{centering}
\end{table}

\subsection{Energy matrices}\label{AppendixA2}
For $z=2$ we have: \small
\begin{equation}
\varepsilon_{\sigma}^{(\xi)}=\left(\begin{array}{ccccc}
-\mu-\sigma & -2\sigma J & 0 & 0 & 0\\
0 & -\mu-\sigma h & -2\sigma J & 0 & 0\\
0 & 0 & -\mu-\sigma h & -2\sigma J & 0\\
0 & 0 & 0 & -\mu-\sigma h & -2\sigma J\\
0 & \sigma J/2 & 0 & -5\sigma J/2 & -\mu-\sigma h
\end{array}\right)\;,\label{eq:EnergyMatrix_Xi}
\end{equation}
\begin{eqnarray}
\varepsilon_{\sigma}^{(\eta)} & = & \varepsilon_{\sigma}^{(\xi)}+U\cdot\boldsymbol{I}_{5\times5}\;,\label{eq:EnergyMatrix_Eta}
\end{eqnarray}
\normalsize where $\sigma=(\uparrow,\downarrow)$ or $\sigma=(+1,-1)$
and $\boldsymbol{I}_{5\times5}$ is the $5\times5$ identity matrix.
The eigenvalues of these matrices are:
\begin{eqnarray}
E_{n}^{(\xi,\sigma)} & = & \left(\begin{array}{c}
-\mu-\sigma\\
-\mu-\sigma h-2\sigma J\\
-\mu-\sigma h-\sigma J\\
-\mu-\sigma h+\sigma J\\
-\mu-\sigma h+2\sigma J
\end{array}\right)\;,\\
E_{n}^{(\eta,\sigma)} & = & \left(\begin{array}{c}
U-\mu-\sigma\\
U-\mu-\sigma h-2\sigma J\\
U-\mu-\sigma h-\sigma J\\
U-\mu-\sigma h+\sigma J\\
U-\mu-\sigma h+2\sigma J
\end{array}\right)\;.
\end{eqnarray}

\subsection{The spectral density matrices}\label{AppendixA3}
The matrices $\Omega_{\sigma}^{(s)}$ are defined as the matrices
of rank $(2z+1)\times(2z+1)$ whose columns are the eigenvectors of
the energy matrices $\varepsilon_{\sigma}^{(s)}$. It is worth noting
that the $\Omega_{\sigma}^{(s)}$ do not depend on $\sigma$ and $s$.
For $z=2$, from \eqref{eq:EnergyMatrix_Xi} and \eqref{eq:EnergyMatrix_Eta}
we have:
\begin{equation}
\Omega_{\sigma}^{(s)}=\left(\begin{array}{ccccc}
1 & 1 & 2^{4} & 2^{4} & 1\\
0 & 1 & 2^{3} & -2^{3} & -1\\
0 & 1 & 2^{2} & 2^{2} & 1\\
0 & 1 & 2 & -2 & -1\\
0 & 1 & 1 & 1 & 1
\end{array}\right)\;.
\end{equation}
The spectral density matrices $\rho_{\sigma}^{(s,n)}$ are calculated
by means of \eqref{eq:SpectralDensityMatrix} in terms of the matrices
$\Omega_{\sigma}^{(s)}$ and $I_{\sigma}^{(s)}$. The relevant elements
of $\rho_{\sigma}^{(s,n)}$ for $z=2$ have the expressions:
\small
\begin{eqnarray}
\rho_{\sigma;1,1}^{(s,1)} & = & I_{\sigma;1,1}-5I_{\sigma;1,3}^{(s)}+4I_{\sigma;1,5}^{(s)}\;,\\
\rho_{\sigma;1,1}^{(s,2)} & = & \frac{\left(-I_{\sigma;1,2}^{(s)}-I_{\sigma;1,3}^{(s)}+4I_{\sigma;1,4}^{(s)}+4I_{\sigma;1,5}^{(s)}\right)}{6}\;,\\
\rho_{\sigma;1,1}^{(s,3)} & = & \frac{4\left(I_{\sigma;1,2}^{(s)}+2I_{\sigma;1,3}^{(s)}-I_{\sigma;1,4}^{(s)}-2I_{\sigma;1,5}^{(s)}\right)}{3}\;,\\
\rho_{\sigma;1,1}^{(s,4)} & = & \frac{4\left(-I_{\sigma;1,2}^{(s)}+2I_{\sigma;1,3}^{(s)}-I_{\sigma;1,4}^{(s)}-2I_{\sigma;1,5}^{(s)}\right)}{3}\;,\\
\rho_{\sigma;1,1}^{(s,5)} & = & \frac{\left(I_{\sigma;1,2}^{(s)}-I_{\sigma;1,3}^{(s)}-4I_{\sigma;1,4}^{(s)}+4I_{\sigma;1,5}^{(s)}\right)}{6}\;.
\end{eqnarray}
\normalsize

\subsection{The normalization matrix}\label{AppendixA4}
We recall that the normalization matrix is defined as $I_{\sigma}^{(s)}=\langle\left\{ \psi_{\sigma}^{(s)}(\boldsymbol{i},t),\psi_{\sigma}^{(s)\dagger}(\boldsymbol{i},t)\right\} \rangle$.
By considering that this matrix is Hermitian and by recalling the
recursion rule \eqref{eq:RecursionRules}, it is easy to see that
all the matrix elements can be expressed in terms of the elements
of the first raw. By means of the algebra, these elements have the
expressions:
\begin{equation}
\begin{array}{l}
I_{\sigma;1,p}^{(\xi)}=\kappa^{(p)}-\lambda_{-\sigma}^{(p)}\\
I_{\sigma;1,p}^{(\eta)}=\lambda_{-\sigma}^{(p)}
\end{array}\;(p=1,\dots,2z+1)\;,
\end{equation}
where:
\begin{equation}
\kappa^{(p)}=\langle\left[n_{3}^{\alpha}(i)\right]^{p-1}\rangle\;,\;\lambda_{\sigma}^{(p)}=\langle n_{\sigma}\left[n_{3}^{\alpha}(i)\right]^{p-1}\rangle\;.
\end{equation}
For $z=2$ the explicit expression of $I_{\sigma}^{(s)}$ is:
\begin{equation}
I_{\sigma}^{(s)}=\left(\begin{array}{ccccc}
I_{\sigma;1,1}^{(s)} & I_{\sigma;1,2}^{(s)} & I_{\sigma;1,3}^{(s)} & I_{\sigma;1,4}^{(s)} & I_{\sigma;1,5}^{(s)}\\
I_{\sigma;2,1}^{(s)} & I_{\sigma;2,2}^{(s)} & I_{\sigma;2,3}^{(s)} & I_{\sigma;2,4}^{(s)} & I_{\sigma;25}^{(s)}\\
I_{\sigma;3,1}^{(s)} & I_{\sigma;3,2}^{(s)} & I_{\sigma;3,3}^{(s)} & I_{\sigma;3,4}^{(s)} & I_{\sigma;3,5}^{(s)}\\
I_{\sigma;4,1}^{(s)} & I_{\sigma;4,2}^{(s)} & I_{\sigma;4,3}^{(s)} & I_{\sigma;5,4}^{(s)} & I_{\sigma;4,5}^{(s)}\\
I_{\sigma;5,1}^{(s)} & I_{\sigma;5,2}^{(s)} & I_{\sigma;5,3}^{(s)} & I_{\sigma;5,4}^{(s)} & I_{\sigma;5,5}^{(s)}
\end{array}\right)\;,
\end{equation}
with:
\begin{equation}
\begin{array}{l}
I_{\sigma;2,5}^{(s)}=-\frac{1}{4}I_{\sigma;1,2}^{(s)}+\frac{5}{4}I_{\sigma;1,4}^{(s)}\;,\\
I_{\sigma;3,5}^{(s)}=-\frac{1}{4}I_{\sigma;1,3}^{(s)}+\frac{5}{4}I_{\sigma;1,5}^{(s)}\;,\\
I_{\sigma;4,5}^{(s)}=-\frac{5}{16}I_{\sigma;1,2}^{(s)}+\frac{21}{16}I_{\sigma;1,4}^{(s)}\;,\\
I_{\sigma;5,5}^{(s)}=-\frac{5}{16}I_{\sigma;1,3}^{(s)}+\frac{21}{16}I_{\sigma;1,5}^{(s)}\;.
\end{array}
\end{equation}

\section{$H_{0}$-representation}\label{AppendixB}
By considering a one-dimensional system, let us write the Hamiltonian
\eqref{eq:Hubbard_Intersite+StrCoupl+Nn} as:
\begin{equation}
\begin{array}{l}
H=H_{0}^{(i)}+H_{I}^{(i)}\;,\\
H_{I}^{(i)}=-2Jn_{3}(i)n_{3}^{\alpha}(i)\;,
\end{array}
\end{equation}
where $i$ is a site of the infinite chain. Since $H_{0}^{(i)}$
and $H_{I}^{(i)}$ commute, for any operator $O$ we can write:
\begin{equation}
\media O=\frac{\media{Oe^{-\beta H_{I}(i)}}_{i}}{\media{e^{-\beta H_{I}(i)}}_{i}}\;,
\end{equation}
where the notation $\media{\dots}_{i}$ denotes the thermal average
with respect tot he reduce Hamiltonian $H_{0}^{(i)}$:
\begin{equation}
\media{\dots}_{i}=\frac{Tr\left\{ \dots e^{-\beta H_{0}^{(i)}}\right\} }{Tr\left\{ e^{-\beta H_{0}^{(i)}}\right\} }\;.
\end{equation}
By using the algebraic properties \eqref{eq:AlgebraicRelations} we
have:
\begin{equation}
e^{-\beta H_{I}^{(i)}}=1+an_{3}(i)R(p)+\left[n(i)-2D(i)\right]Q(p)\;,\label{eq:Exp_betaH_I}
\end{equation}
where:
\begin{eqnarray}
R(p) & = & n_{3}(i_{1})+n_{3}(i_{2})+bn_{3}(i_{2})\left[n(i_{1})-2D(i_{1})\right]+\nonumber \\
 &  & +bn_{3}(i_{1})\left[n(i_{2})-2D(i_{2})\right]\;,\\
Q(p) & = & b\left[n(i_{1})-2D(i_{1})+n(i_{2})-2D(i_{2})\right]+\nonumber \\
 &  & +a^{2}n_{3}(i_{1})n_{3}(i_{2})+b^{2}\left[n(i_{1})-2D(i_{1})\right]+\nonumber \\
 &  & +\left[n(i_{2})-2D(i_{2})\right]\;,
\end{eqnarray}
with:
\begin{equation}
\begin{array}{l}
a=\sinh\left(\beta J\right)\\
b=\cosh\left(\beta J\right)-1
\end{array}\;,
\end{equation}
$i_{p}$ ($p=1,2$) are the first neighbor site of $i$. Now, we observe
that $H_{0}$ describes a system where the original lattice is divided
in two non interacting sub-lattices (the chains to the left and right
of the site $i$). Then, the correlation functions which relates two
sites belonging to different sub-lattices can be written as the product
of the two correlation functions:
\begin{equation}
\media{a(j)b(m)}_{0}=\media{a(j)}_{0}\media{b(m)}_{0}\;,\label{eq:PropertyAverage}
\end{equation}
for $j$ and $m$ belonging to different sub-lattices. By using these
properties and the algebraic relations \eqref{eq:AlgebraicRelations}
we have:
\begin{eqnarray}
\langle e^{-\beta H_{I}^{(i)}}\rangle & = & 1+2aG_{2}X_{2}\left[1+b(X_{1}-2X_{3})\right]+\nonumber \\
 &  & +(G_{1}-2G_{3})\bigr[a^{2}X_{2}^{2}+2b\left(X_{1}-2X_{3}\right)+\nonumber \\
 &  & +b^{2}(X_{1}-2X_{3})^{2}\bigr]\;,
\end{eqnarray}
where we define:
\begin{equation}
\begin{array}{ll}
G_{1}\equiv\media{n(i)}_{0} & \quad X_{1}\equiv\media{n(i+1)}_{0}=\media{n(i-1)}_{0}\\
G_{2}\equiv\media{n_{3}(i)}_{0} & \quad X_{2}\equiv\media{n_{3}(i+1)}_{0}=\media{n_{3}(i-1)}_{0}\\
G_{3}\equiv\media{D(i)}_{0} & \quad X_{3}\equiv\media{D(i+1)}_{0}=\media{D(i-1)}_{0}
\end{array}\;.
\end{equation}
In order to calculate the quantities $\media{n(i)}_{i}$, $\media{n_{3}(i)}_{i}$,
$\media{D(i)}_{i}$ let us consider the retarded Green's functions:
\begin{equation}
\begin{cases}
G_{\sigma}^{(\xi,0)}(t-t')=\media{R\left[\xi_{\sigma}(i,t)\xi_{\sigma}^{\dagger}(i,t')\right]}_{i}\\
G_{\sigma}^{(\eta,0)}(t-t')=\media{R\left[\eta_{\sigma}(i,t)\eta_{\sigma}^{\dagger}(i,t')\right]}_{i}
\end{cases}\;.
\end{equation}
In the $H_{0}$-representation the Hubbard operators satisfy the equations
of motion:
\begin{equation}
\begin{cases}
i\frac{\partial}{\partial t}\xi(i)=\left[\xi(i),\mathcal{H}_{0}(i)\right]=-\left(\mu+h\sigma_{3}\right)\xi(i)\\
i\frac{\partial}{\partial t}\eta(i)=\left[\eta(i),\mathcal{H}_{0}(i)\right]=-\left(\mu+h\sigma_{3}-U\right)\eta(i)
\end{cases}\;.
\end{equation}
By means of these equations we obtain:
\begin{equation}
\begin{cases}
G_{\sigma}^{(\xi,0)}(\omega)=\frac{1-\media{n_{\bar{\sigma}}(i)}_{0}}{\omega+\mu+\sigma h+i\delta}\\
G_{\sigma}^{(\eta,0)}(\omega)=\frac{\media{n_{\bar{\sigma}}(i)}_{0}}{\omega+\mu+\sigma h-U+i\delta}
\end{cases}\;,\label{eq:GFs}
\end{equation}
and for the equal-time correlation functions:
\begin{equation}
\begin{cases}
C_{\sigma}^{(\xi,0)}=\media{\xi_{\sigma}(i)\xi_{\sigma}^{\dagger}(i)}_{0}=\frac{1-\media{n_{\bar{\sigma}}(i)}_{0}}{1+e^{\beta\left(\mu+\sigma h\right)}}\\
C_{\sigma}^{(\eta,0)}=\media{\eta_{\sigma}(i)\eta_{\sigma}^{\dagger}(i)}_{0}=\frac{\media{n_{\bar{\sigma}}(i)}_{0}}{1+e^{\beta\left(\mu+\sigma h-U\right)}}
\end{cases}\;.\label{eq:CFs}
\end{equation}
Recalling the algebraic relations:
\begin{equation}
\begin{array}{l}
\xi_{\sigma}\xi_{\sigma}^{\dagger}+\eta_{\sigma}\eta_{\sigma}^{\dagger}=1-n_{\sigma}\;,\\
\eta_{\sigma}\eta_{\sigma}^{\dagger}=n_{-\sigma}-n_{\uparrow}n_{\downarrow}\;,
\end{array}
\end{equation}
we obtain from \eqref{eq:CFs}:
\begin{equation}\label{eq:GiCoefficients}
\begin{array}{l}
G_{1}\equiv\media{n(i)}_{i}=\frac{e^{\beta\mu}\left(1+e^{2\beta h}+e^{\beta(\mu+h-U)}\right)}{e^{\beta h}+e^{\beta\mu}+e^{\beta(\mu+2h)}+e^{\beta(2\mu+h-U)}}\\
G_{2}\equiv\media{n_{3}(i)}_{i}=\frac{e^{\beta(\mu+2h)}-e^{\beta\mu}}{e^{\beta h}+e^{\beta\mu}+e^{\beta(\mu+2h)}+e^{\beta(2\mu+h-U)}}\;.\\
G_{3}\equiv\media{D(i)}_{i}=\frac{e^{\beta(2\mu+h-U)}}{e^{\beta h}+e^{\beta\mu}+e^{\beta(\mu+2h)}+e^{\beta(2\mu+h-U)}}
\end{array}
\end{equation}
For a homogeneous solution, the internal parameters $X_{1}$, $X_{2}$,
$X_{3}$ are determined by means of the self-consistent equations:
\begin{equation}
\begin{cases}
\media{n(i)}=\media{n(i_{p})}\\
\media{n_{3}(i)}=\media{n_{3}(i_{p})}\\
\media{D(i)}=\media{D(i_{p})}
\end{cases}\;.\label{eq:TranslationalInvarianceRequest}
\end{equation}
By using \eqref{eq:Exp_betaH_I} and the property \eqref{eq:PropertyAverage},
it is easy to show that: \small
\begin{eqnarray}
\media{n(i)e^{-\beta\mathcal{H}_{I}(i)}}_{i} & = & G_{1}+2aG_{2}X_{2}\left[1+b\left(X_{1}-2X_{3}\right)\right]+\nonumber \\
 &  & +\left(G_{1}-2G_{3}\right)\cdot\Bigl[a^{2}X_{2}^{2}+2b\cdot\nonumber \\
 &  & \cdot\left(X_{1}-2X_{3}\right)+b^{2}\left(X_{1}-2X_{3}\right)^{2}\Bigr]\label{eq:n(i)0}
\end{eqnarray}
\begin{eqnarray}
\media{n(i_{1})e^{-\beta\mathcal{H}_{I}(i)}}_{i} & = & X_{1}+aG_{2}X_{2}\left[1+X_{1}+2b\left(X_{1}-2X_{3}\right)\right]+\nonumber \\
 &  & +\left(G_{1}-2G_{3}\right)\Bigr[a^{2}X_{2}^{2}+b\left(1+X_{1}\right)\cdot\nonumber \\
 &  & \cdot\left(X_{1}-2X_{3}\right)+b^{2}\left(X_{1}-2X_{3}\right)^{2}\Bigr]\label{eq:n(i+1)0}
\end{eqnarray}
\begin{eqnarray}
\media{n_{3}(i)e^{-\beta\mathcal{H}_{I}(i)}}_{i} & = & G_{2}+2a\left(G_{1}-2G_{3}\right)X_{2}\Bigr[1+b\left(X_1\right.\nonumber \\
 &  & \left.-2X_{3}\right)\Bigr]+G_{2}\Bigr[a^{2}X_{2}^{2}+2b\left(X_{1}-2X_{3}\right)\nonumber\\
 &  & +b^{2}\left(X_{1}-2X_{3}\right)^{2}\Bigr]\label{eq:n3(i)0}
\end{eqnarray}
\begin{eqnarray}
\media{n_{3}(i_{1})e^{-\beta\mathcal{H}_{I}(i)}}_{i} & = & X_{2}+\left(G_{1}-2G_{3}\right)X_{2}\Bigl[b+\left(a^{2}+b+b^{2}\right)\nonumber \\
 &  & \cdot\left(X_{1}-2X_{3}\right)\Bigr]+aG_{2}\Bigl[X_{2}^{2}\left(1+b\right)+\nonumber\\
 &  & +\left(X_{1}-2X_{3}\right)+b\left(X_{1}-2X_{3}\right)^{2}\Bigr]\label{eq:n3(i+1)0} \\
\media{D(i)e^{-\beta\mathcal{H}_{I}(i)}}_{i} & = & G_{3}\label{eq:D(i)0}
\end{eqnarray}
\begin{eqnarray}
\media{D(i+1)e^{-\beta\mathcal{H}_{I}(i)}}_{i} & = & X_{3}\Bigl[1+aG_{2}X_{2}+b\left(G_{1}-2G_{3}\right)\nonumber \\
 &  & \left(X_{1}-2X_{3}\right)\Bigr]\;.\label{eq:D(i+1)0}
\end{eqnarray}
\normalsize Putting \eqref{eq:n(i)0}-\eqref{eq:D(i+1)0} into \eqref{eq:TranslationalInvarianceRequest}
we obtain the equations:
\begin{eqnarray}
F_{1} & = & G_{1}-X_{1}+aG_{2}X_{2}(1-X_{1})+b(G_{1}-2G_{3})\nonumber \\
 &  & (1-X_{1})(X_{1}-2X_{3})=0\;,\label{eq:SCeq1}\\
F_{2} & = & G_{2}-X_{2}+(G_{1}-2G_{3})X_{2}\nonumber \\
 &  & \left\{ 2a-b+-\left[b+(a-b)^{2}\right](X_{1}-2X_{3})\right\} +\nonumber \\
 &  & +G_{2}\bigr\{ aX_{2}^{2}(a-b-1)+(X_{1}-2X_{3})\nonumber \\
 &  & \left[2b-a+b(b-a)(X_{1}-2X_{3})\right]\bigl\}=0\;,\label{eq:SCeq2}\\
F_{3} & = & G_{3}-X_{3}[1+aG_{2}X_{2}+b(G_{1}-2G_{3})\nonumber \\
 &  & (X_{1}-2X_{3})]=0\;.\label{eq:SCeq3}
\end{eqnarray}
The parameters $\kappa^{(p)}$ and $\lambda_{\sigma}^{(p)}$ can be
straightforwardly calculated. For example:
\begin{eqnarray}
\kappa^{(1)} & = & 1\;,\\
\lambda_{\sigma}^{(1)} & = & \frac{1}{2\media{e^{-\beta H_{I}^{(i)}}}_{i}}\Bigl\{\left(G_{1}+\sigma G_{2}\right)+2a\Bigl[G_{2}+\nonumber \\
 &  & +\sigma\left(G_{1}-2G_{3}\right)\Bigr]X_{2}\left[1+b\left(X_{1}-2X_{3}\right)\right]+\nonumber \\
 &  & +\left(G_{1}-2G_{3}+\sigma G_{2}\right)\Bigl[a^{2}X_{2}^{2}+\nonumber \\
 &  & +2b\left(X_{1}-2X_{3}\right)+b^{2}\left(X_{1}-2X_{3}\right)^{2}\Bigr]\Bigr\}\;,
\end{eqnarray}
\begin{eqnarray}
\kappa^{(2)} & = & \frac{1}{2\media{e^{-\beta H_{I}^{(i)}}}_{i}}\Bigl\{ X_{2}+aG_{2}\bigl\{\left(1+b\right)X_{2}^{2}+\nonumber \\
 &  & +\left[1+b\left(X_{1}-2X_{3}\right)\right]\left(X_{1}-2X_{3}\right)\bigr\}+\nonumber \\
 &  & +\left(G_{1}-2G_{3}\right)X_{2}\Bigl[b+\nonumber \\
 &  & +\left(a^{2}+b+b^{2}\right)\left(X_{1}-2X_{3}\right)\Bigr]\Bigr\}\;,
\end{eqnarray}
\begin{eqnarray}
\lambda_{\sigma}^{(2)} & = & \frac{1}{2\media{e^{-\beta H_{I}^{(i)}}}_{i}}\Bigl\{\left(G_{1}+\sigma G_{2}\right)X_{2}+\nonumber \\
 &  & +a\left[G_{2}+\sigma\left(G_{1}-2G_{3}\right)\right]\bigr\{\left(1+b\right)X_{2}^{2}+\nonumber \\
 &  & +\left[1+b\left(X_{1}-2X_{3}\right)\right]\left(X_{1}-2X_{3}\right)\bigr\}+\nonumber \\
 &  & +\left(G_{1}-2G_{3}+\sigma G_{2}\right)\Bigl[bX_{2}+\nonumber \\
 &  & \left(a^{2}+b+b^{2}\right)X_{2}\left(X_{1}-2X_{3}\right)\Bigr]\;,
\end{eqnarray}
\begin{eqnarray}
\kappa^{(3)} & = & \frac{1}{2\media{e^{-\beta H_{I}^{(i)}}}_{i}}\Bigl\{ X_{1}-2X_{3}+X_{2}^{2}+G_{2}\left(f_{1}+2f_{3}\right)+\nonumber \\
 &  & +\left(G_{1}-2G_{3}\right)\left(f_{2}+f_{4}\right)\Bigr\}\;,
\end{eqnarray}
\begin{eqnarray}
\lambda_{\sigma}^{(3)} & = & \frac{1}{4\media{e^{-\beta H_{I}^{(i)}}}}\Bigr\{\left(G_{1}+\sigma G_{2}\right)\left(X_{1}-2X_{3}+X_{2}^{2}\right)+\nonumber \\
 &  & +\left[G_{2}+\sigma\left(G_{1}-2G_{3}\right)\right]\left(f_{1}+2f_{3}\right)+\nonumber \\
 &  & +\left(G_{1}-2G_{3}+\sigma G_{2}\right)\left(f_{2}+f_{4}\right)\Bigr\}\;.
\end{eqnarray}


\end{document}